\PassOptionsToPackage{colorlinks,breaklinks}{hyperref}  
\documentclass[nopacs,epjc3]{svjour3}

\usepackage{a4p}
\usepackage{cite}
\usepackage{graphicx}
\usepackage{subfigure}
\usepackage{amsmath}
\usepackage{atlasphysics}
\usepackage{rotating}
\usepackage{multirow}

\usepackage[switch]{lineno}

\usepackage[preprint]{atlascover}
\AtlasTitle{
Search for new phenomena in final states with an energetic jet and  large missing transverse momentum 
in $pp$ collisions at $\sqrt{s}=8~$TeV with the ATLAS detector
}
\PreprintIdNumber{CERN-PH-EP-2014-299}  
\AtlasAbstract{
  Results of a search for new phenomena in
  final states with an energetic jet and large missing transverse momentum are reported. The
  search uses 20.3~fb${}^{-1}$ of $\sqrt{s}=8~$TeV data collected in
  2012 with the ATLAS detector at the LHC.  
Events are required to have 
 at least one jet with $\pt > 120$~GeV and  no leptons.  Nine signal regions 
are considered with increasing missing transverse momentum  
requirements between $\met >150$~GeV and 
$\met > 700$~GeV. 
Good agreement is observed between
  the number of events in data and Standard Model expectations.
The results are translated into exclusion limits 
on models with large extra spatial dimensions,   
pair production of weakly interacting dark matter 
candidates, and production of very light gravitinos in 
a gauge-mediated supersymmetric model.
In addition, limits on the 
production of an invisibly decaying Higgs-like boson 
leading to similar  topologies in the final state are presented.
}
\AtlasJournal{Eur. Phys. J. C}  

\begin{document}

\title{
Search for new phenomena in final states with an energetic jet and large missing transverse momentum 
 in $pp$ collisions at $\sqrt{s}=8~$TeV with the ATLAS detector}
\titlerunning{Search for new phenomena in monojet-like final states}
\author{ATLAS Collaboration}
\institute{CERN, 1211 Geneva 23, Switzerland \\ \email{atlas.publications@cern.ch}}
\date{Received: / Accepted:}
\maketitle

\begin{abstract}

  Results of a search for new phenomena in
  final states with an energetic jet and large missing transverse momentum are reported. The
  search uses 20.3~fb${}^{-1}$ of $\sqrt{s}=8~$TeV data collected in
  2012 with the ATLAS detector at the LHC.
Events are required to have
 at least one jet with $\pt > 120$~GeV and  no leptons.  Nine signal regions
are considered with increasing missing transverse momentum
requirements between $\met >150$~GeV and
$\met > 700$~GeV.
Good agreement is observed between
  the number of events in data and Standard Model expectations.
The results are translated into exclusion limits
on models with either large extra spatial dimensions,  
pair production of weakly interacting dark matter
candidates, or production of very light gravitinos in
a gauge-mediated supersymmetric model. 
In addition, limits on the
production of an invisibly decaying Higgs-like boson
leading to similar topologies in the final state are presented.
\end{abstract}

\newcommand{\pho}{\phantom{0}}
\newcommand{\bslash}{\ensuremath{\backslash}}
\newcommand{\BibTeX}{{\sc Bib\TeX}}
\newcommand{\ptj}{p_{\rm T}^{\rm jet}}
\newcommand{\ptjet}{p_{\rm  T}}
\newcommand{\ptmi}{{\vec{p}}_{\rm T}^{\rm \ miss}}
\newcommand{\ptjetem}{p_{\rm T}^{\rm jet,em}}
\newcommand{\etajet}{\eta^{\rm jet}}
\newcommand{\phijet}{\phi}
\newcommand{\rapjet}{y}
\newcommand{\pthat}{\hat{p}_{\rm  T}}
\newcommand{\akt}{\hbox{anti-${k_t}$} }
\newcommand{\njet}{N_{\rm jet}}
\newcommand{\pttrk}{\pt^{\rm \ track}}
\newcommand{\etatrk}{\eta^{\rm \ track}}
\newcommand{\ttb}{t\bar{t}}
\newcommand{\zee}{Z/\gamma^*(\rightarrow e^+e^-)}
\newcommand{\zmm}{Z/\gamma^*(\rightarrow \mu^+\mu^-)}
\newcommand{\znn}{Z(\rightarrow \nu \bar{\nu})}
\newcommand{\ztt}{Z/\gamma^*(\rightarrow \tau^+ \tau^-)}
\newcommand{\zll}{Z/\gamma^*(\rightarrow \ell^+ \ell^-)}
\newcommand{\wen}{W(\rightarrow e \nu)}
\newcommand{\wmn}{W(\rightarrow \mu \nu)}
\newcommand{\wtn}{W(\rightarrow \tau \nu)}
\newcommand{\wln}{W(\rightarrow \ell \nu)}
\newcommand{\ete}{E_{\rm T}^{e}}
\newcommand{\etae}{\eta^{e}}
\newcommand{\ptm}{p_{\rm T}^{\mu}}
\newcommand{\etam}{\eta^{\mu}}
\newcommand{\mee}{M_{e^+ e^-}}
\newcommand{\mmm}{M_{\mu^+ \mu^-}}
\newcommand{\mchi}{m_{\chi}}
\newcommand{\gra}{\tilde{G}}
\newcommand{\mgra}{m_{\gra}}
\newcommand{\glu}{\tilde{g}}
\newcommand{\mglu}{m_{\glu}}
\newcommand{\squ}{\tilde{q}}
\newcommand{\msqu}{m_{\squ}}
\newcommand{\msqgl}{m_{\squ/\glu}}
\newcommand{\mtop}{m_{t}}
\newcommand{\mb}{m_{b}}
\newcommand{\mHH}{m_{H}}
\newcommand{\ptH}{p_{\rm T}^{H}}

\newcommand{\light}{u}
\newcommand{\charm}{c}
\newcommand{\bottom}{b}
\newcommand{\pu}{P_{\light}}
\newcommand{\pc}{P_{\charm}}
\newcommand{\pb}{P_{\bottom}}
\newcommand{\ISR}{ISR}

\newcommand\MDM{\ensuremath{{m}_\chi}}
\newcommand\mDM{\ensuremath{{m}_\chi}}
\newcommand\MS{\ensuremath{{M}_\star}}
\newcommand\MSexp{\ensuremath{\MS^\mathrm{exp}}}
\newcommand\MSvalid{\ensuremath{\MS^\mathrm{valid}}}
\newcommand\MMedOp[1]{\ensuremath{\MMed^{\mbox{#1}}}}
\newcommand\WMed{\ensuremath{{W}_\mathrm{med}}}
\newcommand\MMed{\ensuremath{{M}_\mathrm{med}}}

\def\chaj{\ensuremath{\mathchoice%
      {\displaystyle\raise.4ex\hbox{$\displaystyle\tilde\chi^\pm_j$}}%
         {\textstyle\raise.4ex\hbox{$\textstyle\tilde\chi^\pm_j$}}%
       {\scriptstyle\raise.3ex\hbox{$\scriptstyle\tilde\chi^\pm_j$}}%
 {\scriptscriptstyle\raise.3ex\hbox{$\scriptscriptstyle\tilde\chi^\pm_j$}}}}

\def\neuj{\ensuremath{\mathchoice%
      {\displaystyle\raise.4ex\hbox{$\displaystyle\tilde\chi^0_j$}}%
         {\textstyle\raise.4ex\hbox{$\textstyle\tilde\chi^0_j$}}%
       {\scriptstyle\raise.3ex\hbox{$\scriptstyle\tilde\chi^0_j$}}%
 {\scriptscriptstyle\raise.3ex\hbox{$\scriptscriptstyle\tilde\chi^0_j$}}}}


\section{Introduction}
\label{sec:intro}

Events with an energetic jet and large missing transverse momentum in
the final state constitute a clean and distinctive signature in
searches for new physics beyond the Standard Model (SM) at colliders.  
Such signatures are referred to as monojet-like in this paper.
In particular, monojet-like (as well as   
monophoton and  mono-$W/Z$) final states have been
studied~\cite{Abbiendi:2000hh,Heister:2002ut,Achard:2003tx,Abdallah:2003np,Abazov:2008kp,Aaltonen:2008hh,Aaltonen:2012jb,Chatrchyan:2011nd,Chatrchyan:2012me,Chatrchyan:2012tea,Aad:2011xw,ATLAS:2012ky,Aad:2012fw,Khachatryan:2014rra,Affolder:2000ef,Aad:2014nra,Aad:2014vea,Aad:2014vka,Aad:2013oja,Khachatryan:2014tva}
in the context of searches for supersymmetry (SUSY), large extra
spatial dimensions (LED),  
and the search for weakly interacting massive
particles (WIMPs) as candidates for dark matter (DM). 


The Arkani-Hamed, Dimopoulos, and Dvali (ADD) model for LED~\cite{ArkaniHamed:1998rs}     
explains the large difference between the electroweak unification scale at $O(10^2$)~GeV and 
the Planck scale   $M_{\rm{Pl}} \sim O(10^{19}$)~GeV     
by postulating the presence of $n$ extra spatial dimensions of size $R$,  
and defining a fundamental Planck scale  in $4+n$ dimensions, $M_D$, given by 
${M_{\rm{Pl}}}^2 \sim {M_D}^{2+n}R^n$. An appropriate 
choice of $R$ for a given $n$ yields a value of $M_D$ at  
the electroweak scale. The extra spatial dimensions are compactified,
 resulting in a Kaluza--Klein tower of massive graviton modes.
If produced in high-energy collisions in association with an energetic jet,
these graviton modes escape detection leading to a monojet-like signature in the final state.

\def\ptmiss{\ensuremath{\mathbf{p}_{\mathrm{T}}^{\mathrm{miss}}}}

A non-baryonic DM component in the universe is commonly used to
explain a range of astrophysical measurements~(see, for example,
Ref.~\cite{Bertone:2004pz} for a review).  Since none of the
known SM particles are adequate DM candidates, the existence of a new
particle is often hypothesized.
Weakly interacting massive particles 
are one such class of particle
candidates that can be searched for at the
LHC~\cite{Steigman:1984ac}. They are expected to couple to SM
particles through a generic weak interaction, which could be the 
weak interaction of the SM or a new type of interaction. Such a new
particle would result in the correct relic density values for
non-relativistic matter in the early universe~\cite{Kolb:1990vq}, as
measured by the PLANCK~\cite{Ade:2013zuv} and WMAP~\cite{WMAP9}
satellites, if its mass is between a few GeV and a TeV
and if it has electroweak-scale interaction cross sections. Many new
particle physics models such as SUSY~\cite{Miyazawa:1966,Ramond:1971gb,Golfand:1971iw,Neveu:1971rx,Neveu:1971iv,Gervais:1971ji,Volkov:1973ix,Wess:1973kz,Wess:1974tw}  
also predict WIMPs.

\begin{figure}[h!]
\centering
\mbox{
\subfigure[]{\includegraphics[width=0.38\textwidth]{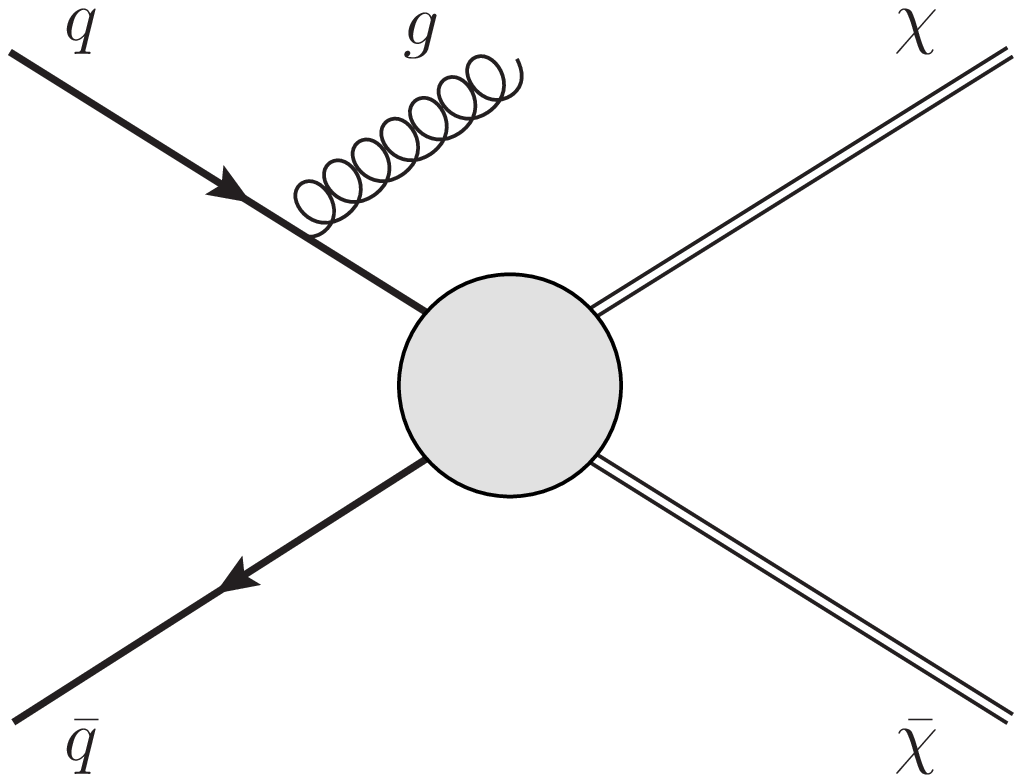}}
\subfigure[]{\includegraphics[width=0.38\textwidth]{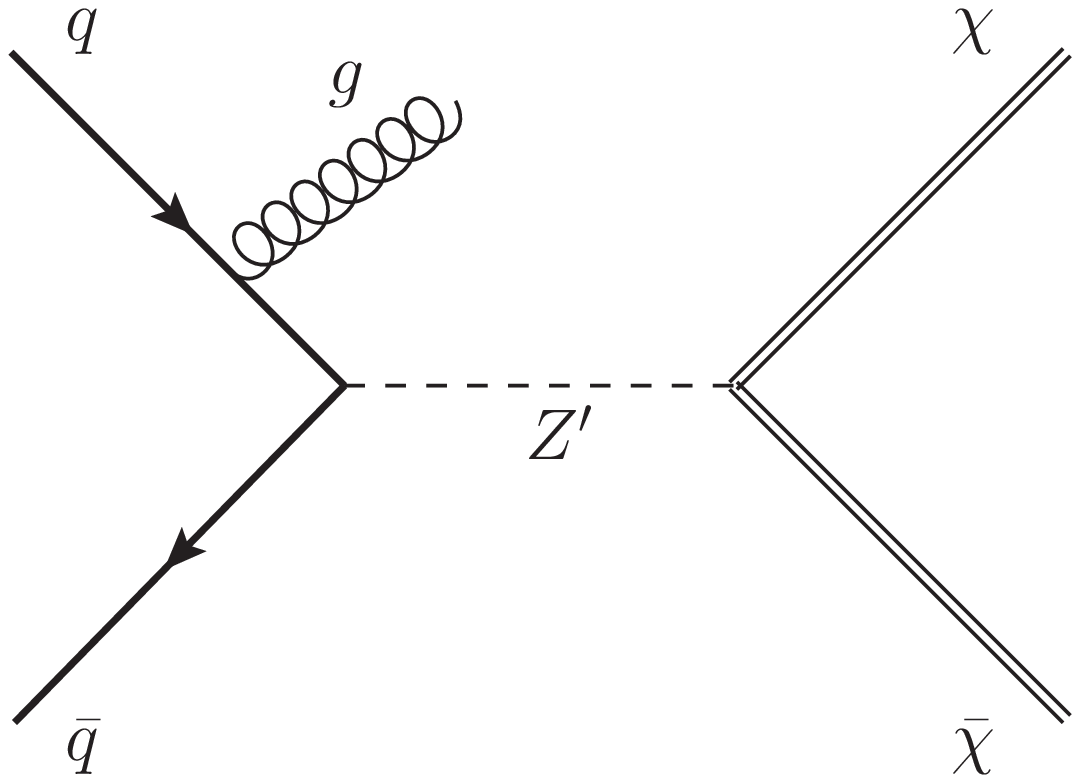}}
}
\caption{Feynman diagrams for the production of weakly interacting massive particle  pairs $\chi\bar{\chi}$ associated with a jet from initial-state radiation of a gluon, $g$. 
(a) A contact interaction described with effective operators. (b) 
  A simplified model with a $Z^\prime$ boson.}
\label{fig:wimp:sketch}
\end{figure}

Because WIMPs 
interact so weakly that they do not deposit energy in the calorimeter, their
production leads to signatures with missing transverse momentum
(${\ptmi}$), 
the magnitude of which is called \met. 
Here, WIMPs are assumed to be produced in pairs, 
and the events are identified via the presence of an energetic jet from 
initial-state radiation~(ISR)~\cite{Birkedal:2004xn,Beltran:2010ww,Rajaraman:2011wf,Fox:2011pm}  
yielding large \met .
 
The interaction of WIMPs with SM particles is described as a contact
interaction using an effective field theory (EFT) approach, 
mediated by a single new heavy particle or particles with mass too
large to be produced directly at the LHC (see
Fig.~\ref{fig:wimp:sketch}(a)).
\begin{table}[h!]
\caption{Effective interactions coupling WIMPs to
  Standard Model quarks or gluons, following the formalism in 
  Ref.~\cite{Goodman:2010ku}, where $M_\star$ is the suppression scale of the interaction. Operators starting with a D
  describe Dirac fermion WIMPs, the ones starting with a C are for
  scalar WIMPs and $G^a_{\mu\nu}$ is the colour field-strength tensor.}
\centering
\begin{tabular}{|c|ccc|}
\hline
Name & Initial state & Type & Operator\\[0.4cm]\hline
C1 &$qq$&scalar&$\frac{m_q}{M^2_\star}{\chi^\dagger}\chi\bar{q}q$\\[0.4cm]
C5 &$gg$&scalar&$\frac{1}{4M^2_\star}{\chi^\dagger}\chi\alpha_{\rm s}(G^a_{\mu\nu})^2$\\[0.4cm]
D1 &$qq$&scalar&$\frac{m_q}{M^3_\star}\bar{\chi}\chi\bar{q}q$\\[0.4cm]
D5 &$qq$&vector&$\frac{1}{M^2_\star}\bar{\chi}\gamma^{\mu}\chi\bar{q}\gamma_{\mu} q$\\[0.4cm]
D8 &$qq$&axial-vector&$\frac{1}{M^2_\star}\bar{\chi}\gamma^{\mu}\gamma^5\chi\bar{q}\gamma_{\mu}\gamma^5 q$\\[0.4cm]
D9 &$qq$&tensor&$\frac{1}{M_\star^2}\bar{\chi}\sigma^{\mu\nu}\chi\bar{q}\sigma_{\mu\nu}q$\\[0.4cm]
D11&$gg$&scalar& $\frac{1}{4M_\star^3}\bar{\chi}\chi\alpha_{\rm s}(G^a_{\mu\nu})^2$\\[0.4cm] 
\hline
\end{tabular}
\label{table:wimp:operators}
\end{table}
It is assumed here that the DM particle is either a Dirac fermion or a scalar
$\chi$; the only difference for Majorana fermions is that certain
interactions are not allowed and that the cross sections for the
allowed interactions are larger by a factor of four. Seven
interactions are considered~(see Table~\ref{table:wimp:operators}), namely
those described by the operators C1, C5, D1, D5, D8, D9, D11,
following the naming scheme in Ref.~\cite{Goodman:2010ku}. These
operators describe different bilinear quark couplings to WIMPs,
$q\bar{q}\ra \chi\bar{\chi}$, except for C5 and D11, which describe
the coupling to gluons, $gg\ra \chi\bar{\chi}$. 
The operators for Dirac fermions and scalars in
Ref.~\cite{Goodman:2010ku} fall into six categories with
characteristic \met\ spectral shapes. 
The  representative set of operators for these six categories are 
C1, C5, D1, D5, D9, and D11, while D8
falls into the same category as D5 but is listed explicitly in
Table~\ref{table:wimp:operators} because it is often used to convert 
LHC results into limits on DM pair production. In the operator
definitions in Table~\ref{table:wimp:operators}, $M_*$ is the
suppression scale of the interaction, after integrating out
the heavy mediator particles. 
The use of a contact interaction to produce WIMP pairs via heavy
mediators is considered conservative because it rarely overestimates
cross sections when applied to a specific scenario for physics beyond the SM.  Cases where
this approach is indeed optimistic are studied in
Refs.~\cite{Fox:2011pm,Friedland:2011za, Busoni:2013lha,
  Busoni:2014sya, Busoni:2014haa, 
  Buchmueller:2013dya 
}. Despite the caveats related to the validity of the EFT approach
(see~\ref{sec:valid}), this formalism is used here, as it
provides a framework for comparing LHC results to existing direct or indirect
DM searches. Within this framework, interactions of SM and DM
particles are described by only two parameters, the suppression scale
\MS{} and the DM particle mass $m_\chi$.
Besides the EFT operators, the pair production of WIMPs is also
investigated within a so-called simplified model, where a pair
of WIMPs couples to a pair of quarks explicitly via a new mediator
particle, a new vector boson $Z^\prime$ (see Fig.~\ref{fig:wimp:sketch}(b)).


In gauge-mediated SUSY-breaking (GMSB) scenarios~\cite{Dine:1981gu, AlvarezGaume:1981wy,Nappi:1982hm,Dine:1993yw,Dine:1994vc,Dine:1995ag},
the gravitino $\gra$ (spin-3/2 superpartner of the graviton)
is often the lightest supersymmetric particle  and a
potential candidate for DM.  Its mass is related to the SUSY-breaking scale $\sqrt{F}$ and $M_{\rm{Pl}}$ via $\mgra \propto
F/M_{\rm{Pl}}$~\cite{Giudice:1998bp}.
At hadron colliders, 
in low-scale SUSY-breaking scenarios 
with  very light gravitinos, the cross section for 
associated production of gravitino--squark ($pp \rightarrow \gra\squ + X$) and gravitino--gluino ($pp \rightarrow \gra\glu  + X$) processes are
relatively large~\cite{Klasen:2006kb}, since the cross section  depends on $\mgra$ as $\sigma \sim 1/\mgra^2$.
The decay of the 
gluino or squark into a gravitino and a gluon ($\glu \rightarrow \gra g$) or a gravitino and a quark ($\squ \rightarrow \gra q$), respectively, dominates~\cite{Klasen:2006kb}. The  
final state is characterized by the presence of a pair of gravitinos that escape detection 
and an energetic jet,   leading to 
a monojet-like topology. 
Previous studies at  colliders~\cite{lepsusy,Affolder:2000ef} considered the  production of gravitinos in association with a 
photon or a jet  and assumed 
extremely heavy squarks and gluinos. Within this approximation,  
a lower limit for the  gravitino mass of  $\mgra > 1.37 \times 10^{-5}$~eV was established. 
 

The study of the properties of the Higgs boson discovered 
by the ATLAS and CMS experiments~\cite{Aad:2012tfa,Chatrchyan:2012ufa}  
does not exclude a sizeable branching ratio for its  decay to invisible particles. It also 
opens up the question of whether a Higgs-like scalar field plays an important role in describing the interaction between dark and ordinary matter in the universe. In particular, 
a sizeable branching ratio to invisible particles could be interpreted in terms of the production of DM.  
Results from LEP~\cite{LEPhiggs} excluded an invisibly decaying Higgs boson, produced in association with a $Z$ boson, for a boson mass ($\mHH$) below 114.4~GeV.
The strongest direct bounds  from the LHC experiments  
on the branching ratio for the Higgs  invisible decay mode~\cite{Aad:2014iia,Chatrchyan:2014tja}
set upper limits of 58$\%$--65$\%$ at 95$\%$ confidence level (CL),   
based on the final state in which the Higgs boson 
is produced either in association with a $Z$ boson or via vector-boson fusion processes.
In this analysis,  the monojet-like final state is used to search for the production of an 
invisibly decaying boson with SM Higgs-like properties
and a mass in the range between 115~GeV and 300~GeV.     


The paper is organized as follows. The ATLAS detector is described in the next section. 
Section~\ref{sec:mc} provides details
of the simulations used in the analysis for background and signal processes.
Section~\ref{sec:recons} discusses the  reconstruction of jets, leptons and  
$\met$, while   
Sect.~\ref{sec:evt} 
describes the event selection.
The estimation of background contributions  and  
the study of systematic uncertainties are discussed in Sects.~\ref{sec:backg} and~\ref{sec:syst}. 
The results are presented in  Sect.~\ref{sec:results}, and are interpreted in terms 
of the search for ADD LED, WIMP pair production, the production of very light gravitinos in GMSB scenarios, and the production
of an invisibly decaying Higgs-like boson. 
Finally, Sect.~\ref{sec:sum} is devoted to the conclusions.

\section{Experimental setup}
\label{sec:atlas}

The ATLAS detector~\cite{Aad:2008zzm} covers almost the whole solid angle\footnote{ATLAS uses a right-handed
  coordinate system with its origin at the nominal interaction point
  (IP) in the centre of the detector and the $z$-axis along the beam
  pipe. The $x$-axis points from the IP to the centre of the LHC ring,
  and the $y$-axis points upward.  The azimuthal angle $\phi$ is
measured around the beam axis, and the polar angle $\theta$ is measured with respect to the $z$-axis.
We define transverse energy as 
$E_{\rm T} = E \ {\rm sin}  \theta$,
transverse momentum as $\pt = p \ {\rm sin} \theta$,
and pseudorapidity as $\eta = - {\rm ln}[{\rm tan}(\theta/2)]$.}
around the collision point with layers of tracking detectors, calorimeters and muon
chambers. The ATLAS inner detector (ID) has full coverage in $\phi$
and covers the pseudorapidity range $|\eta|<2.5$. 
It consists of a silicon pixel detector, a silicon microstrip detector, and a straw tube tracker which also measures transition radiation for particle identification, all immersed in a 2 T axial  magnetic field produced by a solenoid.

High-granularity liquid-argon (LAr) electromagnetic sampling calorimeters, with excellent 
energy and position resolution, cover the pseudorapidity
range $|\eta|<$~3.2. The hadronic calorimetry in the range $|\eta|<$~1.7 is provided by a scintillator-tile calorimeter, consisting of a large barrel and two smaller extended barrel cylinders, one on either side of
the central barrel. In the endcaps ($|\eta|>$~1.5), LAr hadronic
calorimeters match the outer $|\eta|$ limits of the endcap electromagnetic calorimeters. The LAr
forward calorimeters provide both the electromagnetic and hadronic energy measurements, and extend
the coverage to $|\eta| < 4.9$.

 The muon spectrometer measures the deflection of muons 
in the magnetic field provided by
large superconducting air-core toroid magnets
in the pseudorapidity range  $|\eta|<2.7$, instrumented with separate trigger and high-precision tracking chambers. 
Over most of the $\eta$ range, a measurement of the track coordinates in the principal bending direction of the magnetic 
field is provided
 by monitored drift tubes. At large pseudorapidities, cathode strip chambers  with higher granularity are used in the innermost plane over $2.0 < |\eta| < 2.7$.
The muon trigger system covers the pseudorapidity range $|\eta| < 2.4$.

The data are collected using an online  three-level trigger system~\cite{Aad:2012xs} that selects events  of interest and reduces 
the event rate from several MHz to about 400 Hz for recording and offline processing.

\section{Monte Carlo simulation}
\label{sec:mc}

Monte Carlo (MC) simulated event samples are used to 
compute detector acceptance and reconstruction efficiencies,
determine signal and background contributions, and estimate systematic uncertainties on the final results. 

\subsection{Background simulation}


The expected background to the monojet-like signature is dominated by $\znn$+jets and   
$W$+jets production (with $\wtn$+jets  being the dominant among the $W$+jets backgrounds), and  includes small contributions from $\zll$+jets ($\ell = e, \mu, \tau$), multijet, $t\bar{t}$, single-top, 
and diboson ($WW,WZ,ZZ$,$W\gamma$,$Z\gamma$) processes.

Samples of simulated $W$+jets and $Z$+jets production events are generated using 
 {SHERPA}-1.4.1~\cite{sherpa}, including  leading-order (LO) matrix elements for up to five partons in the final state and  
assuming massive $b$/$c$-quarks,  with CT10~\cite{ct10} parton distribution functions (PDF) of the proton. 
The MC expectations are initially 
normalized to next-to-next-to-leading-order (NNLO) perturbative QCD (pQCD) predictions according to 
DYNNLO~\cite{wcrosssection1, wcrosssection}
using  MSTW2008 90$\%$ CL NNLO PDF sets~\cite{mstw}.
%
%
The production of top-quark pairs ($\ttb$) is simulated using the 
{MC@NLO}-4.06~\cite{Frixione:2002ik,Frixione:2003ei} 
MC generator with parton showers and underlying-event modelling as implemented in 
{HERWIG}-6.5.20~\cite{Marchesini:1992:HMC,herwig} plus {JIMMY}~\cite{Butterworth:1996zw}.  
Single-top production samples are generated with {MC@NLO}~\cite{Frixione:2005vw} 
for the $s$- and $Wt$-channel~\cite{Frixione:2008yi},  
 while  {AcerMC}-v3.8~\cite{Kersevan:2002dd} is used for single-top
production in the $t$-channel.  
A top-quark mass of 172.5~\GeV\ is  used consistently. The AUET2C  
and AUET2B~\cite{ATLAStunes} set of optimised parameters for the underlying event description are used 
for $\ttbar$ and  single-top processes,  which use CT10 and CTEQ6L1~\cite{Pumplin:2002vw} PDF, respectively. 
Approximate NNLO+NNLL (next-to-next-to-leading-logarithm) pQCD cross sections, as determined in TOP++2.0~\cite{Czakon:2011xx}, are used in the normalization of the 
 $\ttb$ ~\cite{Czakon:2013goa} and $Wt$~\cite{Kidonakis:2010ux} samples.
Multijet and $\gamma$+jet samples are generated using the {PYTHIA}-8.165 program~\cite{pythia2} with
CT10  PDF. 
Finally, diboson samples ($WW$, $WZ$, $ZZ$, $W\gamma$ and $Z\gamma$ production) are generated using
 {SHERPA} with CT10 PDF and are normalized 
to NLO pQCD predictions~\cite{Campbell:2011bn}. 

\subsection{Signal simulation}

Simulated samples for the ADD LED model with 
different number of extra dimensions in the range $n=$2--6 and $M_D$ in the range $2$--$5$~TeV 
are generated using {PYTHIA}-8.165  with
CT10  PDF. Renormalization and factorization scales are set to 
$\sqrt{1/2 \times m_{G}^2 + p_{\rm T}^2}$,
where $m_G$ is the graviton mass and $p_{\rm T}$ 
denotes the transverse momentum of the recoiling parton.


The effective field theory of WIMP pair production is implemented in
{MADGRAPH5}-v1.5.2~\cite{Alwall:2011uj}, taken from
Ref.~\cite{Goodman:2010ku}. The WIMP pair production plus one or two
additional partons from ISR is simulated in two ways. For all
operators, samples are generated requiring at least one parton with
a minimum $\pt$ of $80\GeV$.
Studies
simulating up to three additional partons along with the WIMP pair
showed no difference in kinematic distributions when compared to the
samples with up to two additional partons.

Only initial states of gluons and the four lightest quarks are
considered, assuming equal coupling strengths for all quark flavours
to the WIMPs. The mass of the charm quark is most relevant for the cross
sections of the operator D1 (see Table~\ref{table:wimp:operators}) and
it is set to $1.42\GeV$. The generated events are interfaced to {
PYTHIA}-6.426~\cite{pythia} for parton showering and hadronization. The {MLM}
prescription~\cite{Mangano:2006rw} is used for matching the
matrix-element calculations of {MADGRAPH5} to the parton shower
evolution of {PYTHIA}-6. The samples are subsequently reweighted to the
{MSTW2008LO}~\cite{mstw} PDF set using {
LHAPDF}~\cite{Butterworth:2014efa}.  The {MADGRAPH5} default
choice for the renormalization and factorization scales is used.  The
scales are set to the geometric average of ${m^2 + \pt^2}$ for the two WIMPs, 
where $m$ is the mass of the particles. Events with WIMP masses
between 10~GeV and 1300~GeV are simulated for six different effective
operators (C1, C5, D1, D5, D9, D11). The WIMPs are taken to be either
Dirac fermions (\emph{D} operators) or scalars (\emph{C} operators),
and the pair-production cross section is calculated at LO. To study
the transition between the effective field theory and a physical
renormalizable model for Dirac fermion WIMPs coupling to Standard
Model particles via a new mediator particle $Z^\prime$, a
simplified model is generated in {MADGRAPH5}. 
For each WIMP mass point, mediator particle masses $\MMed$ between 50
GeV and 30 TeV are considered, each for two values of the mediator
particle width ($\Gamma=\MMed/3$ and $\MMed/8\pi$).


Simulated samples for gravitino production in association with a gluino or a squark in the final state, 
$pp \rightarrow \tilde{G}\tilde{g} + X$ and $pp \rightarrow \tilde{G}\tilde{q} + X$,   
are generated using LO
matrix elements  in {MADGRAPH4.4}~\cite{Mawatari:2011jy}  interfaced with {PYTHIA}-6.426 and using CTEQ6L1 PDF. 
The narrow-width approximation for the gluino and squark decays 
$\tilde{g} \rightarrow g \tilde{G}$ and $\tilde{q} \rightarrow q \tilde{G}$ is assumed. 
The renormalization and factorization scales are set to the average of the mass of the final-state particles
involved in the hard interaction $(\mgra + \msqgl)/2 \simeq \msqgl/2$. 
Values for $\mgra$ in the range between $10^{-3}$~eV and $10^{-5}$~eV 
are considered for squark and gluino masses in the range 50~GeV to 2.6~TeV.

%
%

Finally, MC simulated samples for the production of a Higgs  boson are generated  including the  
$gg \to  H$, $VV \to H$ ($V = W,Z$), and $VH$ production channels.  Masses for the boson in the range between
115~GeV and 300~GeV are considered.  
This Higgs boson is assumed to be produced as predicted in the Standard Model but unlike the SM Higgs 
it may decay into invisible particles at a significant rate.
The signal is modelled using {POWHEG}-r2262~\cite{Alioli:2008tz,Nason:2009ai,Oleari:2011ey}, 
which calculates separately the $gg \to H$ and $VV \to H$ production mechanisms with NLO pQCD
matrix elements. The description of the Higgs boson $\pt$ spectrum in the $gg \to H$  process follows the 
calculation in Ref.~\cite{deFlorian:2011xf}, which includes NLO + NNLL corrections. 
The effects of finite quark masses are also taken into account~\cite{Bagnaschi:2011tu}. 
For $gg \to H$ and $VV \to H$ processes,  {POWHEG} is interfaced to  {PYTHIA}-8.165  for showering and hadronization.
For $ZH$ and $WH$ processes,  {POWHEG} interfaced to  {HERWIG++}~\cite{Bahr:2008pv}  is used and the 
$Z/W$ bosons are forced to decay to a pair of quarks. The invisible decay of the Higgs-like boson is simulated 
by forcing the boson to decay to two $Z$ bosons, which are then forced to decay to neutrinos. 
Signal samples are generated with renormalization
and factorization scales set to $\sqrt{(\mHH)^2 + (\ptH)^2}$.
The Higgs boson production cross sections, as well as their uncertainties, are taken from 
Refs.~\cite{LHCHiggsCrossSectionWorkingGroup:2011ti,LHCHiggsCrossSectionWorkingGroup:2012vm}. For the $gg \to H$ process,  cross-section 
calculations at NNLO+NNLL accuracy ~\cite{Harlander:2002wh,Anastasiou:2002yz,Ravindran:2003um,Catani:2003zt} in pQCD are used and 
NLO electroweak corrections~\cite{Aglietti:2004nj,Actis:2008ug} are included. 
The cross sections 
for $VV \to H$ processes are calculated with full NLO pQCD and electroweak corrections~\cite{Ciccolini:2007jr,Ciccolini:2007ec,Arnold:2008rz}.
The cross sections for the associated production ($WH$ and $ZH$) are calculated at 
NNLO~\cite{Brein:2003wg} in pQCD, and include  NLO electroweak corrections~\cite{Ciccolini:2003jy}.  

Differing pileup (multiple proton--proton interactions in the same or neighbouring bunch-crossings) 
conditions as a function of the instantaneous luminosity are taken into account by overlaying simulated 
minimum-bias events generated with   {PYTHIA}-8 onto the hard-scattering process. 
The MC-generated samples are processed either with a full ATLAS detector simulation~\cite{:2010wqa} based on
the GEANT4 program~\cite{Agostinelli:2002hh} or a fast simulation 
of the 
response of the electromagnetic and hadronic calorimeters~\cite{FastCaloSim} and of the trigger system.
The results based on fast simulation are validated against fully simulated samples and the difference is 
found to be negligible. 
The simulated events are
reconstructed and analysed with the same analysis chain as for the data, using the same trigger and event 
selection criteria.

\section{Reconstruction of physics objects}
\label{sec:recons}

Jets are defined using the $\akt$ jet algorithm~\cite{paper:antikt} with the radius parameter 
$R = 0.4$.
Energy depositions 
reconstructed as clusters in the calorimeter 
are the inputs to the jet algorithm.                   
The measured 
jet $\pt$   
is corrected for detector effects,  including the non-compensating character of the calorimeter, by weighting
 energy deposits arising from electromagnetic and hadronic showers differently.   
In addition, jets are corrected for 
contributions from pileup,  
as described in Ref.~\cite{Aad:2011he}.  Jets with  corrected $p_{\rm T} > 30$~GeV and $|\eta| <4.5$ are considered
 in the analysis.  
Jets with $|\eta| < 2.5$ containing a $b$-hadron are identified using
a neural-net-based algorithm~\cite{ATLAS-CONF-2012-043}
with an efficiency of 80$\%$ and a rejection factor of 30 (3) against 
jets originating from fragmentation of light quarks or gluons (jets containing a $c$-hadron), as determined  
using $\ttbar$ simulated events.

The presence of leptons (muons or electrons) in the final state  is used in the analysis 
to define control samples and to reject background contributions in the signal regions (see Sects.~\ref{sec:evt} and~\ref{sec:backg}). 
Muon candidates are formed by combining information from the muon spectrometer and inner 
tracking detectors as described in
Ref.~\cite{Aad:2014rra} and are 
required to have  $p_{\rm T} > 7$~GeV and  $|\eta| < 2.5$.
In addition, muons are required to be isolated: the sum of the transverse momenta of the
tracks not associated with the muon in a cone of size  
$\Delta R=\sqrt{(\Delta \eta)^2 + (\Delta \phi)^2} = 0.2$
around the muon direction 
is required to be less than 1.8~GeV.   
The muon $\pt$ requirement is increased to $\pt > 20$~GeV to define the $\wmn$+jets and $\zmm$+jets control regions.

Electron candidates are initially required to have 
$p_{\rm T} > 7$~GeV and $|\eta| <2.47$, and to pass the medium
 electron shower shape and track selection criteria described 
in Ref.~\cite{Aad:2014fxa}, which are  reoptimized for 2012 data.
Overlaps between identified electrons and jets in the final state are
resolved. Jets are discarded if their separation $\Delta R$ from
an identified electron is less than~0.2. 
The electron $\pt$ requirement is increased to  $\pt > 20$~GeV and the 
transition region between calorimeter sections $1.37 < |\eta| < 1.52$ is excluded 
to  reconstruct $Z$ and $W$  boson candidates in the $\zee$+jets and $\wen$+jets control regions, respectively.
The electron requirements are further tightened for the $\wen$+jets control sample to 
constrain the irreducible $\znn$+jets background contribution (see below). In this case, electrons are selected to pass  
tight~\cite{Aad:2014fxa} 
 electron shower shape and track selection criteria, their $\pt$ threshold is raised to 25~GeV, and they  
are required to be isolated:  the sum of the transverse momenta of the
tracks not associated with the electron in a cone of radius $\Delta R =  0.3$ around the electron direction 
is required to be less than 5$\%$ of the electron $\pt$. 
An identical isolation criterion, based on the calorimeter energy deposits not associated with the electron, is 
also applied.

The $\met$ is reconstructed using all energy deposits in the calorimeter up 
to pseudorapidity $|\eta| =  4.9$. 
Clusters associated with either electrons or photons with $\pt >10$~GeV and those
associated with jets with $\pt >20$~GeV make use of the corresponding calibrations for these objects. Softer jets 
and clusters not associated with these objects are calibrated using both calorimeter and tracking information~\cite{Aad:2012re}.

\section{Event selection}
\label{sec:evt}

The data sample considered in this paper 
corresponds  
to a total integrated luminosity of $20.3 \ \rm fb^{-1}$. 
The uncertainty on the integrated luminosity is $2.8\%$, as estimated following the same methodology as 
detailed in Ref.~\cite{Aad:2013ucp}.
The data were selected online using 
a trigger logic that selects events with  
$\met$ above 80~GeV, as computed at the final stage of the three-level
trigger system~\cite{Aad:2012xs}.
With respect to the final analysis requirements,  
the trigger selection is fully  efficient for $\met > 150$~GeV, 
as determined using a data sample with muons in the final state.
Table~\ref{tab:sr} summarizes the different event selection criteria applied in the signal
regions. The following preselection criteria are applied.

\begin{itemize}

\item Events are required to have a reconstructed primary vertex for the interaction  
 consistent with the beamspot envelope and to have at least two  associated  
tracks with $\pt > 0.4$~GeV; when more than one such vertex is found, the vertex with the largest summed $\pt^2$
of the associated tracks is chosen.

\item Events are required to have $\met > 150$~GeV and at least one  
jet with $\ptjet > 30$~GeV and $|\eta| < 4.5$  in  the final state.

\item The analysis selects events with
a leading jet with $\pt > 120$~GeV and $|\eta| < 2.0$.
Monojet-like topologies in the final state
are selected by requiring the leading-jet $\pt$ and the
$\met$ to satisfy $\pt/\met > 0.5$.  An additional requirement on
  the azimuthal separation  $\Delta\phi(\textrm{jet},{\ptmi}) > 1.0$
  between the direction of the missing transverse momentum and that of each of the selected jets is
  imposed. This requirement reduces the multijet background contribution where the
  large $\met$  originates  mainly from jet energy mismeasurement.

\item Events are rejected if they contain any jet with $\ptjet > 20$~GeV and 
$|\eta| < 4.5$ that presents 
an electromagnetic fraction in the calorimeter,  calorimeter sampling fraction, or  
charged fraction~\footnote{The charged fraction is defined as
$f_{{\textrm{ch}}}=\sum p_{\rm T}^{{\textrm{track,jet}}} /p_{\rm T}^{{\textrm{jet}}}$,
where $\sum p_{\rm T}^{{\textrm{track,jet}}}$ is the scalar
sum of the transverse momenta of tracks associated with the primary vertex
within a cone of radius $R=0.4$ around the jet axis, and
$p_{\rm T}^{{ \textrm{jet}}}$ is the transverse momentum as determined
from calorimetric measurements.} 
(for jets with $|\eta|<2.5$) 
inconsistent with the requirement that they originate
from a proton--proton collision~\cite{Aad:2013zwa}.
In the case of the leading (highest $\pt$) jet in the event,
the requirements are tightened  to reject remaining contributions from
beam-related backgrounds and cosmic rays.
Events are also rejected if any of the jets is reconstructed close to known 
partially instrumented regions of the calorimeter. 
Additional requirements 
based on the timing and the pulse shape of the cells in the calorimeter
are applied to 
suppress coherent noise and electronic noise bursts in the calorimeter producing
anomalous energy deposits~\cite{ATLAS-CONF-2012-020}; these requirements have a negligible effect on the signal efficiency. 

\item Events with muons or electrons with $\pt > 7$~GeV are vetoed. In addition, events with isolated tracks with 
$\pt > 10$~GeV and $|\eta| < 2.5$ are vetoed to reduce background from non-identified leptons ($e$, $\mu$ or $\tau$) in the final state. The track isolation is defined such that there must be no additional track 
with $\pt > 3$~GeV within a cone of radius 0.4 around it. 

\end{itemize}

\noindent
Different signal regions (SR1--SR9) are 
considered with increasing $\met$ thresholds from 150~GeV to 700~GeV.


\begin{table}[!ht]
\renewcommand{\baselinestretch}{1}
\caption{Event selection criteria applied for 
the selection of monojet-like signal regions, SR1--SR9.}
\begin{center}
\begin{small}
\renewcommand{\baselinestretch}{1.2}
\begin{tabular*}{\textwidth}{@{\extracolsep{\fill}}lccccccccc}\hline
\multicolumn{10}{c}{\footnotesize{Selection criteria}} \\\hline\hline
\multicolumn{10}{c}{{\footnotesize{Preselection}} } \\\hline 
\multicolumn{10}{l}{Primary vertex}\\
\multicolumn{10}{l}{$\met > 150$~GeV }\\
\multicolumn{10}{l}{Jet quality requirements}   \\
\multicolumn{10}{l}{At least one jet with $\pt >30$~GeV and $|\eta|< 4.5$}\\
\multicolumn{10}{l}{Lepton  and isolated track vetoes}\\ \hline
\multicolumn{10}{c}{\footnotesize{Monojet-like selection}}\\\hline
\multicolumn{10}{l}{The leading jet with $\pt > 120$~GeV and $|\eta|<2.0$}\\
\multicolumn{10}{l}{Leading jet $\pt/\met > 0.5$}\\
\multicolumn{10}{l}{$\Delta\phi(\textrm{jet},\ptmi) > 1.0$}\\\hline 
Signal region        & SR1 & SR2 & SR3 & SR4 & SR5 & SR6 & SR7 & SR8 & SR9 \\ 
Minimum $\met$ [GeV] & 150 & 200  & 250 & 300 & 350 & 400 & 500 & 600 & 700 \\ \hline
\end{tabular*}
\end{small}
\end{center}
\label{tab:sr}
\end{table}

\section{Background estimation}
\label{sec:backg}

The $W$+jets and $\znn$+jets backgrounds are estimated using 
MC event samples normalized using data in selected control regions. In particular, the dominant 
$\znn$+jets background contribution is constrained using a combination of 
estimates from $W$+jets and $Z$+jets control regions. 
The remaining SM backgrounds from $\zll$+jets, $t\bar{t}$, single top, and dibosons 
are determined using MC simulated samples, while the multijet background  contribution
is extracted from data.  In the case of the $\ttbar$ background process, which contributes to 
both the signal and $W$+jets control regions, dedicated control samples are defined to validate
the MC normalization and to estimate systematic uncertainties.    
Finally, the potential
contributions from beam-related background and cosmic rays are
estimated in data using jet timing information.
The methodology and the samples used for estimating the background are summarised in Table~\ref{tab:CRs}. 
The details are given in the following sections. 

\begin{table}[!ht]
\caption{
Summary of the methods and control samples used to constrain the different background contributions in the signal regions. 
}
\begin{center}
\begin{footnotesize}
\begin{tabular*}{\textwidth}{@{\extracolsep{\fill}}lcc}\hline
{Background process}     & Method       &  {Control sample}  \\ \hline
$\znn$+jets    & MC and control samples in data  & $\zll$, $\wln$ ($\ell = e,\mu$) \\          
$\wen$+jets    & MC and control samples in data  & $\wen$ (loose) \\             
$\wtn$+jets    & MC and control samples in data  & $\wen$ (loose) \\            
$\wmn$+jets    & MC and control samples in data  & $\wmn$ \\           \hline 
$\zll$+jets ($\ell = e,\mu,\tau$)    &   MC-only & \\
$\ttb$, single top &  MC-only & \\           
Diboson            &  MC-only & \\         
Multijets          &  data-driven & \\
Non-collision      &  data-driven & \\ \hline\hline
\end{tabular*}
\end{footnotesize}
\end{center}
\label{tab:CRs}
\end{table}

\subsection{$W/Z$+jets background}

In the analysis, control samples in data,  
with identified electrons or muons in the final state  and with identical requirements on 
the jet $\ptjet$ and $\met$,  are used to determine 
the  $\wln$+jets ($\ell = e,\mu,\tau$) and $\znn$+jets  electroweak background contributions.
This  reduces significantly the relatively large  theoretical and experimental 
systematic uncertainties, of the order of  $20\%$--$40\%$,  associated with purely MC-based expectations. 
The $\met$-based online trigger used in the analysis 
does not include muon information in the $\met$ calculation. This allows the 
collection of $\wmn$+jets and $\zmm$+jets control samples with the same trigger as for the signal regions. 
This is not the case for the $\wen$+jets and $\zee$+jets control samples used to constrain the $\znn$+jets background (see below). 

A $\wmn$+jets control sample is defined using events with a muon with $\pt >20$~GeV and  $W$
transverse mass
in the range  $40$~GeV~$<  m_{\rm T} <  100$~GeV. 
The transverse mass $m_{\rm T}$  is defined by the lepton ($\ell$) and neutrino ($\nu$) $\pt$ and direction as
$m_T=\sqrt{2\pt^{\ell}\pt^{\nu}(1-\cos(\phi^{\ell}-\phi^{\nu}))}$,
where the $(x,y)$ components of the neutrino momentum are taken to be
the same as the corresponding $\ptmi$ components.
Similarly, a $\zmm$+jets control
sample is selected,  requiring the presence of two muons with $\pt >20$~GeV and 
invariant mass in the range $66$~GeV~$< m_{\mu \mu} < 116$~GeV. In the $\wmn$+jets and  
$\zmm$+jets control regions,  the $\met$ is not corrected for the presence of 
the muons in the final state, which are considered invisible, 
motivated by the fact that these control regions
are used to estimate the irreducible $\znn$+jets background in the signal regions.

\begin{figure}[!ht]
\begin{center}
\mbox{
\subfigure[]{\includegraphics[width=0.485\textwidth]{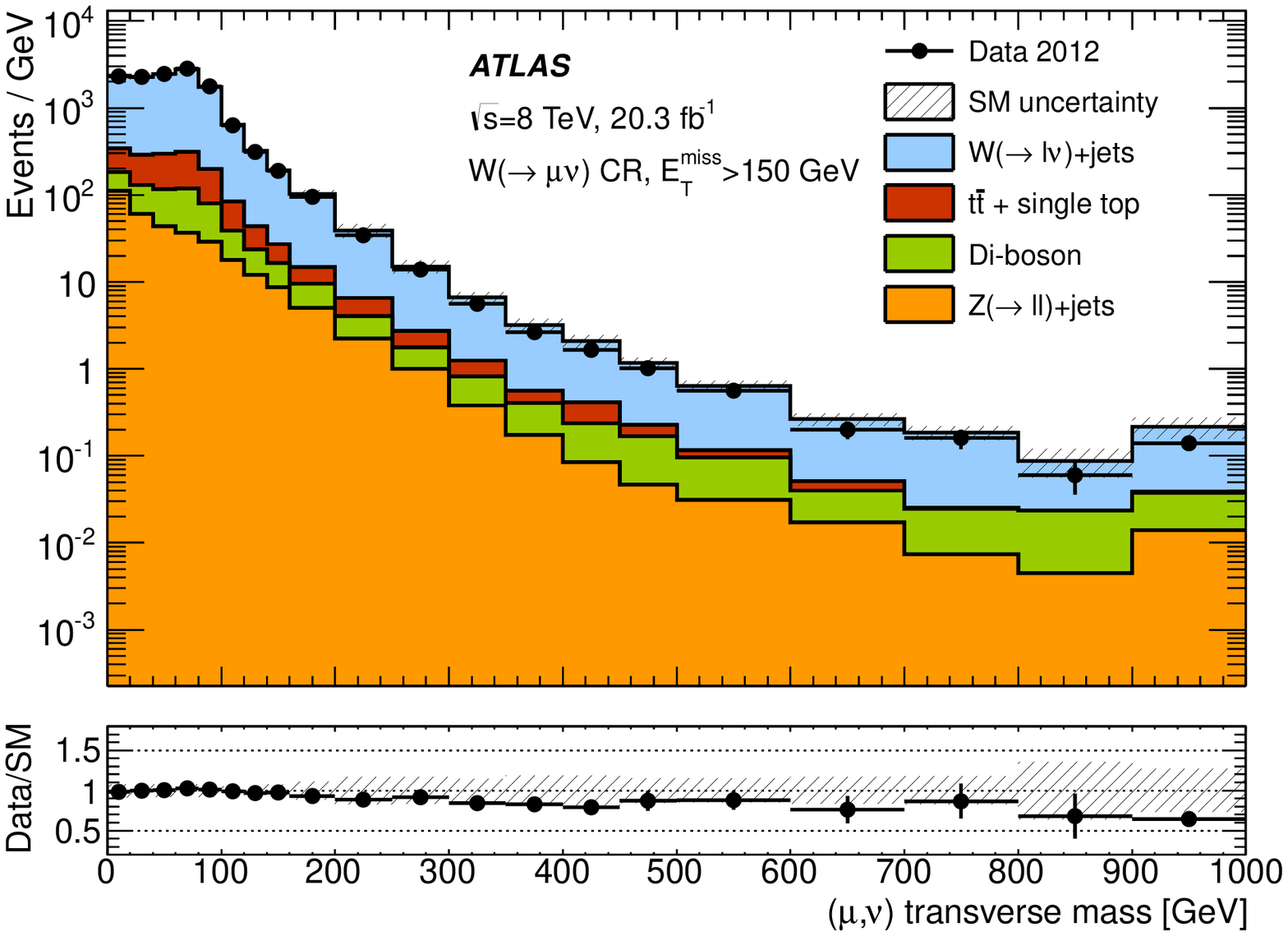}}
\subfigure[]{\includegraphics[width=0.485\textwidth]{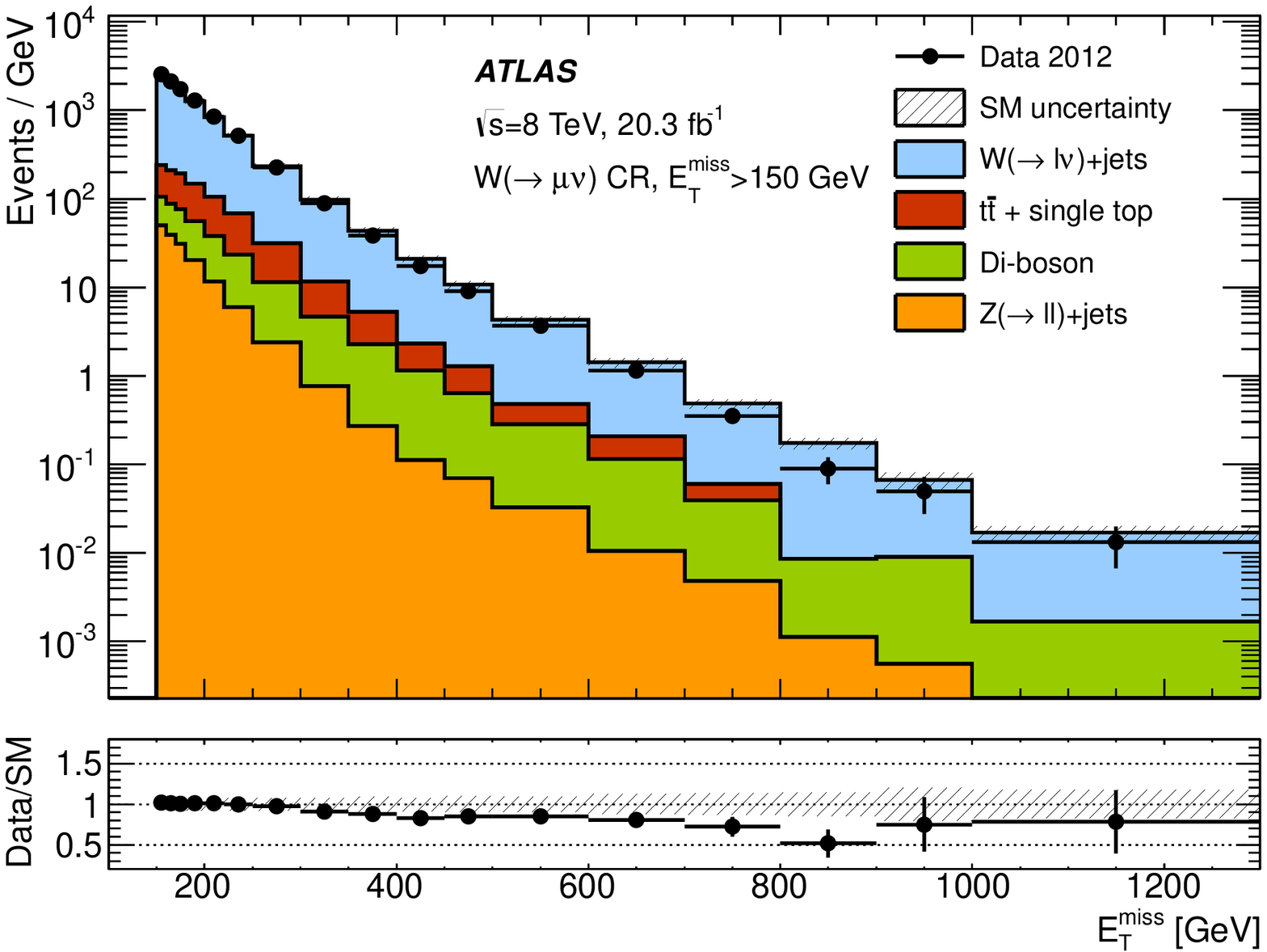}}
}
\mbox{
\subfigure[]{  \includegraphics[width=0.485\textwidth]{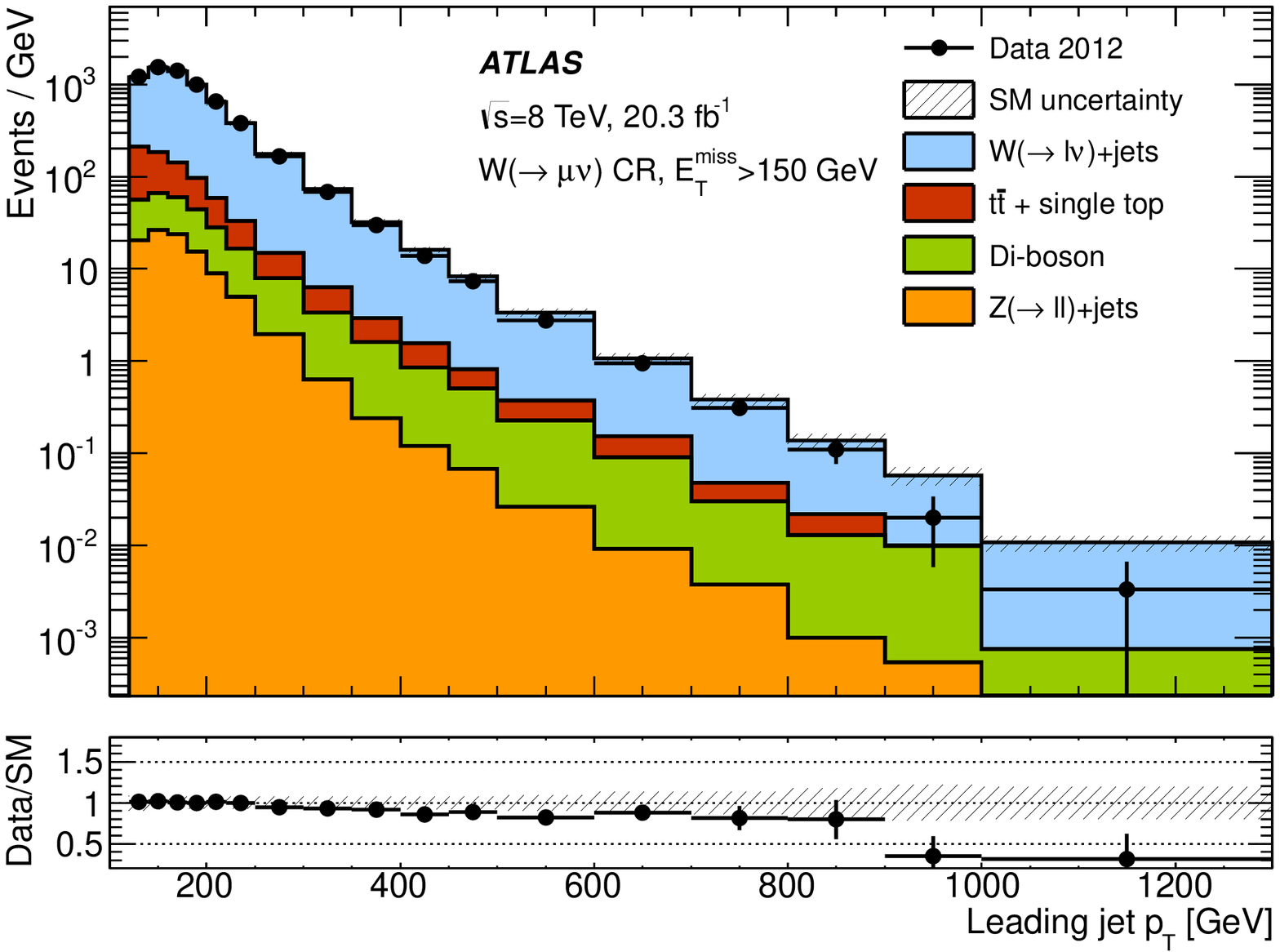}}
\subfigure[]{  \includegraphics[width=0.485\textwidth]{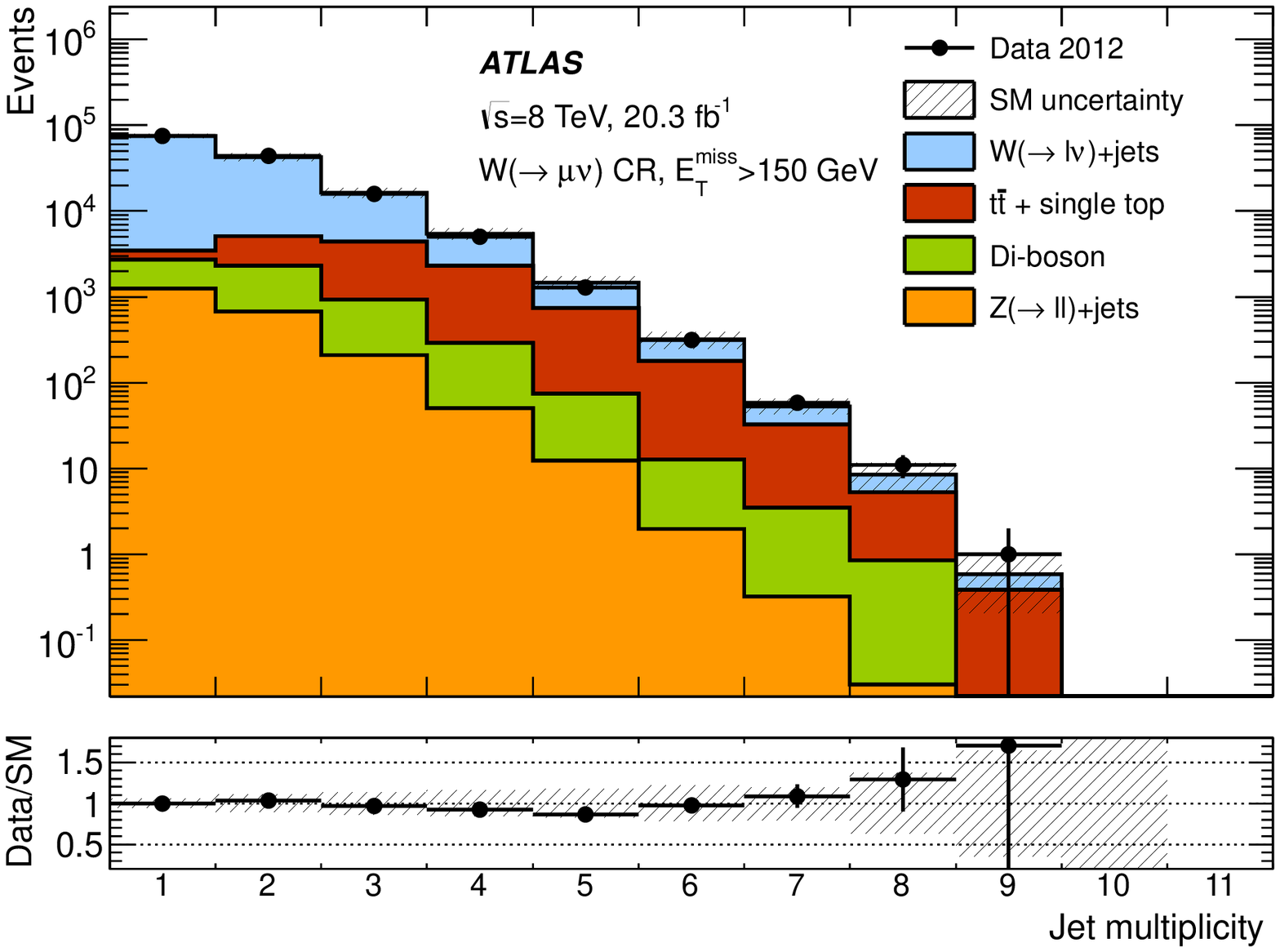}}
}
\end{center}
\caption{
Distributions of the measured (a) transverse mass of the identified muon and the missing transverse momentum, (b) $\met$, (c) leading jet $\pt$ and (d) jet multiplicity distributions
in the $\wmn$+jets control region for the inclusive SR1 
selection, compared to
the background expectations.
The latter include the global normalization factors extracted from the data.
Where appropriate, the last bin of the distribution includes overflows.
The lower panels represent the ratio of data to MC expectations. 
The error bands in the ratios include the statistical and experimental  uncertainties 
on the background expectations.
}
\label{fig:cr1}
\end{figure}

\begin{figure}[!ht]
\begin{center}
\mbox{
\subfigure[]{\includegraphics[width=0.485\textwidth]{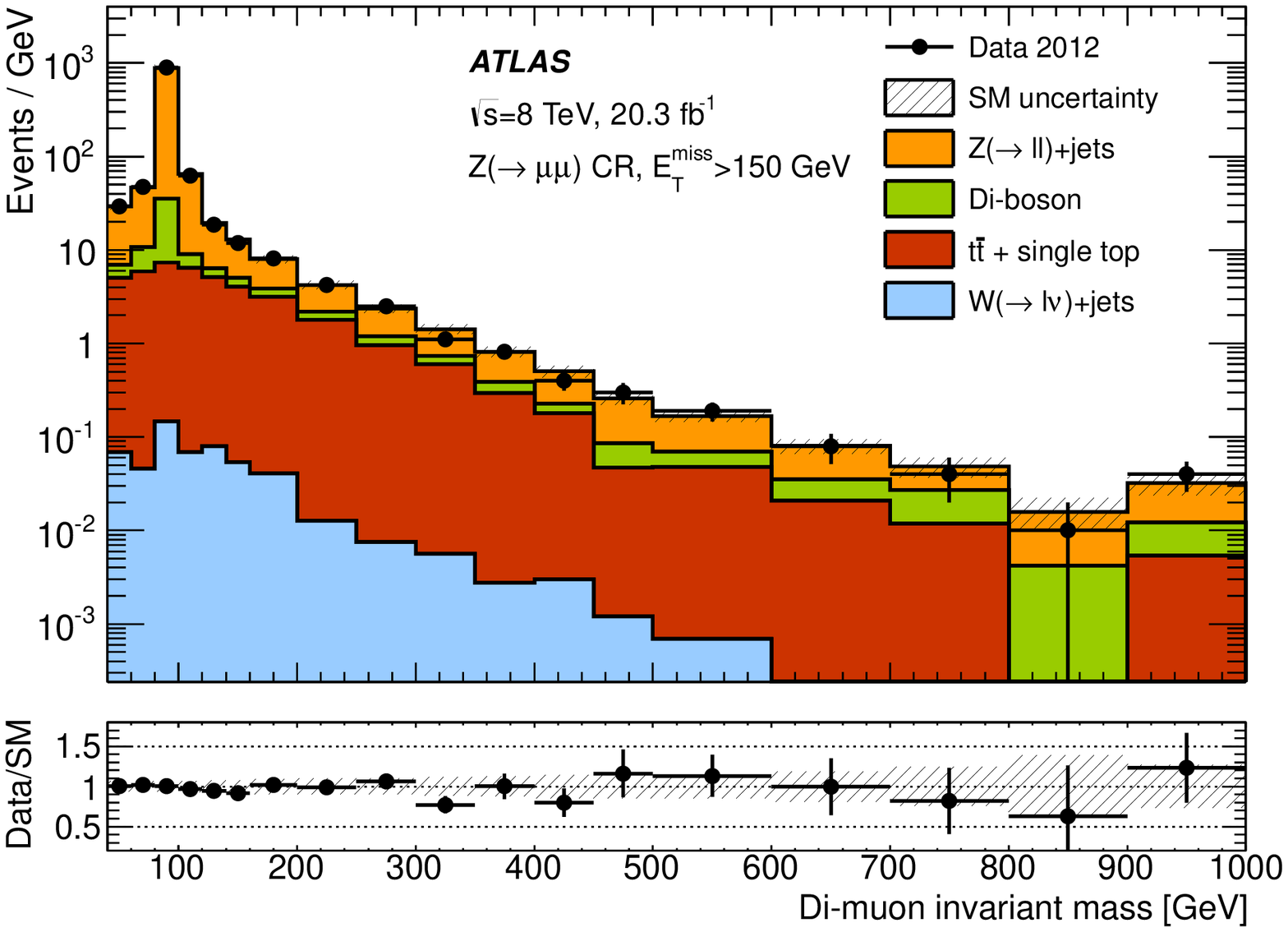}}
\subfigure[]{\includegraphics[width=0.485\textwidth]{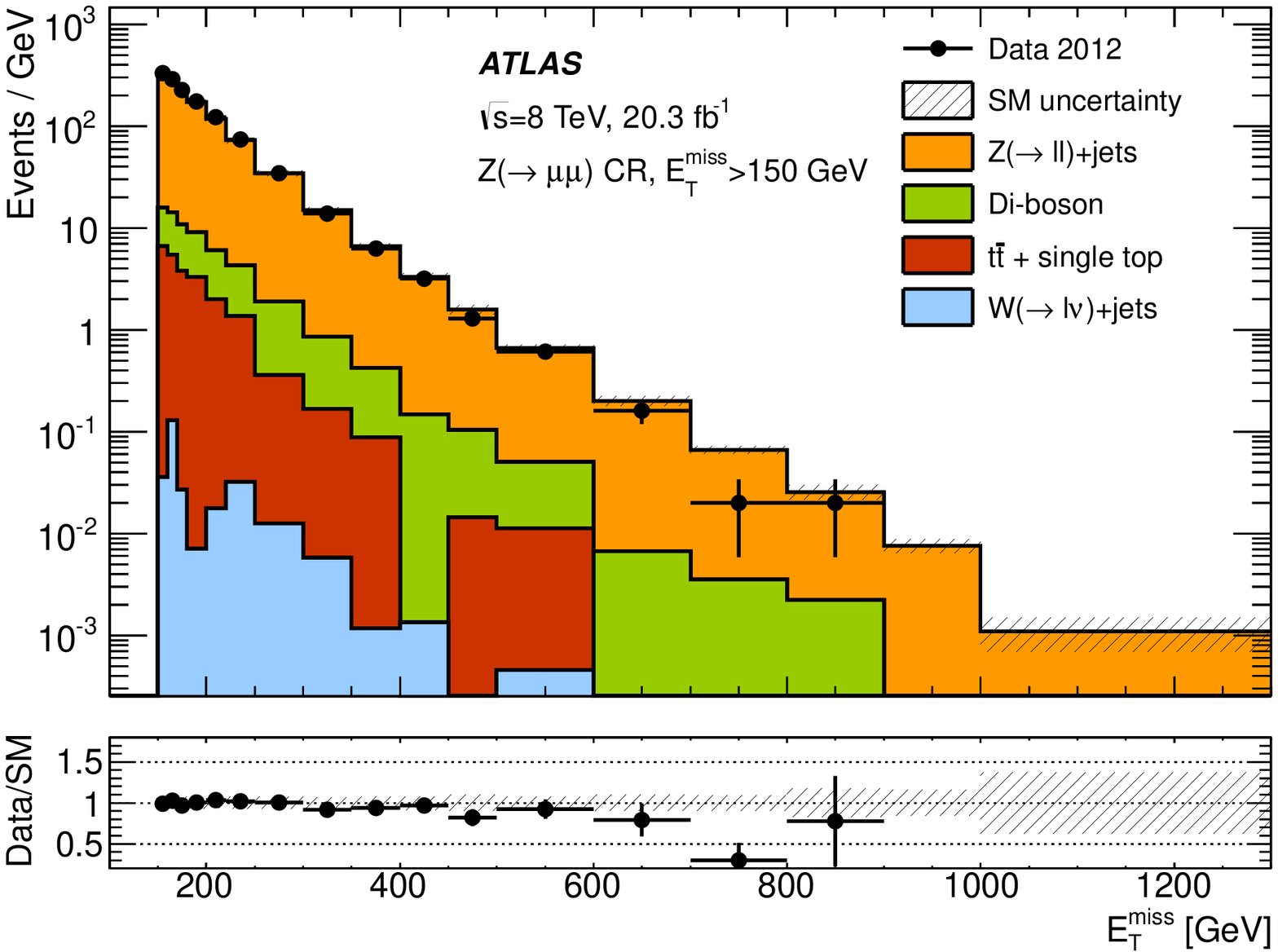}}
}
\mbox{
\subfigure[]{  \includegraphics[width=0.485\textwidth]{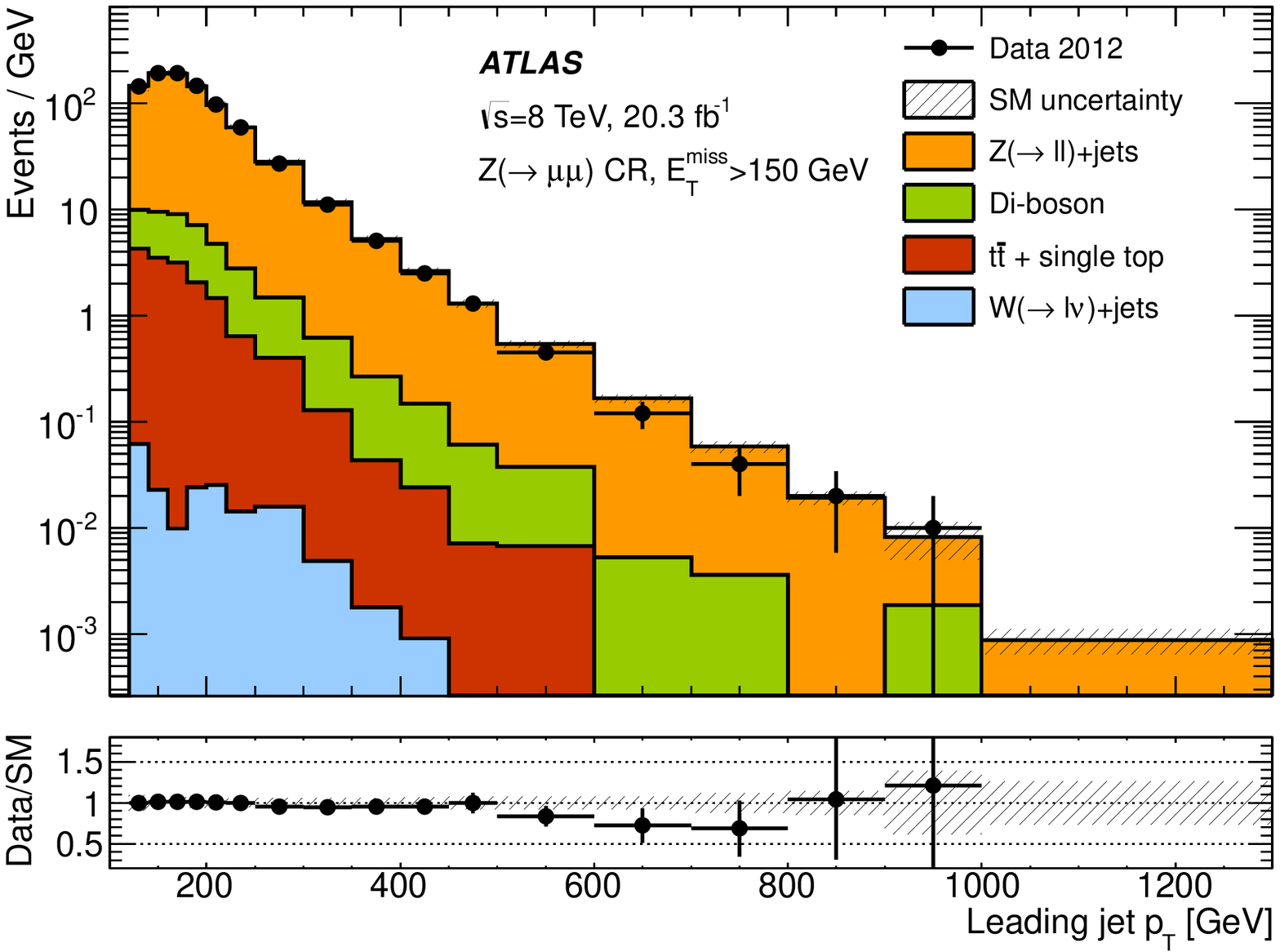}}
\subfigure[]{  \includegraphics[width=0.485\textwidth]{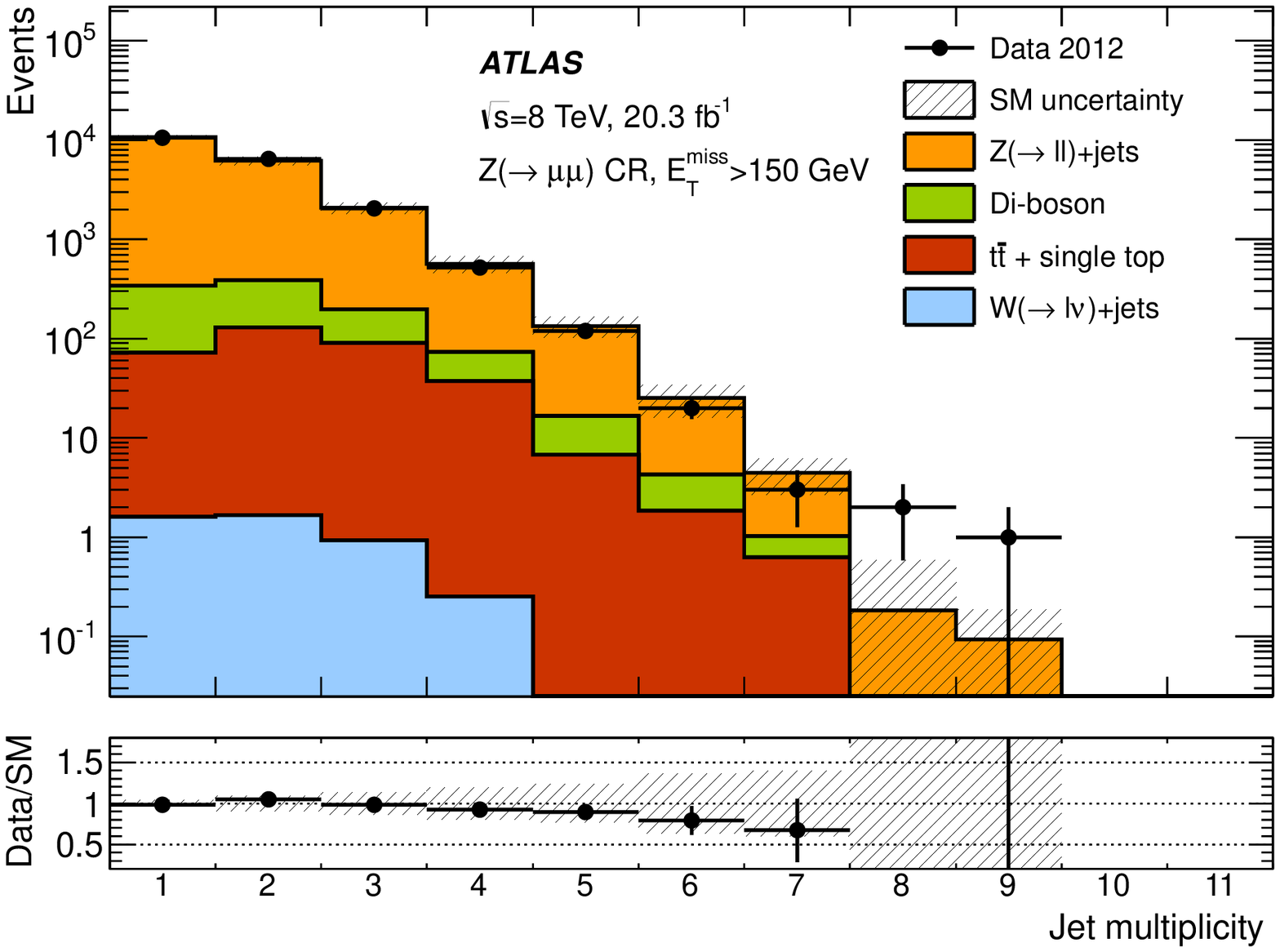}}
}
\end{center}
\caption{
Distributions of the measured (a) dilepton invariant mass, (b) $\met$, (c) leading jet $\pt$ and (d) jet multiplicity distributions
in the $\zmm$+jets control region for the inclusive SR1 
selection, compared to
the background expectations.
The latter include the global normalization factors extracted from the data.
Where appropriate, the last bin of the distribution includes overflows.
The lower panels represent the ratio of data to MC expectations.
The error bands in the ratios include the statistical and experimental  uncertainties 
on the background expectations.
}
\label{fig:cr2}
\end{figure}

The $\wen$+jets and $\zee$+jets control samples  
used to constrain the $\znn$+jets background in the signal regions
are collected using  online triggers that 
select events with an electron in the final state. The $\met$ is corrected by
removing the contributions from the electron energy clusters in the calorimeters. 
In the $\zee$+jets control 
sample, events are selected with exactly two electrons with $\pt > 20$~GeV and dilepton 
invariant mass in the range   $66$~GeV~$< m_{e e} < 116$~GeV. 
In the $\wen$+jets control 
sample a tight selection is applied: 
events are selected to have only a single electron with $\pt > 25$~GeV, transverse mass
in the range $40$~GeV~$<  m_{\rm T} <  100$~GeV, and uncorrected $\met >25$~GeV. The latter requirements
suppress background contamination from multijet processes where jets are misidentified as electrons.  

A separate  $\wen$+jets control
sample, collected with the $\met$-based trigger and looser requirements that increase the number of events,  is defined to constrain the $\wen$+jets and $\wtn$+jets background contributions. In this case,  the electron $\pt$
requirement is reduced to  $\pt > 20$~GeV and no further cuts on electron isolation and  $m_{\rm T}$ are applied. In addition, the $\met$ 
calculation in this case is not corrected for the presence of the electron or $\tau$ leptons in the final state, as they contribute
to the calorimeter-based $\met$ calculation in the signal regions.     

\begin{figure}[!ht]
\begin{center}
\mbox{
\subfigure[]{  \includegraphics[width=0.485\textwidth]{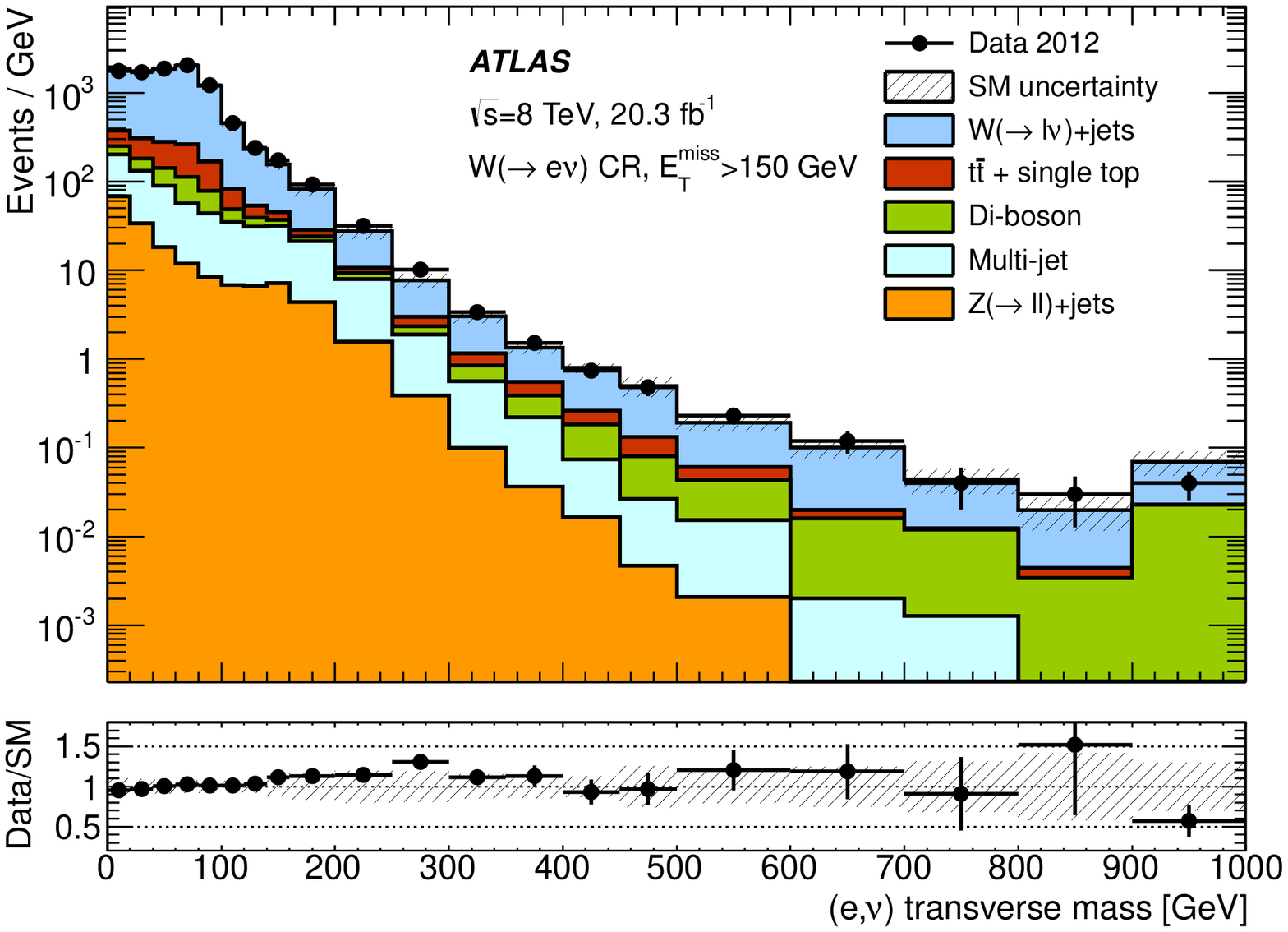}}
\subfigure[]{  \includegraphics[width=0.485\textwidth]{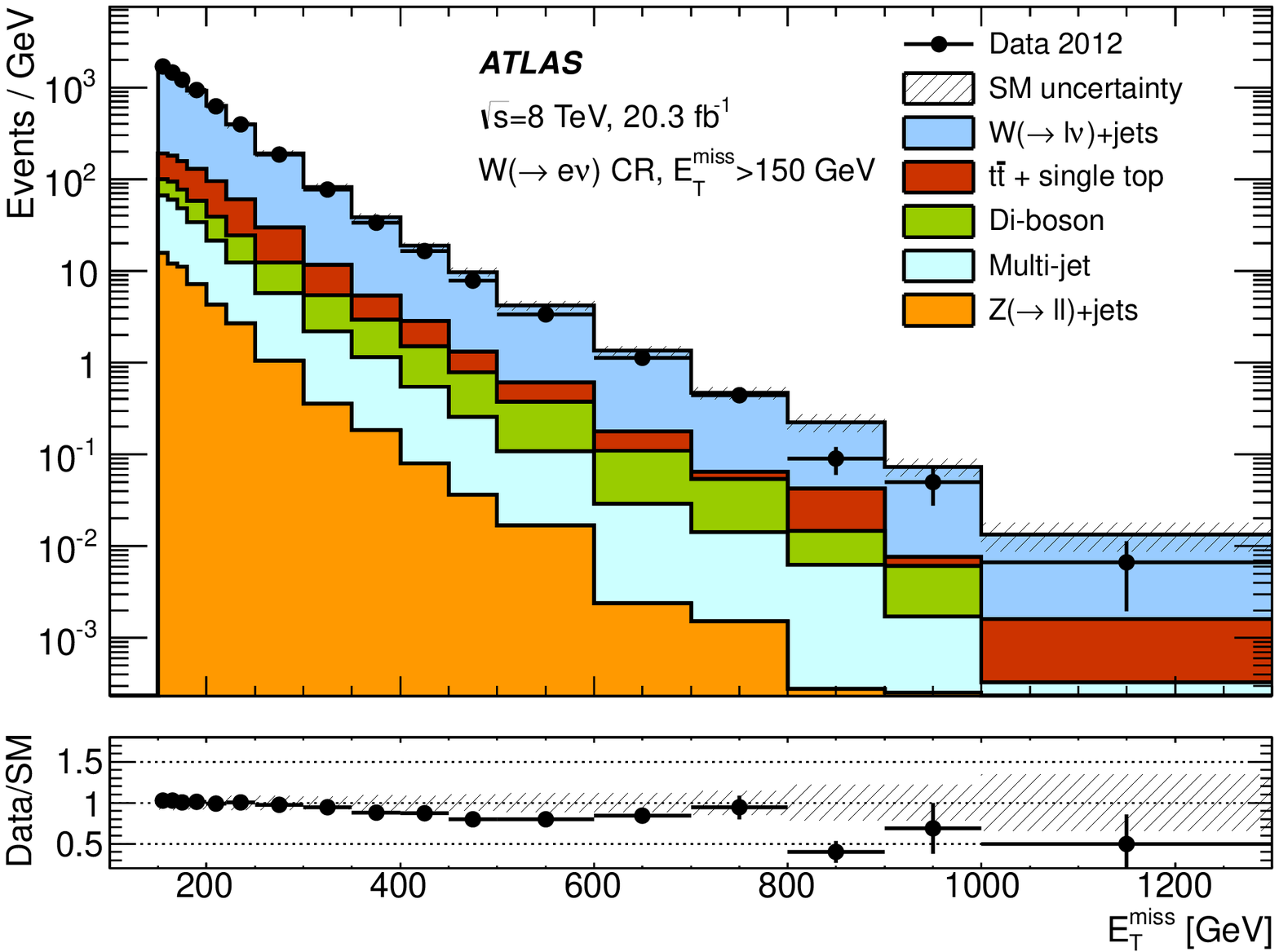}}
}
\mbox{
\subfigure[]{  \includegraphics[width=0.485\textwidth]{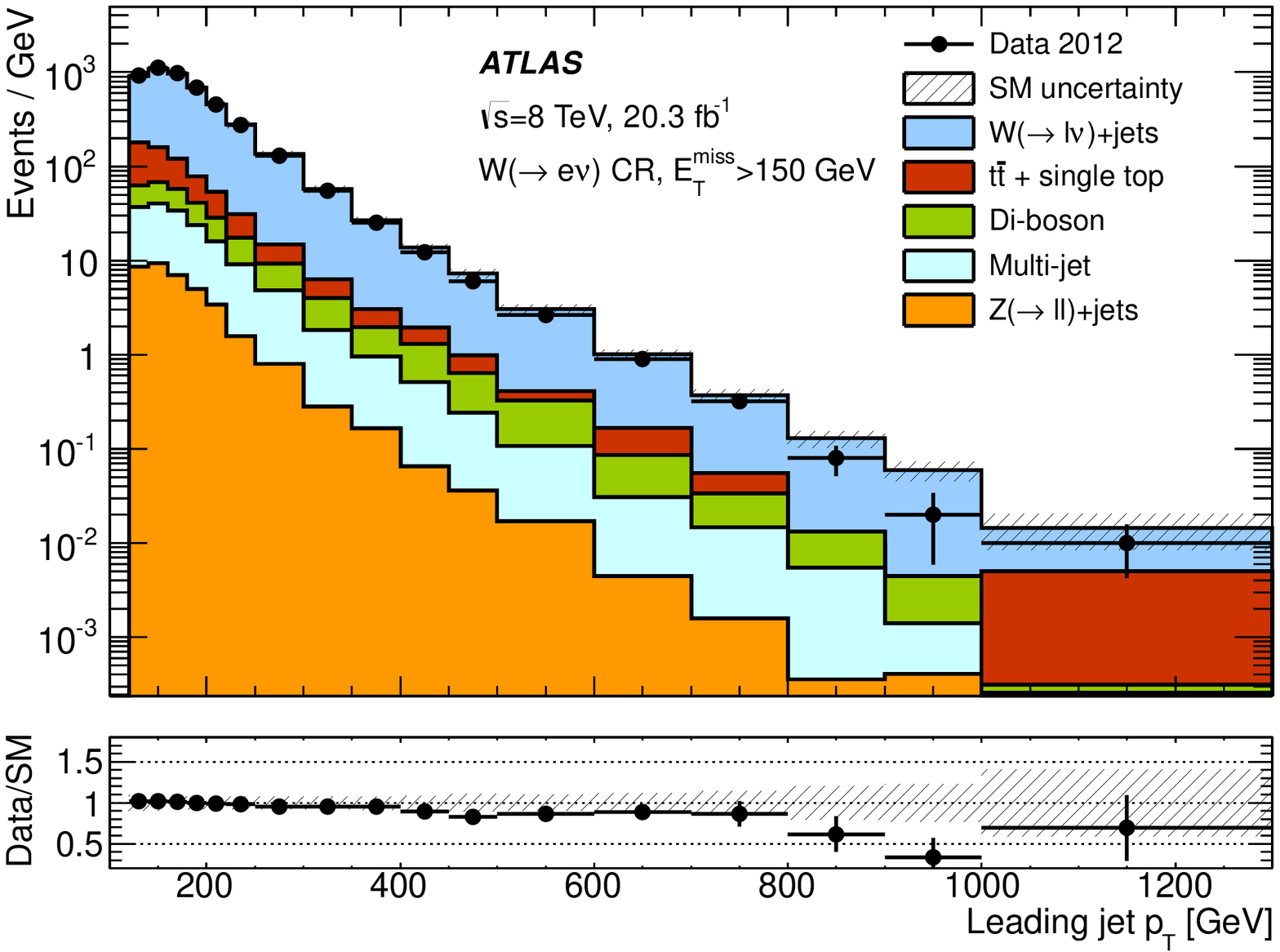}}
\subfigure[]{  \includegraphics[width=0.485\textwidth]{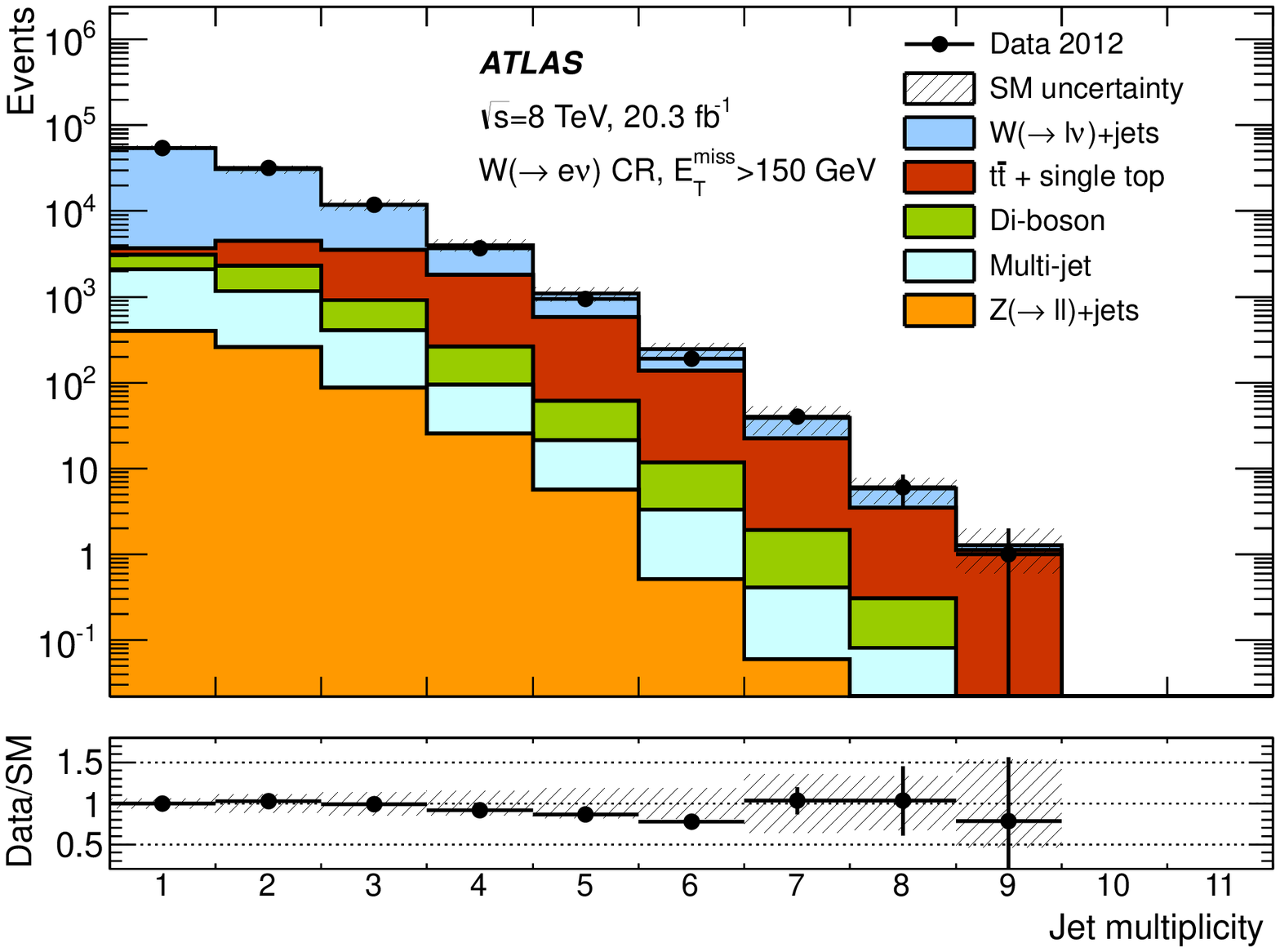}}
}
\end{center}
\caption{
Distributions of the measured (a) transverse mass of the identified electron and the missing transverse momentum, (b) $\met$, (c) leading jet $\pt$ and (d) jet multiplicity distributions
in the $\wen$+jets control region for the inclusive SR1 
selection, compared to
the background expectations.
The latter include the global normalization factors extracted from the data.
Where appropriate, the last bin of the distribution includes overflows.
The lower panels represent the ratio of data to MC expectations.
The error bands in the ratios include the statistical and experimental  uncertainties 
on the background expectations.
}
\label{fig:cr3}
\end{figure}

\begin{figure}[!ht]
\begin{center}
\mbox{
\subfigure[]{  \includegraphics[width=0.485\textwidth]{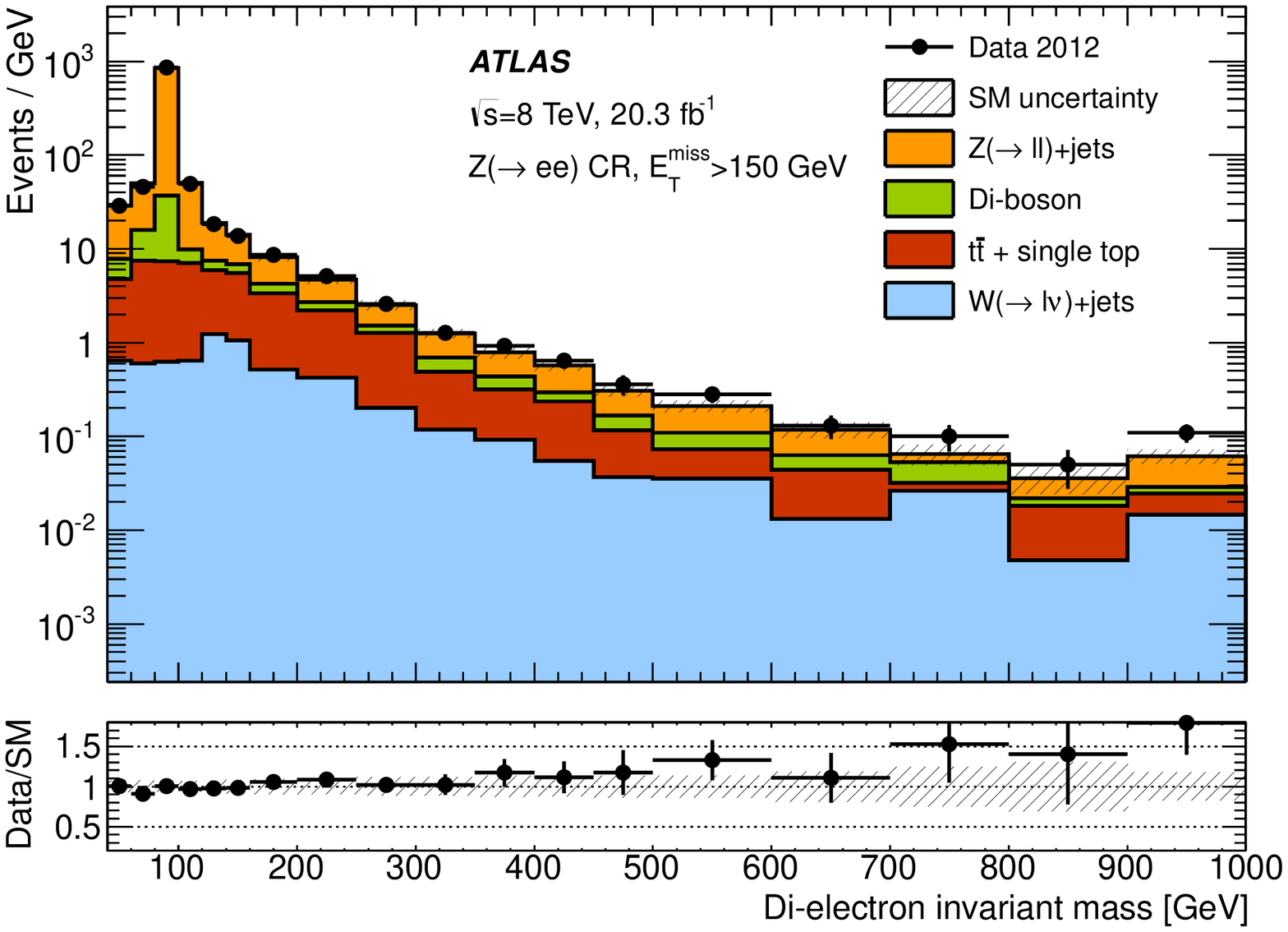}}
\subfigure[]{  \includegraphics[width=0.485\textwidth]{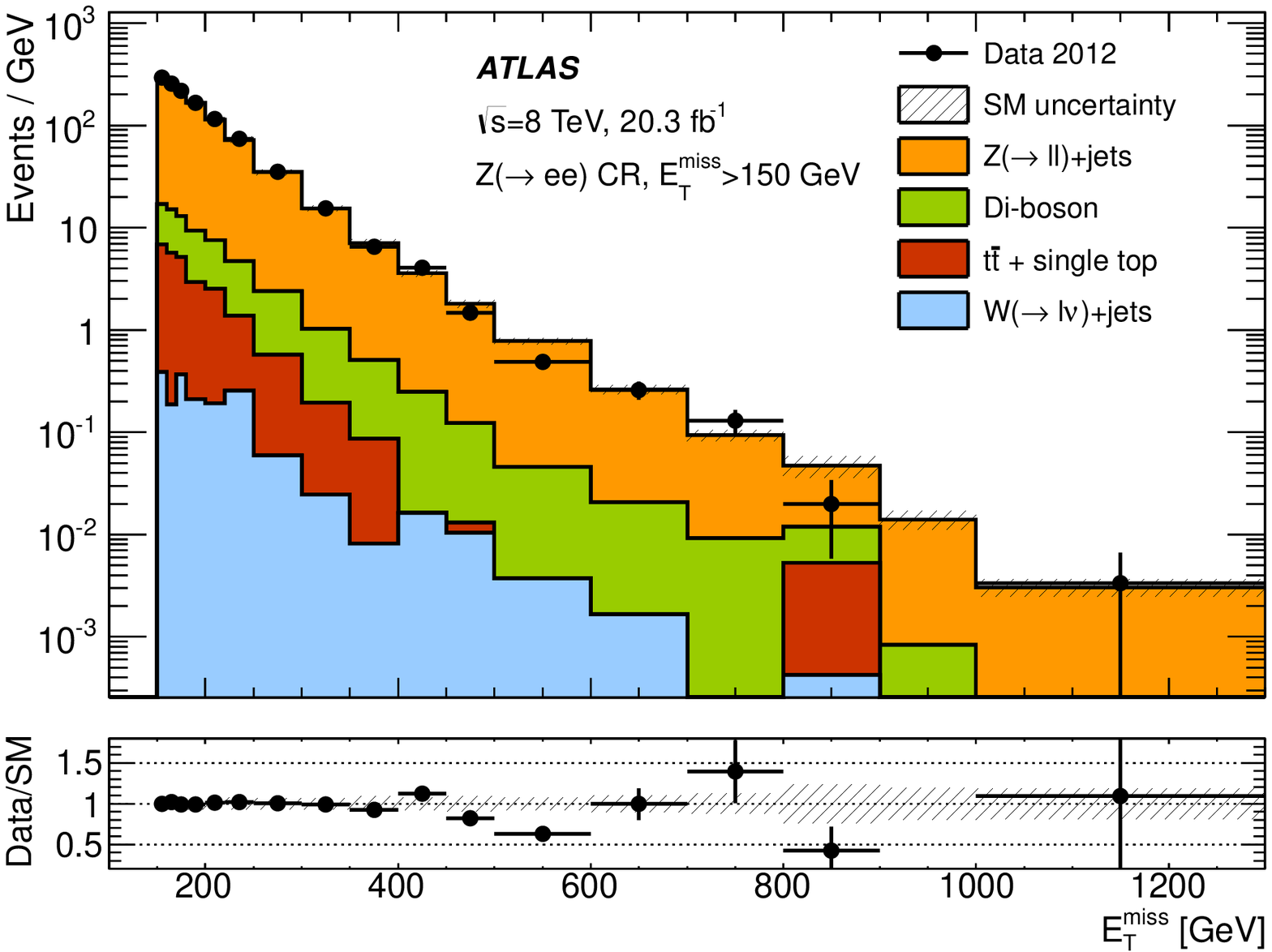}}
}
\mbox{
\subfigure[]{  \includegraphics[width=0.485\textwidth]{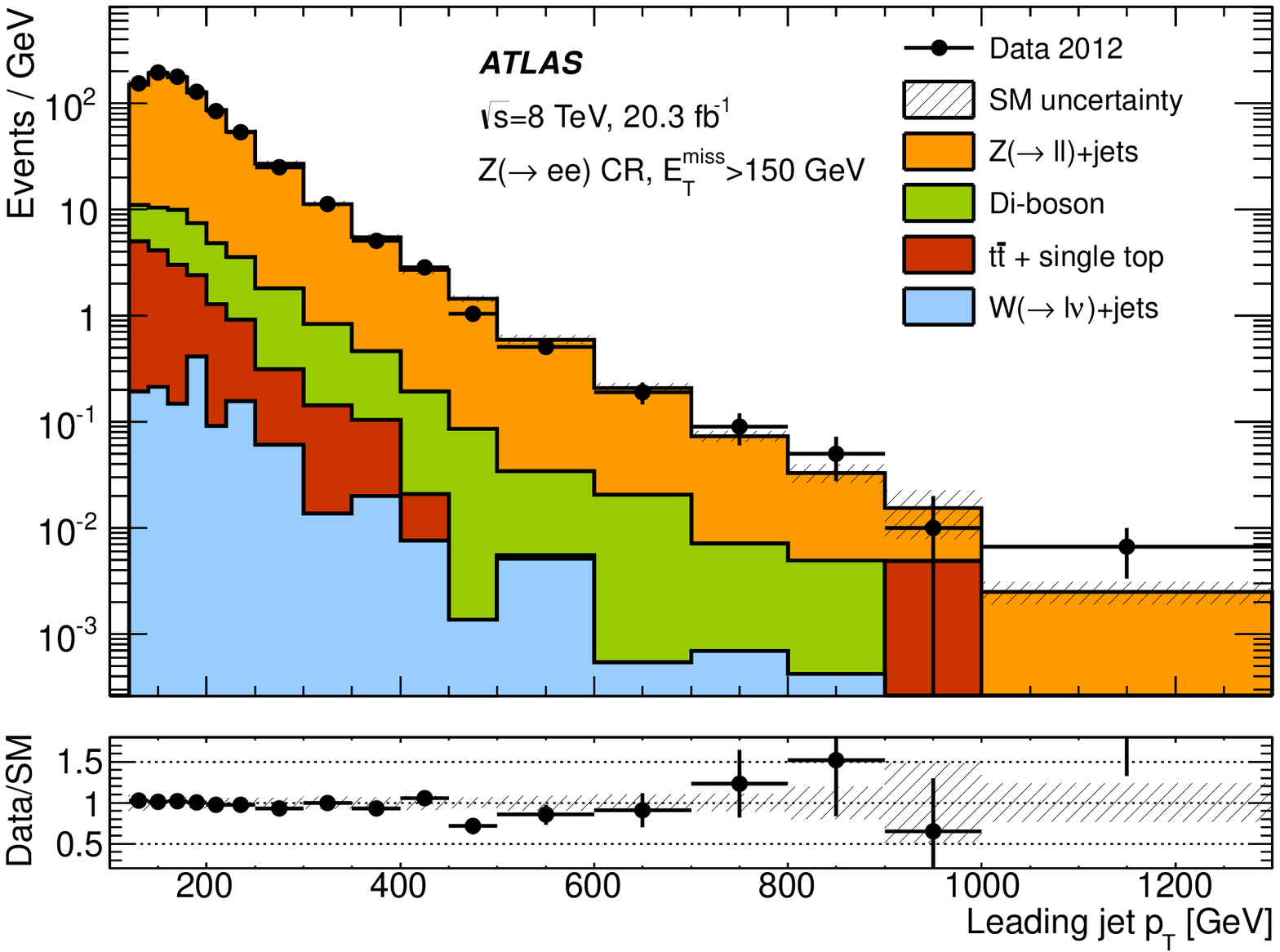}}
\subfigure[]{  \includegraphics[width=0.485\textwidth]{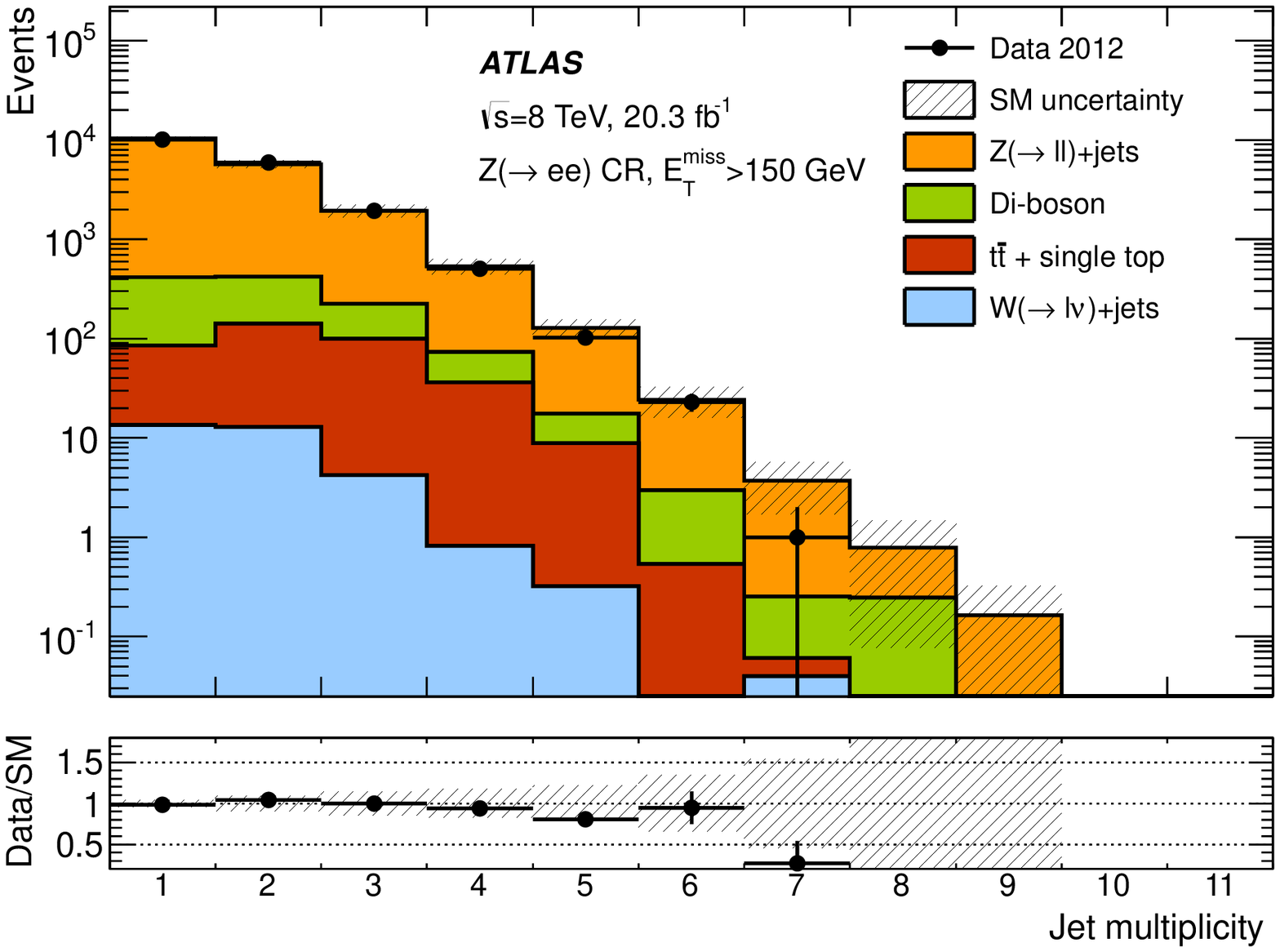}}
}
\end{center}
\caption{
Distributions of the measured (a) dilepton invariant mass, (b) $\met$, (c) leading jet $\pt$ and (d) jet multiplicity distributions
in the $\zee$+jets control region for the inclusive SR1 
selection, compared to
the background expectations.
The latter include the global normalization factors extracted from the data.
Where appropriate, the last bin of the distribution includes overflows.
The lower panels represent the ratio of data to MC expectations.
The error bands in the ratios include the statistical and experimental  uncertainties 
on the background expectations.
}
\label{fig:cr4}
\end{figure}

Figures~\ref{fig:cr1}--\ref{fig:cr4} 
show, for the SR1 monojet-like kinematic selection, some distributions  
in data in the different $W$+jets and $Z$+jets  
control regions compared to MC expectations.   In this case, the MC expectations 
are globally normalized to the data in the control regions, using normalization factors as explained below, so  
that a comparison of the shape of the different distributions in data and MC simulation can be made.  The MC expectations 
provide a fair description of the shapes in data but present harder $\met$ and leading-jet $\pt$ spectra. This 
is mainly attributed to an inadequate modelling of the boson $\pt$ distribution in the $W/Z$+jets MC samples. 

The data in the control regions and MC--based correction factors, determined from the SHERPA simulation, 
are used for each  of the signal selections (SR1--SR9) to estimate the electroweak background 
contributions from $W$+jets and $\znn$+jets processes. 
As an example, the $\wmn$+jets and $\znn$+jets background 
contributions to a given signal region,   
$N^{\wmn}_{\rm{signal}}$  and $N^{\znn}_{\rm{signal}}$, respectively,  
are determined using the  $\wmn$+jets control sample in data 
according to

\begin{equation}
N^{\wmn}_{\rm{signal}} = \frac{(N^{\rm{data}}_{\wmn,{\rm{control}}} - N^{{\rm{non}}-W/Z}_{\wmn,{\rm{control}}})}{N^{{\rm{MC}}}_{\wmn,{\rm{control}}}}  \times
N^{{\rm{MC}} (\wmn)}_{\rm{signal}} \times \xi_{\ell} \times \xi_{\rm{trg}} \times \xi_{\ell}^{\rm{veto}} 
\end{equation}
\noindent

and 

\begin{equation}
N^{\znn}_{\rm{signal}} = \frac{(N^{\rm{data}}_{\wmn,{\rm{control}}} - N^{{\rm{non}}-W/Z}_{\wmn,{\rm{control}}})}{N^{{\rm{MC}}}_{\wmn,{\rm{control}}}}  \times
N^{{\rm{MC}} (\znn)}_{\rm{signal}} \times \xi_{\ell} \times \xi_{\rm{trg}},    
\end{equation}
\noindent

\noindent
where $N^{{\rm{MC}} (\wmn)}_{\rm{signal}}$ and $N^{{\rm{MC}} (\znn)}_{\rm{signal}}$ denote, respectively, the $\wmn$+jets and $\znn$+jets 
background predicted by the MC simulation in the signal region, and $N^{\rm{data}}_{\wmn,{\rm{control}}}$,  
 $N^{\rm{MC}}_{\wmn,{\rm{control}}}$, and $N^{{\rm{non}}-W/Z}_{\wmn,{\rm{control}}}$ denote, in the control region, the 
number of $\wmn$+jets  candidates  in data and  $W/Z$+jets MC simulation,  and 
the  non-$W/Z$ background contribution, respectively.  
The $N^{{\rm{non}}-W/Z}_{\wmn,{\rm{control}}}$ term  refers mainly to top-quark and diboson processes,  but also 
includes  contributions from multijet processes determined using data.  Finally, $\xi_{\rm{\ell}}$, $\xi_{\ell}^{\rm{veto}}$, and $\xi_{\rm{trg}}$ account for possible data--MC differences in the  
lepton identification, lepton vetoes, and trigger efficiencies, respectively; they typically depart 
from unity by less than 1$\%$. 
 The MC-to-data normalization factors  
(the $(N^{\rm{data}}_{\wmn,{\rm{control}}} - N^{{\rm{non}}-W/Z}_{\wmn,{\rm{control}}})/N^{{\rm{MC}}}_{\wmn,{\rm{control}}}$ term in Eq. (1))
for each process  vary between about 0.9 and 0.6 as the required minimum $\met$ increases from 150~GeV to 700~GeV, 
and account for the tendency of the MC expectations for $W/Z$+jets processes to exceed the data in the control regions (see, for example, Fig.~\ref{fig:cr1}).  
Similarly, bin-by-bin correction factors are used to correct the shape of the different 
distributions in the signal regions. 
 
As already mentioned, the different background contributions in the signal regions from 
$\wln$+jets processes (with $\ell = e, \mu$)  are constrained using correction factors obtained from 
the corresponding control regions.  In the case of the $\wtn$+jets contributions, the 
correction factors from the  
$\wen$+jets  control regions are used.  
For each of the signal regions, four separate sets of correction factors are 
considered to constrain the dominant $\znn$+jets background contribution, following Eq. (2),  as determined separately using $\zll$+jets and $\wln$+jets  
control samples. The four resulting  $\znn$+jets background estimations in each signal region are found to be consistent within uncertainties and  
are statistically combined using the BLUE (Best Linear Unbiased Estimate)~\cite{blue} method, which 
takes into account correlations of systematic uncertainties.

\subsection{Multijet background}

The multijet background with large $\met$ mainly originates 
from the misreconstruction of the energy of a 
jet in the calorimeter and to a lesser extent from the presence 
of neutrinos in the final state due to heavy-flavour decays.  
The multijet background is determined from data, using a  {\it jet smearing} method as
described in Ref.~\cite{Aad:2012fqa}, which relies  on the assumption that the $\met$ of multijet events 
is dominated by fluctuations in the detector response to jets measured in the data. 
For the SR1  and SR2 selections, the multijet background constitutes 
about $2\%$ and $0.7\%$ of the total background, respectively, and is   
 below $0.5\%$ for the rest of the signal regions with higher $\met$ thresholds.

\subsection{Non-collision background}

Detector noise, beam-halo and cosmic muons leading to large energy deposits in the calorimeters represent a significant portion of data acquired by $\met$ triggers. 
These non-collision backgrounds resemble the topology of monojet-like final states and require a dedicated strategy to suppress them. The selection described in Sect.~\ref{sec:evt} is expected to maintain the non-collision background below the percent level. 
The rate of the fake jets due to cosmic muons surviving the selection criteria, as measured in dedicated cosmic datasets, is found negligible with respect to the rate of data in the  
monojet-like signal regions. The major source of the non-collision backgrounds is thus beam-halo muons. Since jets due to collisions are expected to be in time with the bunch crossing, 
an assumption is made that all events containing a leading jet within the out-of-time window are due to beam-induced backgrounds. The characteristic shape of the fake jets due to beam-halo muons is extracted from signal-region events identified as beam-induced backgrounds based on the spatial alignment of the signals in the calorimeter and the muon system~\cite{Aad:2013zwa}. 
The level of non-collision background in the signal region is extracted as 
\begin{equation}
N_\mathrm{NCB}^\mathrm{SR}=N_{-10<t<-5}^\mathrm{SR}\times\frac{N^\mathrm{NCB}}{N_{-10<t<-5}^\mathrm{NCB}},
\end{equation}
where $N_{-10<t<-5}^\mathrm{SR}$ denotes the number of events in the signal region with a leading jet in the range $-10 {\ \rm{ns}} < t < -5 {\ \rm{ns}}$, $N_{-10<t<-5}^\mathrm{NCB}$ is the number of identified beam-induced background events there and $N^\mathrm{NCB}$ represents all identified events in the signal region.  
The results of this study indicate that the non-collision background in the 
different signal regions is negligible.

 \section{Systematic uncertainties}
 \label{sec:syst}
 Several sources of systematic uncertainty are considered in the determination of the background
contributions.  
Uncertainties on the absolute jet energy scale and resolution~\cite{Aad:2011he}  
translate  into an uncertainty on
the total background which varies from  $0.2\%$ for SR1 and $1\%$ for SR7 to   $3\%$ for SR9.
Uncertainties on the  $\met$ reconstruction 
introduce an  uncertainty on
the total background which varies from  $0.2\%$ for SR1 and $0.7\%$ for SR7 to   $1\%$ for SR9. 
Uncertainties of the order of 1$\%$--2$\%$ on
the simulated lepton identification and reconstruction efficiencies, energy/momentum scale and resolution, and 
a 0.5$\%$--1$\%$ uncertainty on the track isolation efficiency  translate, altogether, into a  
$1.4\%$, $1.5\%$, and $2\%$ uncertainty in the total background for the SR1, SR7, and  SR9 selections, respectively.
Uncertainties of the order of $1\%$ on  the $\met$ trigger simulation  at low $\met$ and on the efficiency of the lepton triggers used to
define the electron and muon control samples  translate into uncertainties on the total background of about  0.1$\%$ for SR1 and  
become  negligible for the rest of the signal regions. 

The top-quark-related background contributions, as determined from MC simulations (see Sect.~\ref{sec:mc}), are validated in 
dedicated validation regions defined similarly to the $\wen$+jets and $\wmn$+jets control regions with 
$\Delta \phi (\ptmi , \rm{jet}) > 0.5$ and by requiring the presence of 
two $b$-tagged jets in the final state with jet $|\eta| < 2.4$.  The comparison between data and MC expectations 
in those validation regions leads to uncertainties on the top-quark background yields which increase from  
$20\%$  for SR1 to 100$\%$ for SR7 and SR9.  This translates into uncertainties in the total background 
expectations which vary from  $0.7\%$ for SR1 and 2.7$\%$ for SR7 to  4$\%$ for SR9.  
Similarly, uncertainties on the simulated diboson background yields include  
uncertainties in the MC generators and the modelling of parton showers employed,  
variations in the set of parameters that govern the parton showers and the amount of 
initial- and final-state soft gluon radiation, and uncertainties 
due to the choice of renormalization and factorization scales and PDF.  
This introduces an uncertainty on the diboson background
expectation which increases from 20$\%$ for SR1  to 30$\%$ for SR7  and  80$\%$ for SR9.
This  
results in  an uncertainty on the total 
background of 0.7$\%$, 2.3$\%$, and 3$\%$ for the SR1, SR7, and SR9 selections, respectively. 

Uncertainties on the $W/Z$+jets modelling include: 
variations of the renormalization, factorization, and parton-shower matching scales  and PDF 
in the  {SHERPA}  $W/Z$+jets background samples; and uncertainties on the parton-shower model considered.
In addition, the effect of NLO electroweak corrections on the $W$+jets to $Z$+jets ratio is taken into account~\cite{Denner:2012ts,Denner:2011vu,Denner:2009gj}. 
Altogether, this translates into an uncertainty on the total background of about 1$\%$ for SR1 and SR7 and 3$\%$ for SR9.

Uncertainties on the multijet and $\gamma$+jets background contamination of 100$\%$ and $50\%$, respectively, in the $\wen$+jets control region,  
 propagated to the $\znn$+jets background determination in the signal regions,  introduce an additional 1$\%$ uncertainly on the total 
background for the SR9 selection.  The uncertainty on the multijet background contamination  in the signal regions leads to a 2$\%$ and 0.7$\%$ 
uncertainty on the total background for the SR1 and SR2 selections, respectively. 
Finally, the  impact of uncertainty on the total integrated luminosity, which partially cancels in the 
data-driven determination of the SM background, is negligible.

 After including statistical uncertainties on the data and MC expectations in control regions and on the MC expectations in the signal regions, 
 the total background in the signal regions  is  determined with uncertainties that 
vary from  $2.7\%$ for SR1 and 6.2$\%$  for SR7  to $14\%$  for SR9.

%
%

\subsection{Signal systematic uncertainties}
\label{sec:signalsys}

Several  sources of systematic uncertainty on the predicted signal yields are considered 
for each of the models for new physics. The uncertainties are computed separately for each signal region by  
varying the model parameters (see Sect.~\ref{sec:results}).

Experimental uncertainties include: those related to the jet and $\met$ reconstruction, energy
scales and resolutions; those on the proton beam energy, as considered by 
simulating samples with the lower and upper allowed values given in
Ref.~\cite{Wenninger:1546734};  a $1\%$ uncertainty on the trigger efficiency,  affecting only SR1; and 
the $2.8\%$ uncertainty on the integrated luminosity. Other uncertainties related to the track veto or 
the jet quality requirements are negligible ($< 1\%$).

Uncertainties affecting the signal
acceptance times efficiency $A \times \epsilon$, related to the generation of the signal samples, 
include:  uncertainties on the  
modelling of the initial- and final-state
gluon radiation, as determined using simulated samples with 
modified parton-shower parameters, by  factors of two and one half, that enhance or 
suppress the parton radiation; uncertainties due to PDF and variations of the $\alpha_{\rm s}(m_Z)$ value employed,  
as computed from the envelope of CT10, MRST2008LO and NNPDF21LO error sets; 
and the choice of renormalization/factorization scales,  and the parton-shower 
matching scale settings, varied by factors of two and one half.

In addition, theoretical uncertainties on the predicted cross sections, including PDF and renormalization/factorization 
scale  uncertainties, are computed separately for the different models.

\section{Results and interpretation}
\label{sec:results}

\begin{figure}[!ht]
\begin{center}
\mbox{
\subfigure[]{\includegraphics[width=0.485\textwidth]{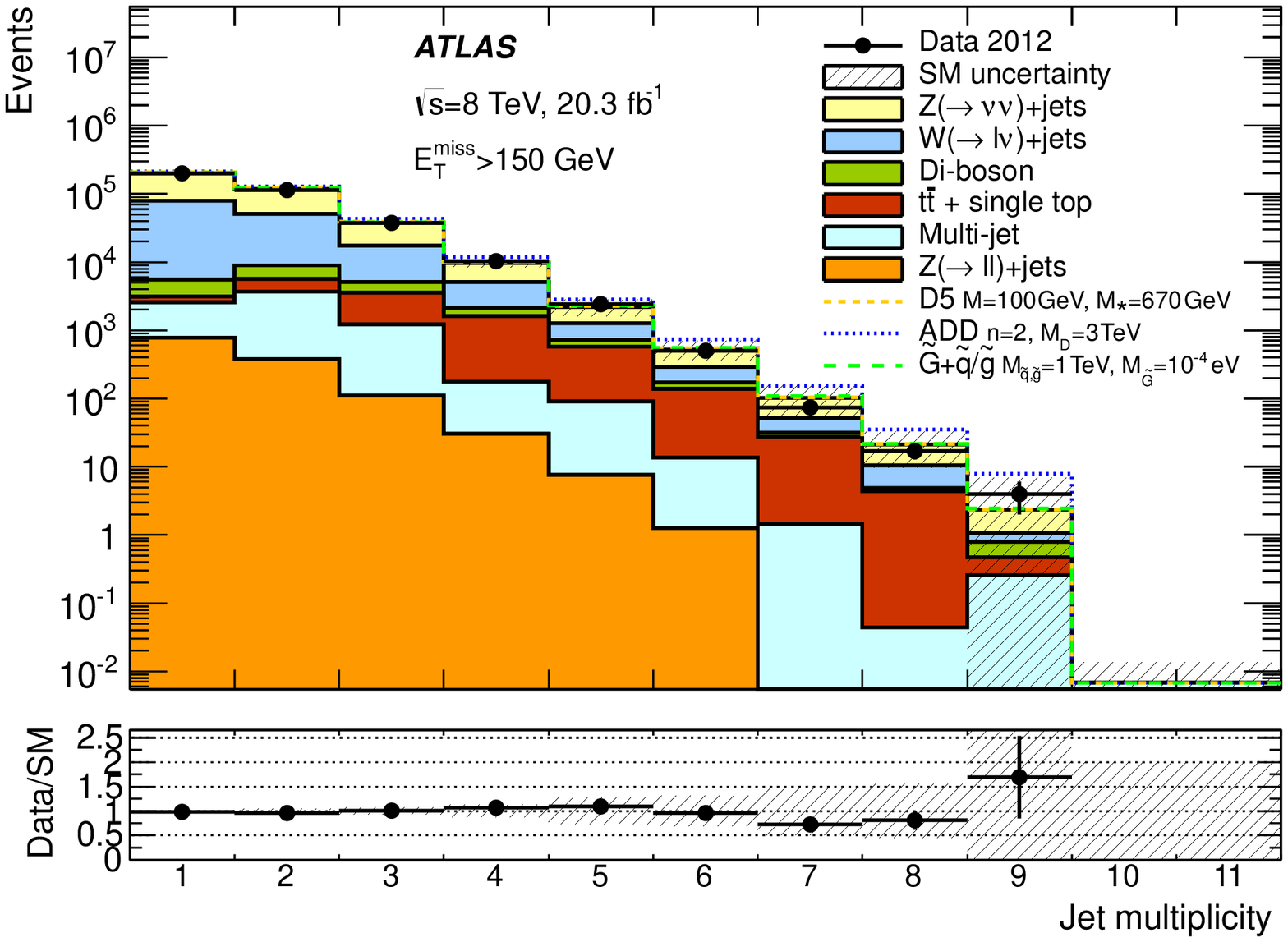}}
\subfigure[]{\includegraphics[width=0.485\textwidth]{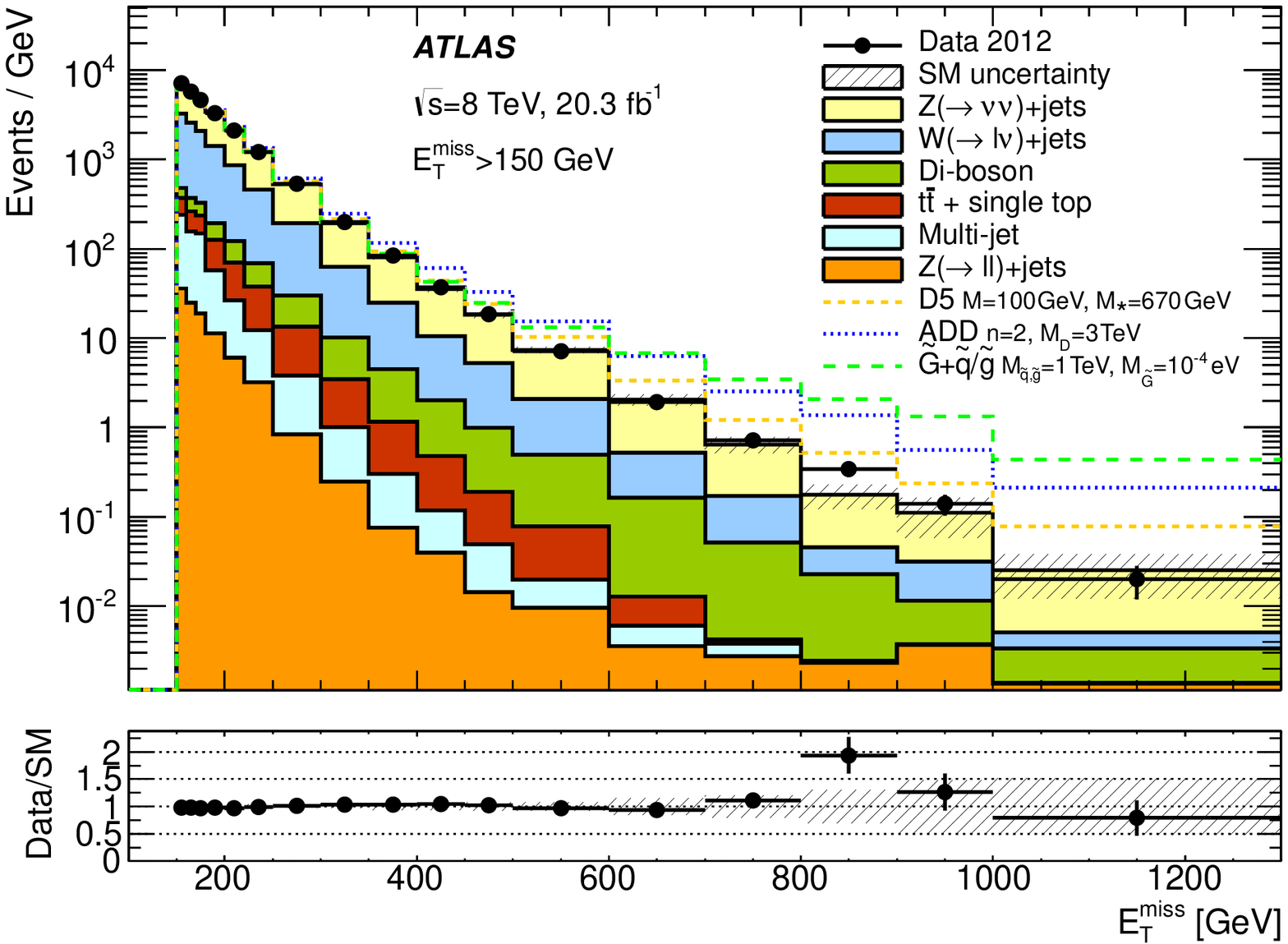}}
}
\mbox{
\subfigure[]{\includegraphics[width=0.485\textwidth]{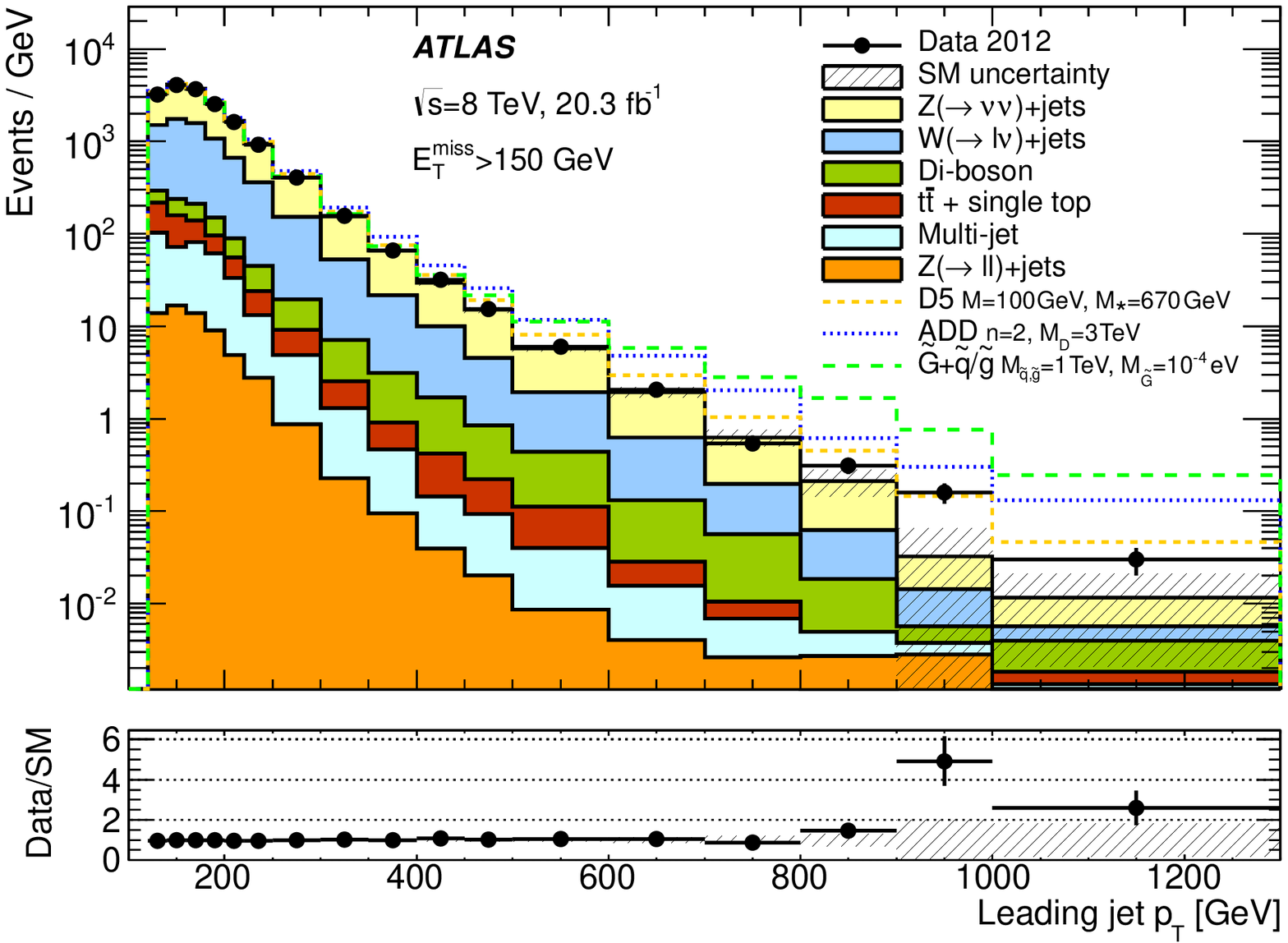}}
\subfigure[]{\includegraphics[width=0.485\textwidth]{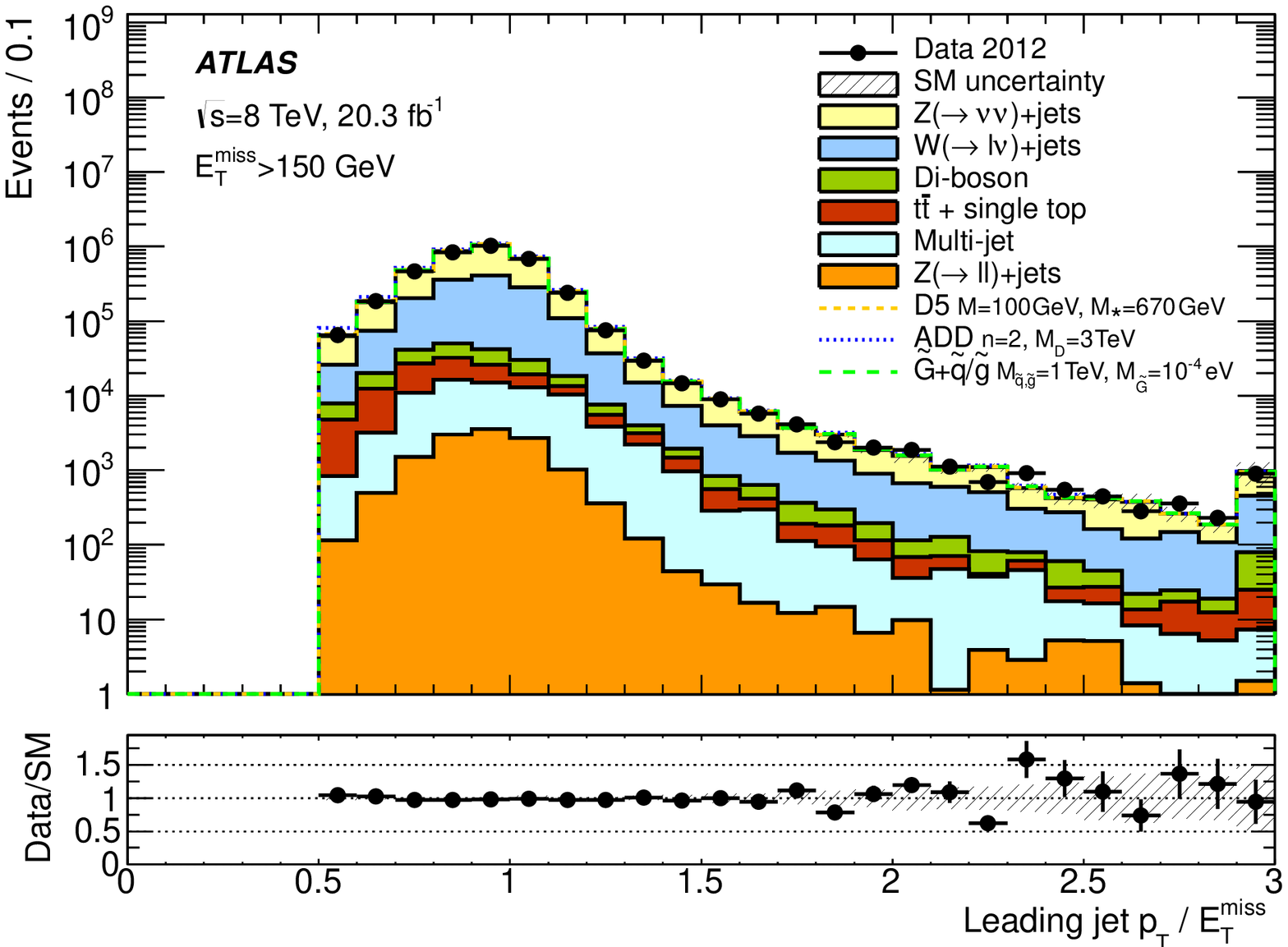}}
}
\end{center}
\caption{
Measured distributions of (a) the jet multiplicity,  (b) $\met$, (c) leading jet $\pt$, and (d) the leading jet $\pt$ to $\met$ ratio  
for the SR1  selection compared to the SM expectations.  The $\znn$+jets contribution is shown as constrained by the $\wmn$+jets control sample.
Where appropriate, the last bin of the distribution includes overflows.
For illustration purposes, the distribution of different ADD, WIMP and GMSB scenarios  are included.
The error bands in the ratios shown in lower panels include both the statistical and systematic uncertainties 
on the background expectations.
}
\label{fig:srfig1}
\end{figure}

\begin{figure}[!ht]
\begin{center}
\mbox{
\includegraphics[width=0.485\textwidth]{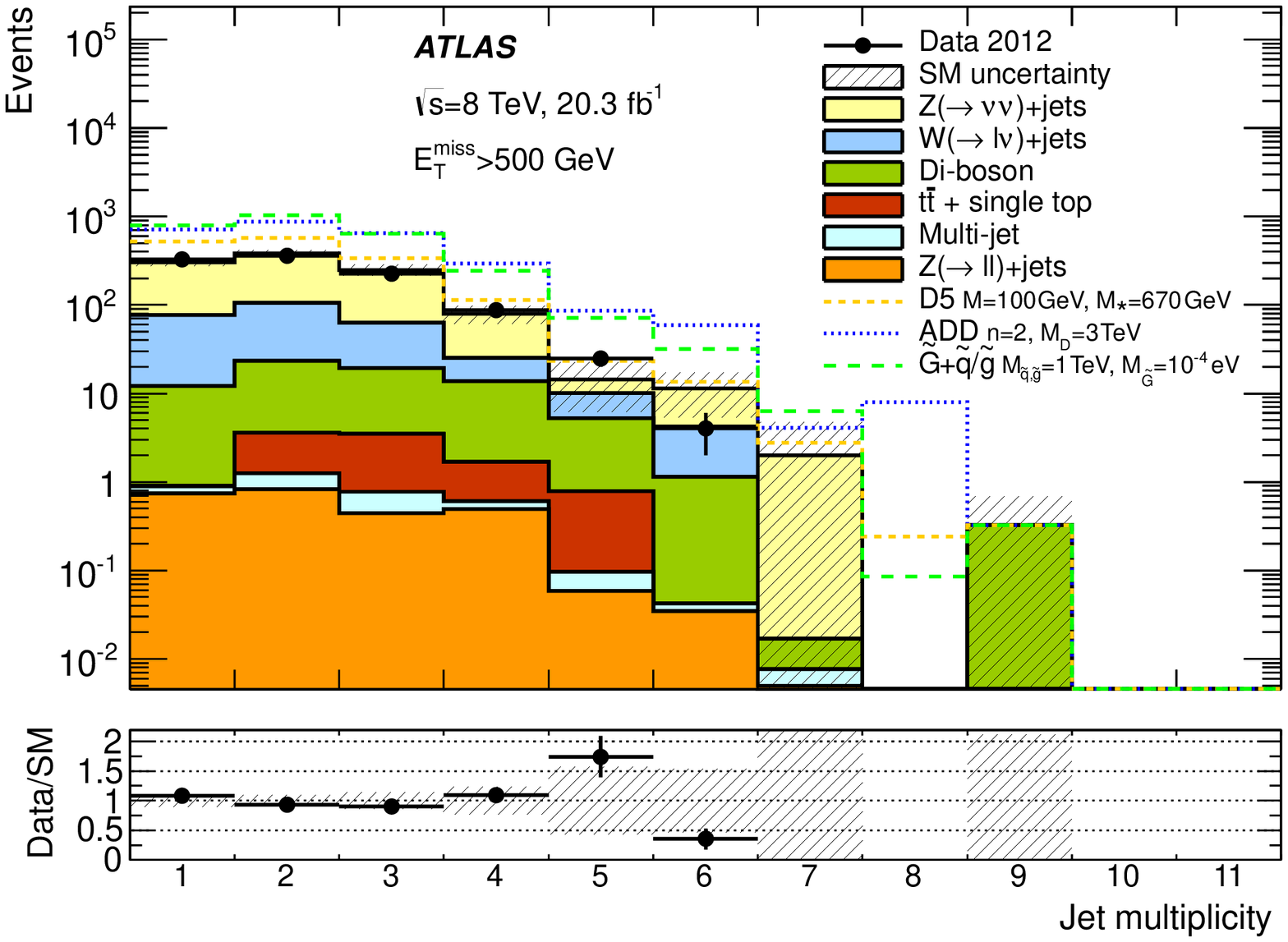}
\includegraphics[width=0.485\textwidth]{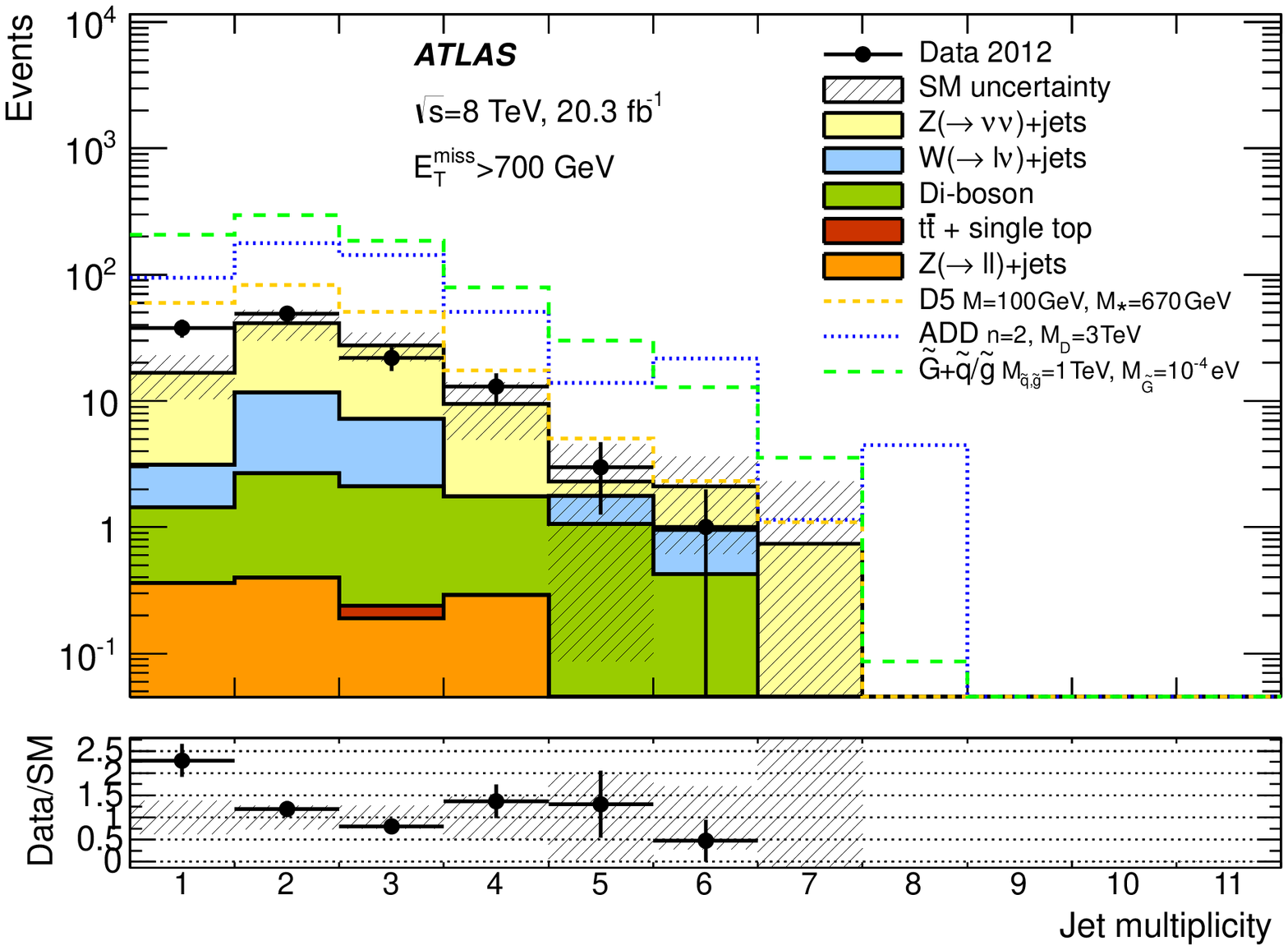}
}
\mbox{
\includegraphics[width=0.485\textwidth]{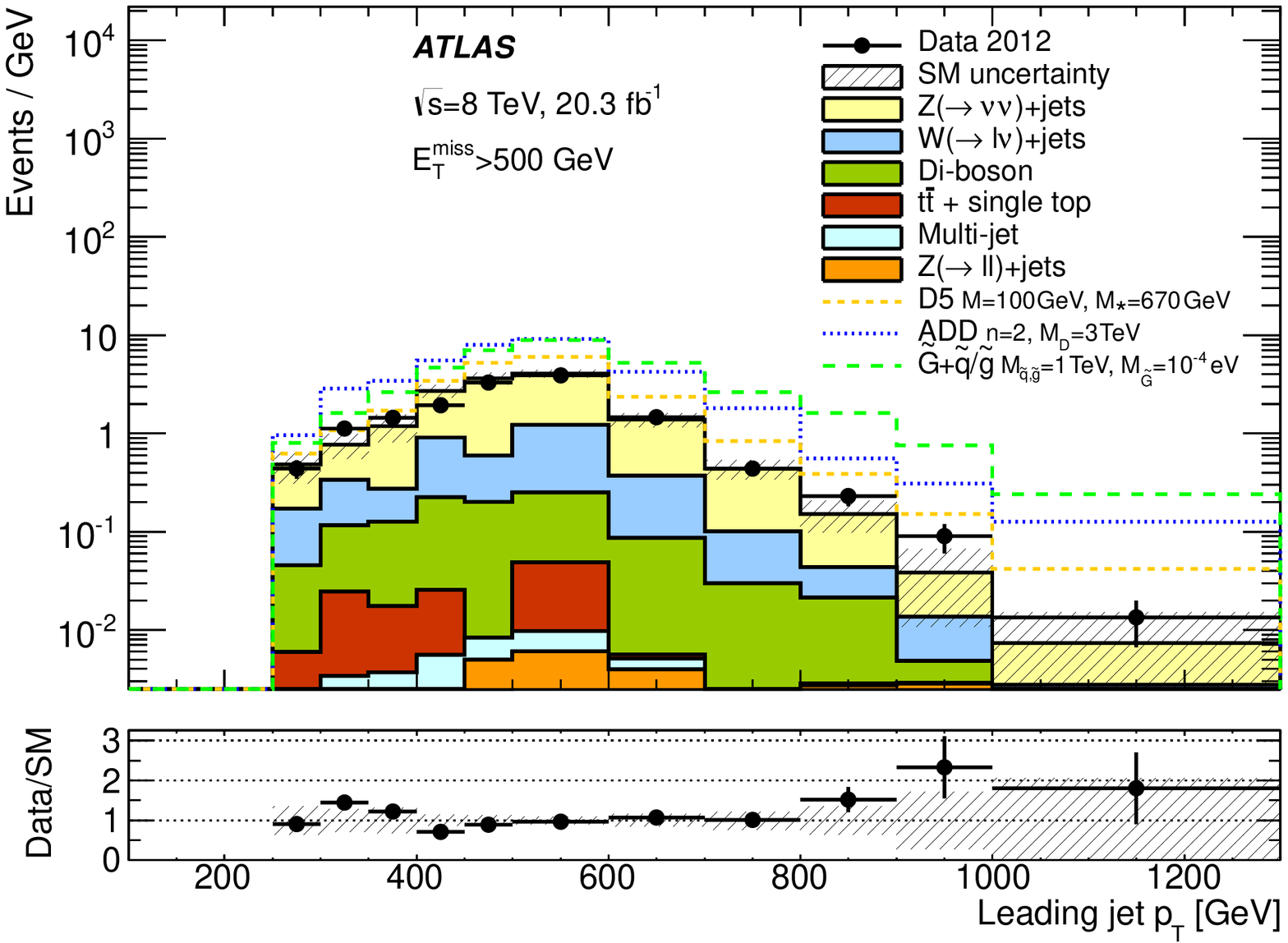}
\includegraphics[width=0.485\textwidth]{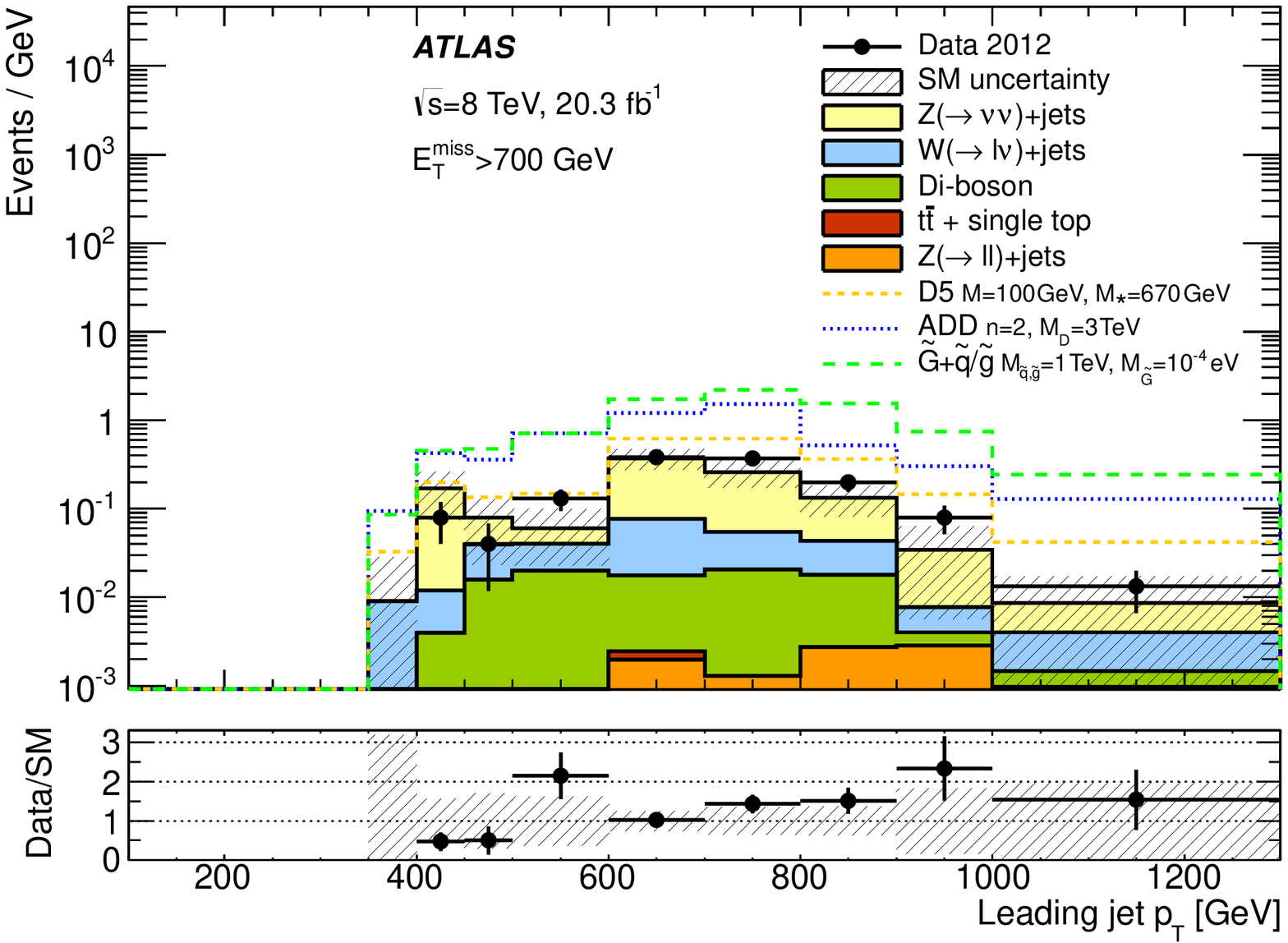}
}
\mbox{
\subfigure[]{\includegraphics[width=0.485\textwidth]{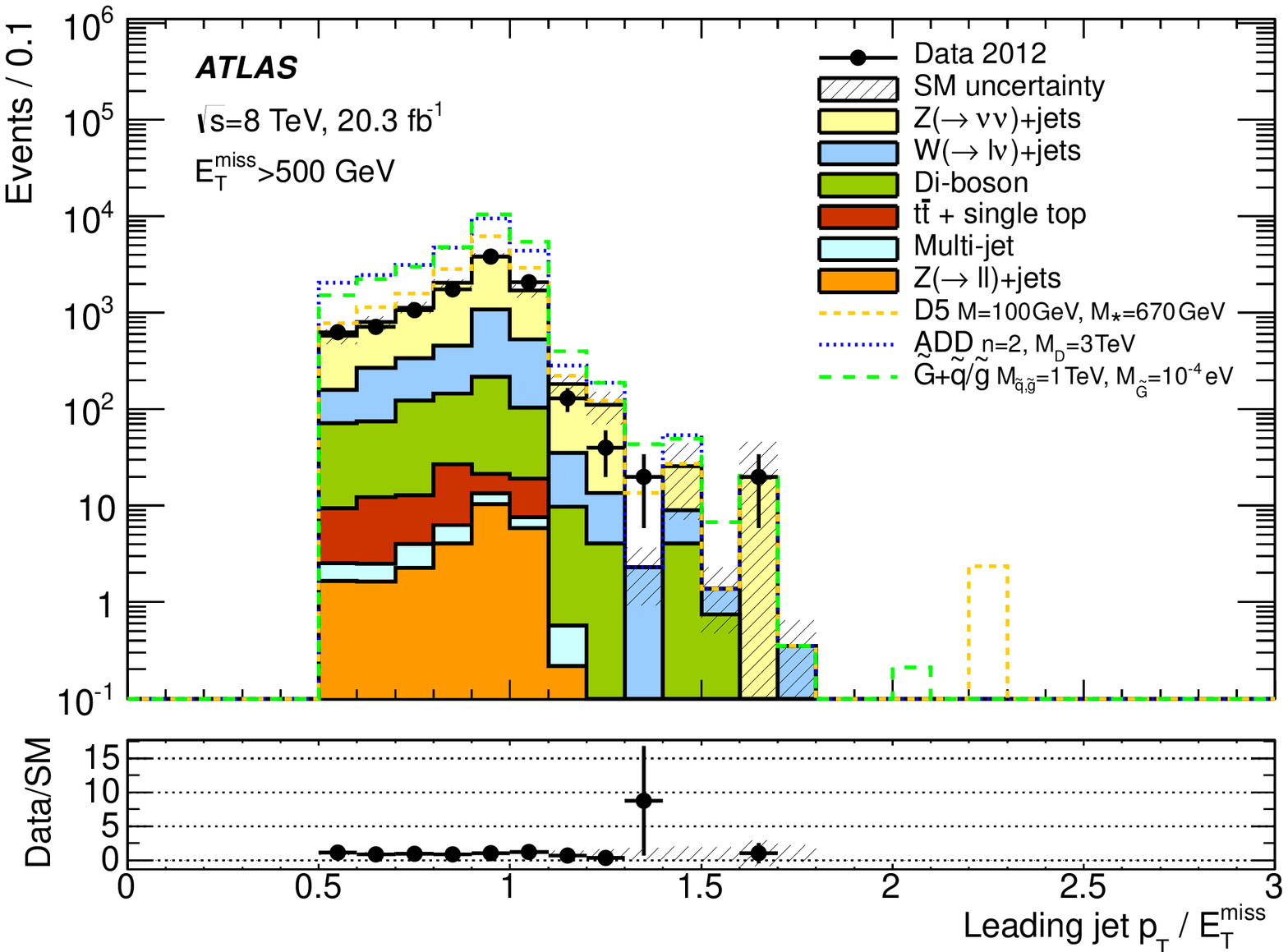}}
\subfigure[]{\includegraphics[width=0.485\textwidth]{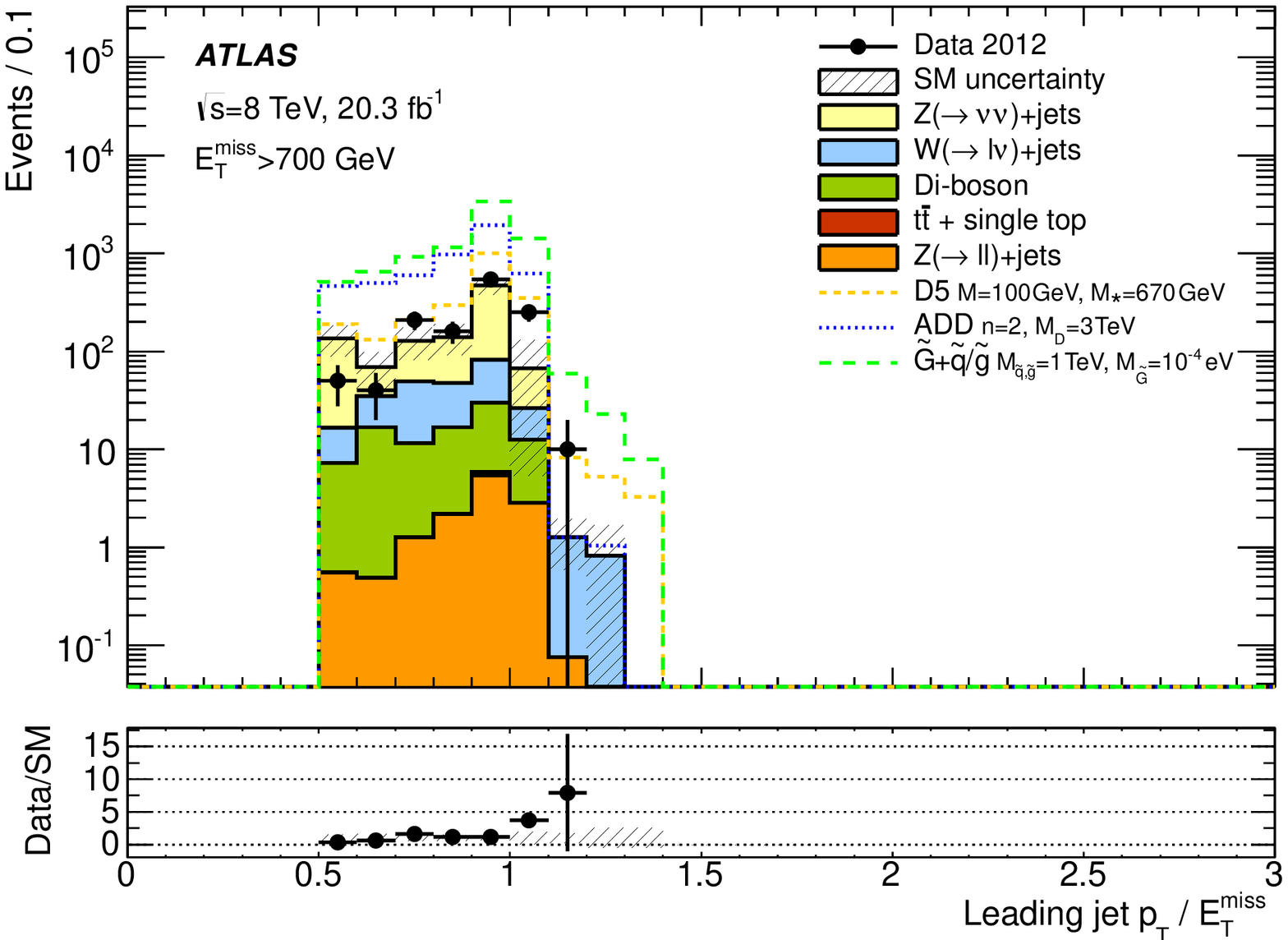}}
}
\end{center}
\caption{
Measured distributions of the jet multiplicity,  leading jet $\pt$, and the leading jet $\pt$ to $\met$ ratio
for (a) SR7  and (b) SR9 selections compared to the SM expectations. The $\znn$+jets    contribution is       
shown as constrained by the $\wmn$+jets control sample.
Where appropriate, the last bin of the distribution includes overflows.
For illustration purposes, the distribution of different ADD, WIMP and GMSB scenarios  are included.
The error bands in the ratios shown in lower panels include both the statistical and systematic uncertainties
on the background expectations.
}
\label{fig:srfig2}
\end{figure}

The data and the expected SM expectations in the different signal regions are presented in 
Tables~\ref{tab:sr12345}~and~\ref{tab:sr6789}.  In general, good agreement is observed between the
data and the SM expectations.  
The largest difference between the number of events in data and the  expectations is observed in the signal region SR9,  
corresponding to a 1.7$\sigma$ deviation with a $p$-value of 0.05, consistent with the background-only hypothesis. 
Figures~\ref{fig:srfig1}  and~\ref{fig:srfig2} show 
several measured distributions in data 
compared to the SM expectations for SR1, and SR7 and SR9, respectively. For illustration 
purposes, the distributions  include the impact of different ADD, WIMP, and GMSB SUSY scenarios.  


\begin{table}[!ht]
\caption{Data and SM background expectation in the signal region for the SR1--SR5 selections.
For the SM expectations both the statistical and systematic uncertainties are included.
In each signal region, the individual uncertainties for the different background processes can be correlated, 
and do not necessarily add in quadrature to the total background uncertainty.
}
\begin{center}
\begin{footnotesize}
\begin{tabular*}{\textwidth}{@{\extracolsep{\fill}}lrrrrr}\hline
{\bf  Signal Region}  & SR1 & SR2 & SR3 & SR4 & SR5\\
Observed events &  364378 & 123228 & 44715 & 18020 & 7988 \\ \hline
SM expectation &  372100 $\pm$ 9900 & 126000 $\pm$ 2900 & 45300 $\pm$ 1100 & 18000 $\pm$ 500 & 8300 $\pm$ 300 \\ \hline
$\znn$        &  217800 $\pm$ 3900  & 80000 $\pm$ 1700  & 30000 $\pm$ 800  & 12800 $\pm$ 410  & 6000 $\pm$ 240 \\
$\wtn$        &  79300 $\pm$ 3300 & 24000 $\pm$ 1200 & 7700 $\pm$ 500 & 2800 $\pm$ 200 & 1200 $\pm$ 110 \\
$\wen$        &  23500 $\pm$ 1700 & 7100 $\pm$ 560 & 2400 $\pm$ 200 & 880 $\pm$ 80 & 370 $\pm$ 40 \\
$\wmn$        &  28300 $\pm$ 1600 & 8200 $\pm$ 500 & 2500 $\pm$ 200 & 850 $\pm$ 80 & 330 $\pm$ 40 \\
$\zmm$        &   530 $\pm$ 220 & 97 $\pm$ 42 & 19 $\pm$ 8 & 7 $\pm$ 3 & 4 $\pm$ 2 \\
$\ztt$        &   780 $\pm$ 320 & 190 $\pm$ 80 & 45 $\pm$ 19 & 14 $\pm$ 6 & 5 $\pm$ 2 \\
$\ttbar$, single top   & 6900 $\pm$ 1400 & 2300 $\pm$ 500 & 700 $\pm$ 160 & 200 $\pm$ 70 & 80 $\pm$ 40 \\      
Dibosons      &  8000 $\pm$ 1700 & 3500 $\pm$ 800 & 1500 $\pm$ 400 & 690 $\pm$ 200 & 350 $\pm$ 120 \\
Multijets     &  6500 $\pm$ 6500 & 800 $\pm$ 800 & 200 $\pm$ 200 & 44 $\pm$ 44 & 15 $\pm$ 15 \\ \hline\hline
\end{tabular*}
\end{footnotesize}
\end{center}
\label{tab:sr12345}
\end{table}

%

\begin{table}[!ht]
\caption{Data and SM background expectation in the signal region for the SR6--SR9 selections.
For the SM expectations both the statistical and systematic uncertainties are included.
In each signal region, the individual uncertainties for the different background processes can be correlated,    
and do not necessarily add in quadrature to the total background uncertainty.
}
\begin{center}
\begin{footnotesize}
\begin{tabular*}{\textwidth}{@{\extracolsep{\fill}}lrrrr}\hline
{\bf  Signal Region}  & SR6 & SR7 & SR8 & SR9 \\
Observed events  & 3813 & 1028 & 318 & 126\\ \hline
SM expectation &  4000 $\pm$ 160 & 1030 $\pm$ 60 & 310 $\pm$ 30 & 97 $\pm$ 14 \\ \hline
$\znn$        &  3000 $\pm$ 150  & 740 $\pm$ 60  & 240 $\pm$ 30  & 71 $\pm$ 13 \\
$\wtn$        &  540 $\pm$ 60 & 130 $\pm$ 20 & 34 $\pm$ 8 & 11 $\pm$ 3\\                                                  
$\wen$        &    170 $\pm$ 20 & 43 $\pm$ 7 & 9 $\pm$ 3 & 3 $\pm$ 1\\
$\wmn$        &    140 $\pm$ 20 & 35 $\pm$ 6 & 10 $\pm$ 2 & 2 $\pm$ 1\\ 
$\zmm$        &   3 $\pm$ 1 & 2 $\pm$ 1 & 1 $\pm$ 1 & 1 $\pm$ 1 \\ 
$\ztt$        &  2 $\pm$ 1 & 0 $\pm$ 0 & 0 $\pm$ 0 & 0 $\pm$ 0 \\
$\ttbar$, single top & 30 $\pm$ 20 & 7 $\pm$ 7 & 1 $\pm$ 1 & 0 $\pm$ 0\\ 
Dibosons      &  183 $\pm$ 70 & 65 $\pm$ 35 & 23 $\pm$ 16 & 8 $\pm$ 7 \\
Multijets     & 6 $\pm$ 6 & 1 $\pm$ 1 & 0 $\pm$ 0 & 0 $\pm$ 0 \\ \hline\hline
\end{tabular*}
\end{footnotesize}
\end{center}
\label{tab:sr6789}
\end{table}

The agreement between the data and the SM expectations for the total number of events  
in the different signal regions is translated into
model-independent 90$\%$  and 95$\%$ confidence level (CL) upper limits 
on the visible cross section, defined as the production cross section times acceptance times efficiency $\sigma \times A \times \epsilon$,   
using the $CL_s$ modified frequentist approach~\cite{Read:2002hq} and 
considering  the systematic 
uncertainties on the SM backgrounds 
and the  uncertainty on the quoted integrated luminosity.
The results are presented in Table~\ref{tab:indep}. 
Values of $\sigma \times A \times \epsilon$  
above 599~fb--2.9~fb (726~fb--3.4~fb) are excluded at 90$\%$ CL (95$\%$ CL)
for  SR1--SR9 selections, respectively.  
Typical event selection efficiencies varying from $88\%$ for SR1 and  $83\%$ for SR3 to  $82\%$ for SR7 and $81\%$ for SR9 are found in simulated $\znn$+jets background processes.


\begin{table}
\caption{
Observed and expected 90$\%$ CL and 95$\%$ CL upper limits on the
product of cross section, acceptance and efficiency, $\sigma \times A \times \epsilon$, for the SR1--SR9 selections.}
\begin{center}
\begin{footnotesize}
\begin{tabular*}{\textwidth}{@{\extracolsep{\fill}}lcc}\hline
\multicolumn{3}{c}{Upper limits  on $\sigma \times A \times \epsilon$ [fb]} \\ \hline 
Signal Region & 90$\%$ CL Observed (Expected) & 95$\%$ CL Observed (Expected) \\ \hline
SR1 & 599 (788)   & 726 (935)   \\
SR2 & 158 (229)   & 194 (271)   \\
SR3 & 74 (89)     & 90  (106)    \\
SR4 & 38 (43)     & 45  (51)    \\
SR5 &  17 (24)    &  21 (29)    \\
SR6 &  10 (14)     &  12  (17)      \\
SR7 &  6.0 (6.0)      &  7.2  (7.2)      \\
SR8 &   3.2 (3.0)     &   3.8 (3.6)     \\
SR9 &   2.9 (1.5)     &   3.4 (1.8)     \\ \hline\hline
\end{tabular*}
\end{footnotesize}
\label{tab:indep}
\end{center}
\end{table}


\subsection{Large extra spatial dimensions}

The results are translated into limits  on the parameters of the ADD model.
The typical $A \times \epsilon$ of the selection criteria vary, as the number of extra dimensions $n$ increases from $n=2$ to $n=6$, 
between 23$\%$ and 33$\%$ for SR1  and between $0.3\%$ and 1.4$\%$ for SR9, and 
are approximately independent of $M_D$. 

The experimental uncertainties related to the jet and $\met$
scales and resolutions introduce, when combined, uncertainties in the signal yields which vary between
2$\%$ and $0.7\%$ for SR1 and between $8\%$ and 5$\%$ for SR9, with increasing $n$.   
The uncertainties on the proton beam energy  
result in uncertainties on the signal cross sections which 
vary between 2$\%$ and 5$\%$ with increasing $n$, and uncertainties on the signal acceptance of 
about 1$\%$ for SR1 and  3$\%$--4$\%$ for SR9.     
The uncertainties related to the modelling of the initial- and final-state 
gluon radiation  
translate into 
uncertainties on the ADD signal acceptance which  
vary with increasing $n$ between 2$\%$ and 3$\%$ in SR1 and between  11$\%$ and 21$\%$ in SR9. 
The  uncertainties due to PDF, affecting both the predicted signal cross section and the signal acceptance, 
result in uncertainties on the signal yields  
which vary with increasing $n$ between 18$\%$ and 30$\%$ for SR1  
and between 35$\%$ and  41$\%$ for SR9.   For the SR1 selection, the uncertainty on the signal acceptance itself 
is about $8\%$--9$\%$,  and increases to about 30$\%$ for  the SR9 selection.
Similarly, the variations of the renormalization and factorization scales 
introduce a 9$\%$ to $30\%$ change in the signal acceptance and a  22$\%$ to 40$\%$ uncertainty 
on the signal yields with increasing $n$ and $\met$ requirements.   

\begin{figure}[h]
\begin{center}
  \includegraphics[width=0.6\textwidth]{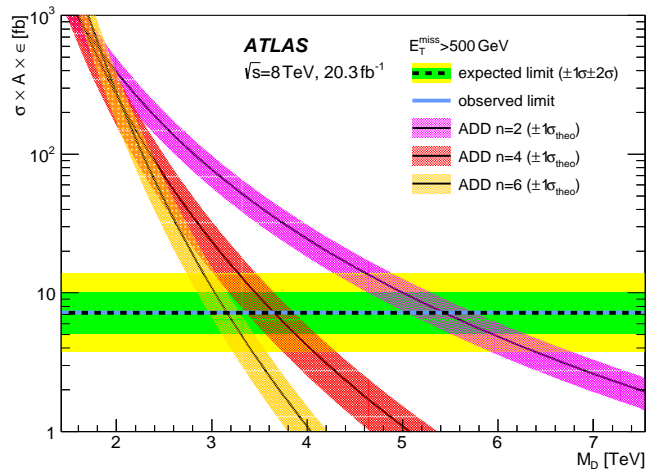}
\end{center}
\caption{
The predicted ADD product of cross section, acceptance and efficiency, $\sigma \times A \times \epsilon$, for the SR7 selection 
as a function of the fundamental Planck scale in $4+n$ dimensions, $M_D$, for
$n=2$, $n=4$, and $n=6$, where  bands represent the uncertainty on the theory. For comparison,
the model-independent observed  (solid line) and expected (dashed  line)  95$\%$ CL limits on
 $\sigma \times A \times \epsilon$ are shown. The shaded areas around the expected limit
indicate the expected $\pm 1\sigma$ and $\pm 2\sigma$ ranges of limits in the absence of a signal. }
\label{fig:add_one}
\end{figure}
\begin{figure}[h]
\begin{center}
  \includegraphics[width=0.6\textwidth]{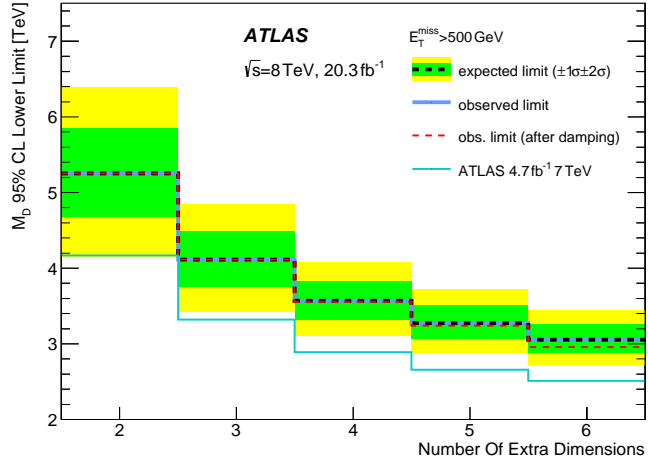}
\end{center}
\caption{
Observed and expected 95$\%$ CL limit on the fundamental Planck scale in $4+n$ dimensions, $M_D$, as a function of the number of extra dimensions.
In the figure the two results overlap.
The 
shaded areas around the 
expected limit  indicate the expected $\pm 1\sigma$ and $\pm 2\sigma$
ranges of limits in the absence of a signal. Finally, the thin dashed line shows the 95$\%$ CL observed limits 
after the suppression  of the events with 
$\hat{s} > M_D^2$ (damping) is applied,  as described in the body of the text.  
The results from this analysis are compared to previous results from ATLAS at 7~TeV~\cite{ATLAS:2012ky} without any damping applied.
}
\label{fig:add}
\end{figure}

The signal region SR7 provides the most stringent expected limits and
is used to obtain the final results.
Figure~\ref{fig:add_one} shows, for the SR7 selection,  the  ADD $\sigma \times A \times \epsilon$ as a function of $M_D$ for
$n=2$, $n=4$, and $n=6$, calculated at LO. For comparison,  the model-independent 95$\%$ CL limit is shown.
Expected and observed 95$\%$ CL lower limits are set on the value of $M_{D}$ as a function of 
the number of extra dimensions considered in the ADD model. 
The $CL_s$ approach is used, 
including statistical and systematic uncertainties. For the latter,
the uncertainties on the signal acceptance times efficiency, 
the background expectations, and the luminosity are considered, and 
correlations between systematic uncertainties on signal and background expectations 
are taken into account. 
In addition, observed limits are computed 
taking into account the $\pm 1\sigma$ LO theoretical uncertainty. 
Values of $M_D$ below 5.25~TeV ($n=2$), 4.11~TeV ($n=3$), 3.57~TeV
($n=4$), 3.27~TeV ($n=5$), and 3.06~TeV ($n=6$) are excluded at
95$\%$~CL, which extend significantly the exclusion from previous results using 7 TeV data~\cite{ATLAS:2012ky}. 
The observed limits decrease by about $6\%$--$8\%$ after considering
the $-1\sigma$ uncertainty from PDF and scale variations in the ADD theoretical predictions (see
Table~\ref{tab:add} and Fig.~\ref{fig:add}).  

As discussed in Ref.~\cite{ATLAS:2012ky}, the  analysis partially probes the phase-space region with $\hat{s} > M_D^2$, where $\sqrt{\hat{s}}$ is the
centre-of-mass energy of the hard interaction. This challenges the validity of model implementation and the lower bounds
on $M_D$, as they depend on  the unknown ultraviolet behaviour of the effective theory.  
For the SR7 selection, the  fraction of signal events with $\hat{s} > M_D^2$ is negligible for $n=2$, but increases with increasing $n$ from 1$\%$ for
$n=3$ and 6$\%$ for $n=4$,  to about 17$\%$ for $n=5$ and 42$\%$ for $n=6$.
The observed 95$\%$ CL limits are recomputed after 
suppressing, with a weighting  factor $M_D^4/\hat{s}^2$, the signal events with $\hat{s} > M_D^2$, here 
referred to as damping.  This results in a decrease of the quoted 95$\%$  CL on $M_D$  which is negligible for $n=2$ and about 
3$\%$ for $n=6$ (see Fig.~\ref{fig:add}).

\begin{table}[ht!]
   \caption{The $95\%$ CL observed and expected
limits on the fundamental Planck scale in $4+n$ dimensions, $M_D$, as a function of the number of extra dimensions $n$ for the SR7 selection and considering LO signal cross sections.
The impact
of the $\pm 1\sigma$   theoretical uncertainty on
the observed limits and the  expected $\pm 1\sigma$ range of limits in the absence of a
signal are also given. Finally, the $95\%$ CL observed limits after damping of signal cross section for  $\hat{s} > M_D^2$ 
(see body of the text)
are quoted between parentheses.
}
\begin{center}
\begin{footnotesize}
\begin{tabular*}{\textwidth}{@{\extracolsep{\fill}}lcccccc}\hline
  \multicolumn{7}{c}{$95\%$ CL  limits on $M_D$ [TeV]} \\ \hline  
   $n$ extra   & \multicolumn{3}{c}{$95\%$ CL observed limit}     &  \multicolumn{3}{c}{$95\%$ CL expected limit}           \\ 
   dimensions &  $+1\sigma$(theory)  &  Nominal (Nominal after damping)  &  $-1\sigma$(theory)    &  $+1\sigma$    &  Nominal  &  $-1\sigma$ \\
   \hline
         2 &      $+$0.31& 5.25 (5.25) & $-$0.38 &$-$0.59 &5.25 &$+$0.58 \\
         3 &      $+$0.25& 4.11 (4.11) & $-$0.33 &$-$0.38 &4.11 &$+$0.36 \\
         4 &      $+$0.20& 3.57 (3.56) & $-$0.29 &$-$0.26 &3.57 &$+$0.25 \\
         5 &      $+$0.17& 3.27 (3.24) & $-$0.25 &$-$0.23 &3.27 &$+$0.21 \\
         6 &      $+$0.13& 3.06 (2.96) & $-$0.19 &$-$0.20 &3.06 &$+$0.18 \\
   \hline
   \hline
\end{tabular*}
  \end{footnotesize}
\end{center}
\label{tab:add}
\end{table}

\subsection{Weakly interacting massive particles}

In the following, the results are converted into limits on the pair
production of WIMPs. As illustrated in Fig.~\ref{fig:wimp:sketch},
this is done both in the EFT framework and in a simplified
model where the WIMP pair couples to Standard Model quarks via a
$Z^\prime$ boson.

For each EFT operator defined in Table~\ref{table:wimp:operators}, the
limits on \MS{} are extracted from those signal regions that exhibit the best
expected sensitivity: these are SR4 for C1, SR7 for D1, D5, D8, and
SR9 for C5, D9, D11. These are translated into  
corresponding 95\% CL limits on the
suppression scale \MS{} as a function of  $m_\chi$. 
\begin{figure}[!h]
\centering
\subfigure[]{\includegraphics[width=0.4\textwidth]{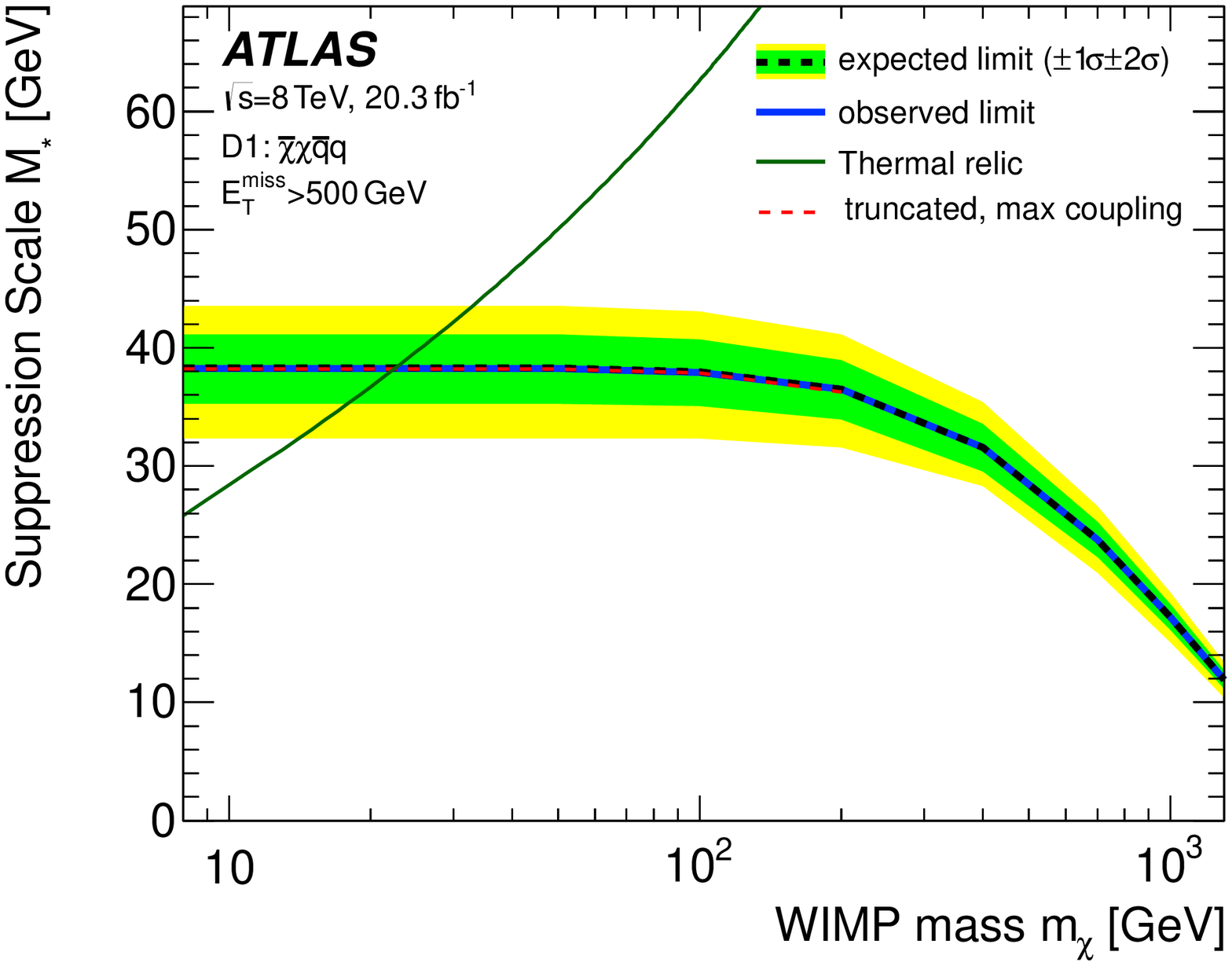}}
\subfigure[]{\includegraphics[width=0.4\textwidth]{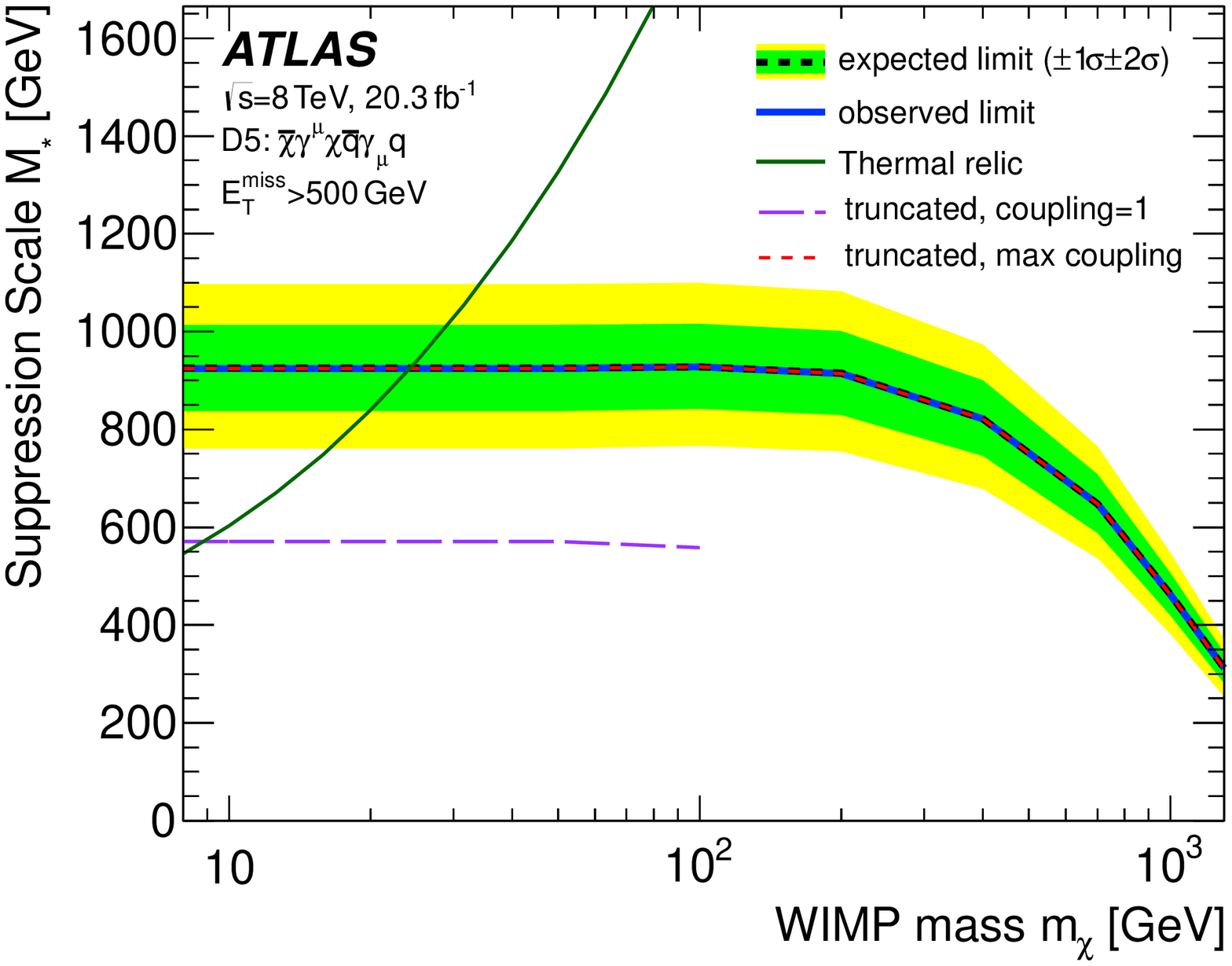}}
\subfigure[]{\includegraphics[width=0.4\textwidth]{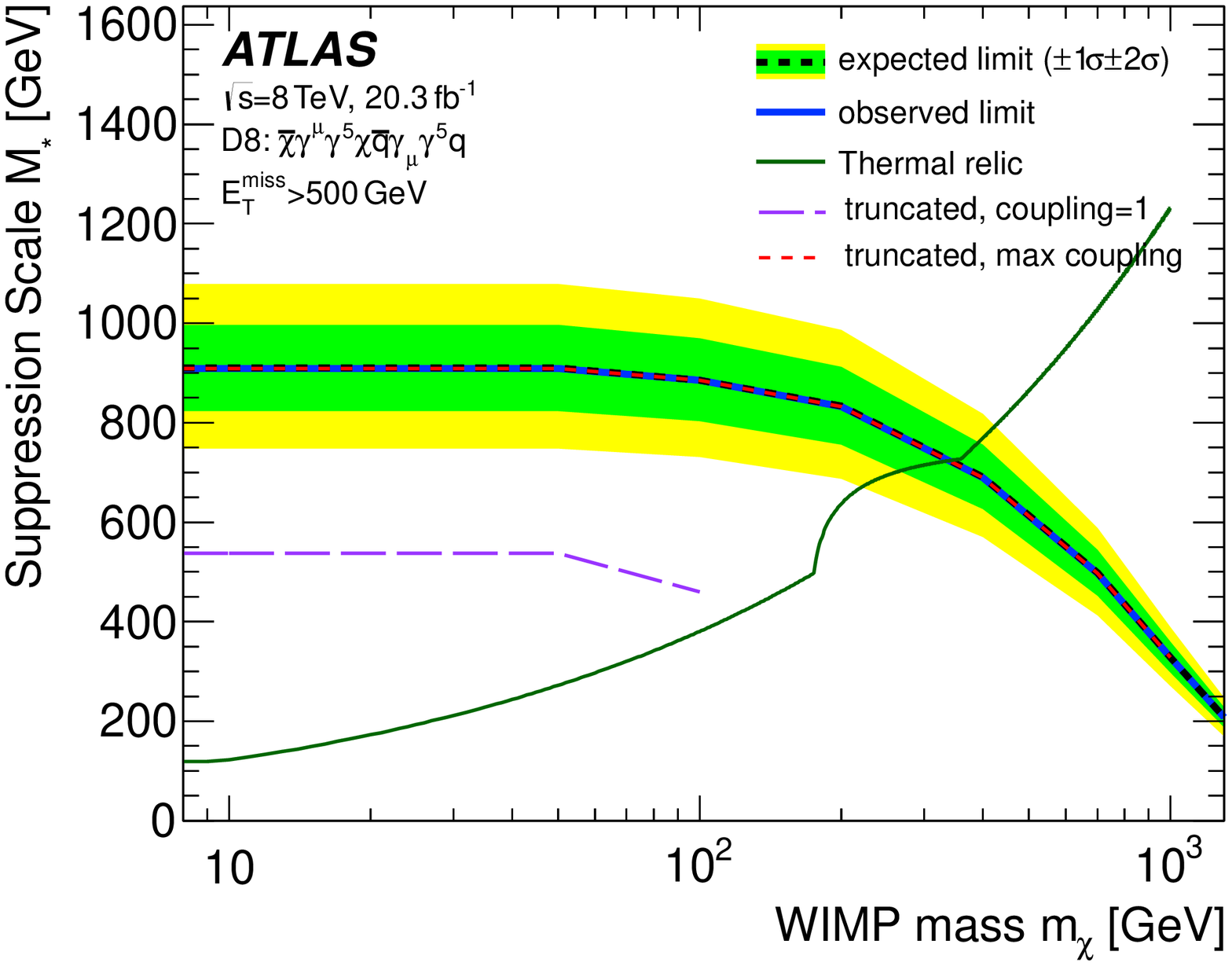}}
\subfigure[]{\includegraphics[width=0.4\textwidth]{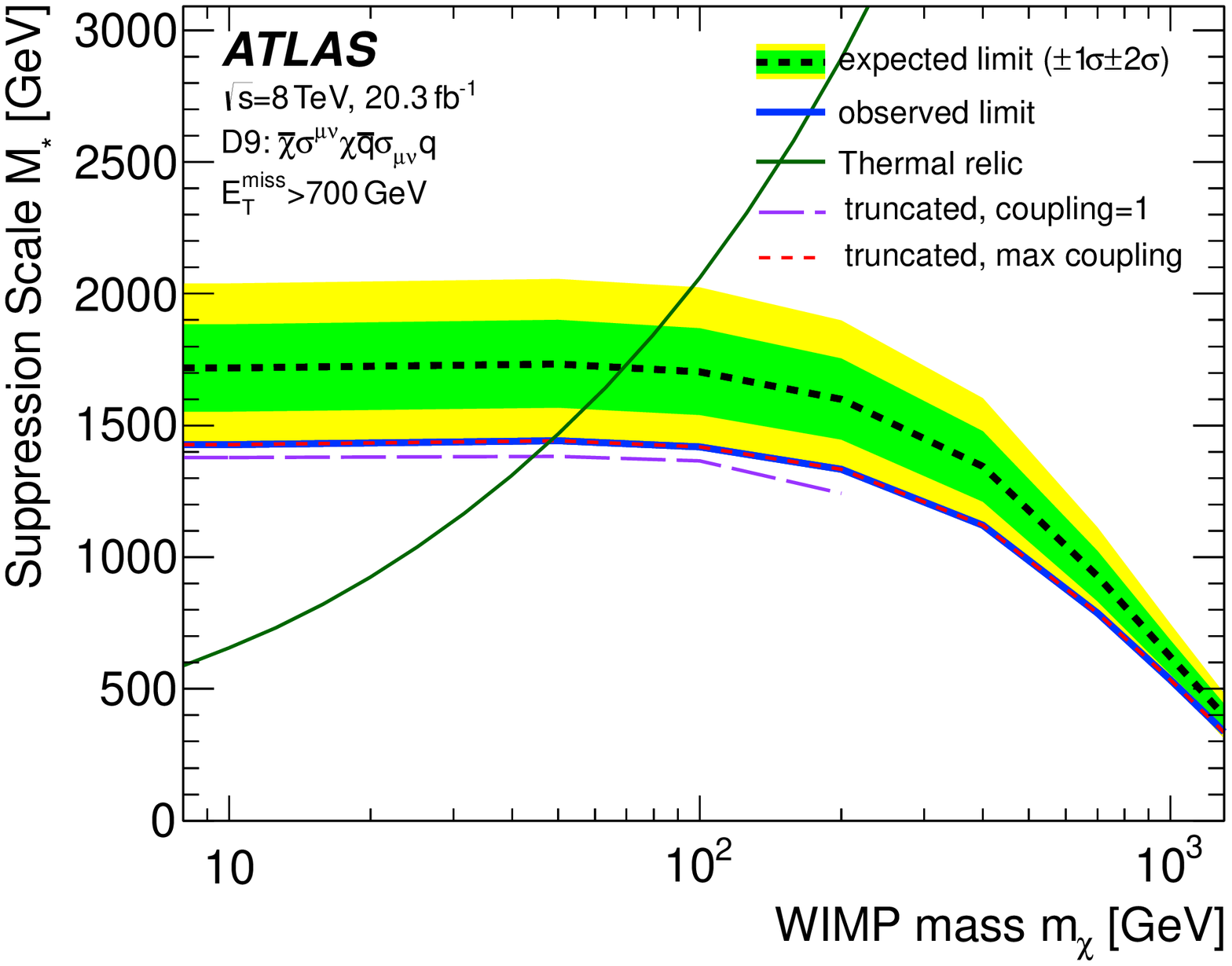}}
\subfigure[]{\includegraphics[width=0.4\textwidth]{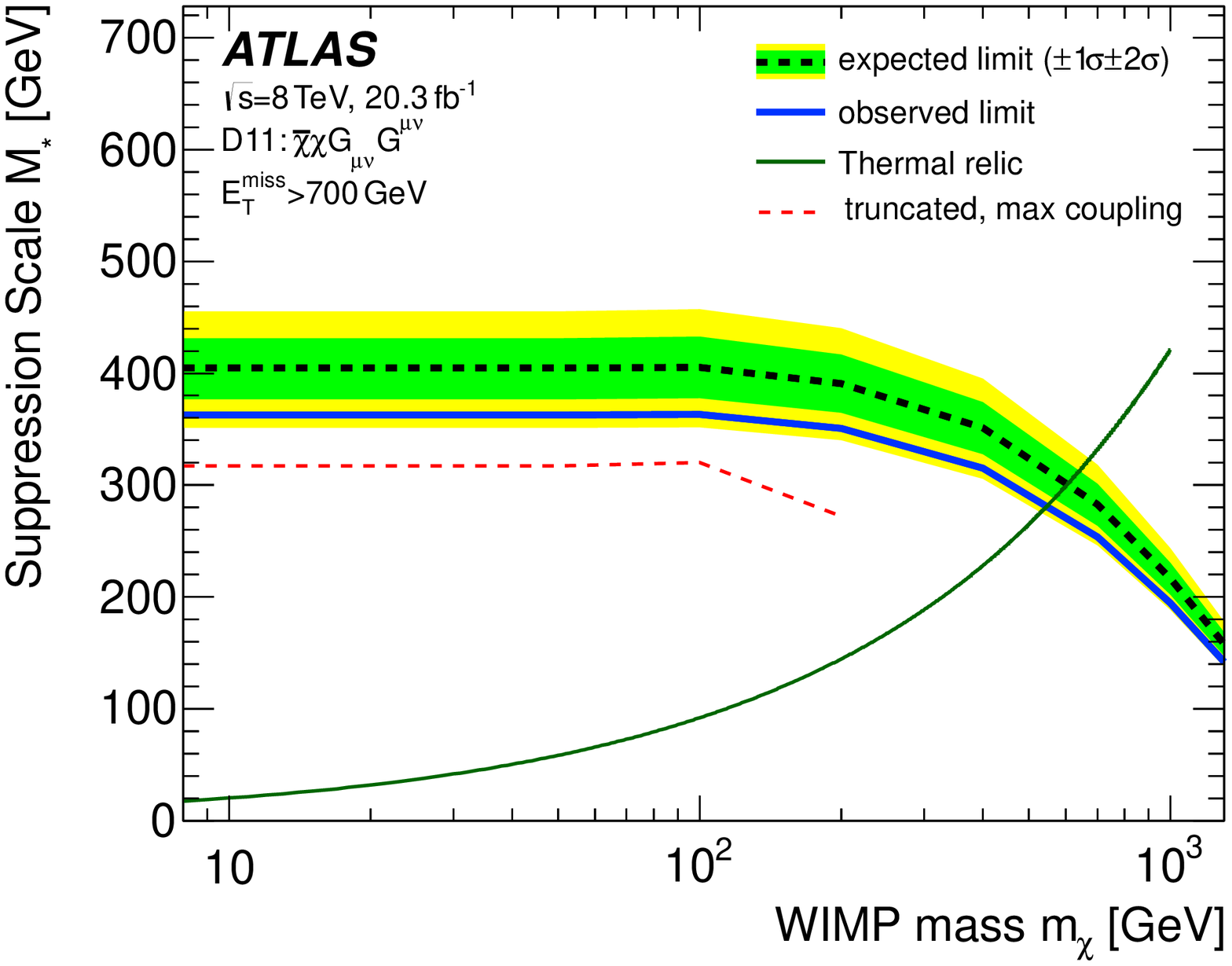}}
\subfigure[]{\includegraphics[width=0.4\textwidth]{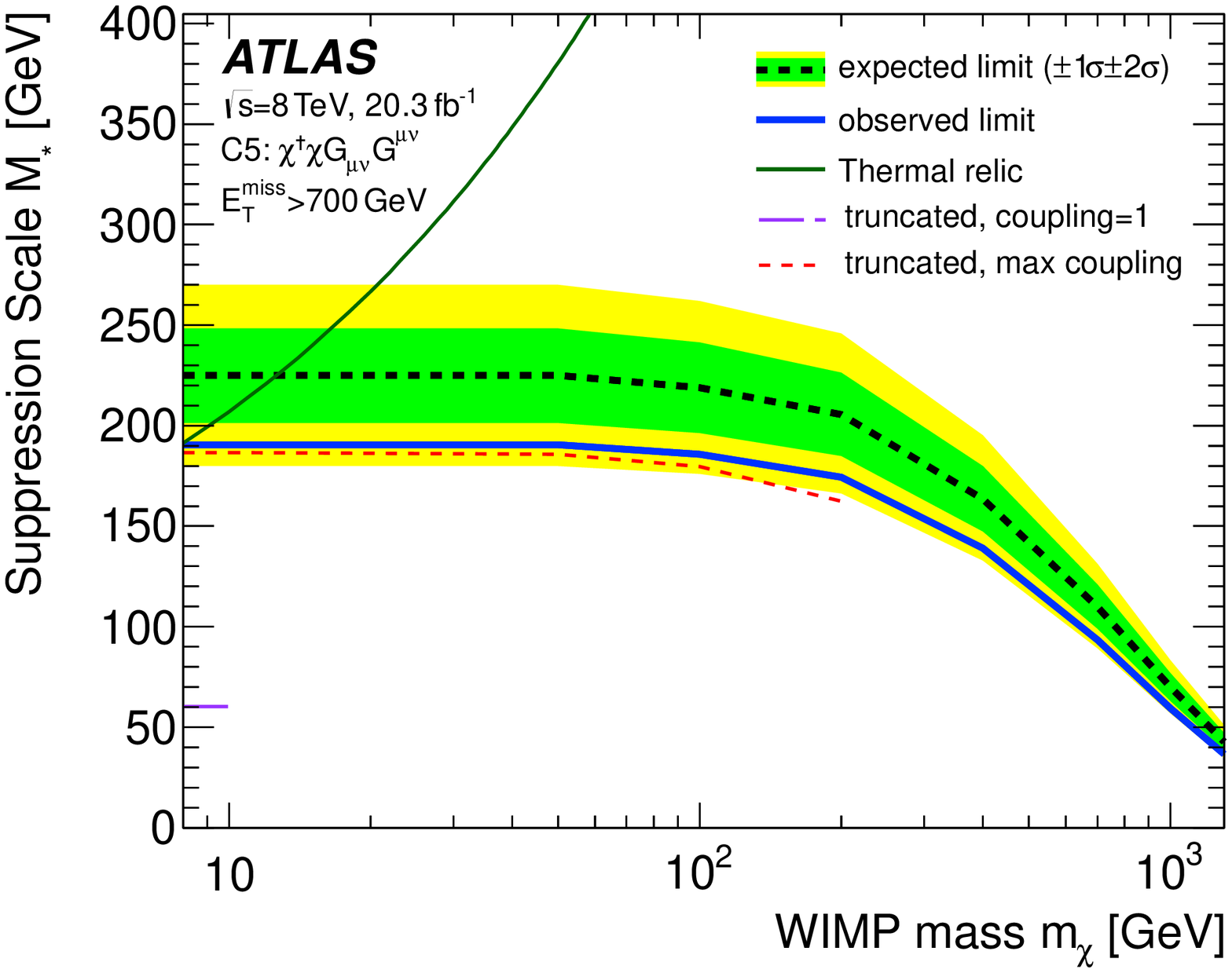}}
\caption{Lower limits at 95\% CL on the suppression scale $M_*$ are shown  
  as a function of the WIMP mass
  $m_\chi$ for (a) D1, (b) D5, (c) D8, (d) D9, (e) D11 and (f) C5 operators, in each case
  for the most sensitive SR (SR7 for D1, D5, D8, SR9 for D9, D11 and C5). The
  expected and observed limits are shown as dashed black and solid
  blue lines, respectively. The rising green lines are the \MS{}
  values at which WIMPs of the given mass result in the relic density
  as measured by WMAP~\cite{WMAP9}, assuming annihilation in the early
  universe proceeded exclusively via the given operator. 
  The purple long-dashed
  line is the 95$\%$ CL observed limit on \MS{} imposing a validity
  criterion with a coupling strength of 1, the red dashed thin lines
  are those for the maximum physical coupling strength (see~\ref{sec:valid} for further details).}
\label{fig:wimp:ms}
\end{figure}
To derive these  lower limits on \MS{}, the  same $CL_s$ 
approach as in the case of the ADD LED model is used.
The uncertainties on the WIMP signal acceptance include: 
a $3\%$ uncertainty from the uncertainty on the beam energy; a $3\%$ 
uncertainty from the variation of the renormalization and factorization 
scales and a $5\%$ uncertainty from the variation of the parton-shower matching scale; a 
$1\%$ to $10\%$ uncertainty   
from uncertainties on jet and $\met$ energy scale and resolution; and  
a $5\%$ to $29\%$ uncertainty due to PDF, depending on the operator and WIMP mass.

Similarly, the uncertainties on the signal cross section are:  a $2\%$
to $17\%$ ($40\%$ to $46\%$) uncertainty due to the variation of the
renormalization and factorization scales in D1, D5 and D9 (C5 and D11)
operators; and a $19\%$ to $70\%$ ($5\%$ to $36\%$) uncertainty due to the
PDF for C5, D11 and D1 (D5 and D9) operators, with increasing WIMP
mass. These theoretical cross-section uncertainties are not considered
when deriving limits and are not displayed in the plots.   
A $2\%$ to $9\%$ uncertainty on the cross section, due to the beam energy uncertainty, is taken into account.

The \MS{} limits for five of the operators are shown in
Fig.~\ref{fig:wimp:ms} down to WIMP masses of 10~\GeV, and could be
extrapolated even to smaller $m_\chi$ values since there is a
negligible change in the cross section or the kinematic distributions at the
LHC for such low-mass WIMPs. The 1$\sigma$ and 2$\sigma$ error bands 
around the expected limit are due to the acceptance
uncertainties (experimental and theoretical). 
The effect of the beam-energy uncertainty on the observed limit is negligible
and is not shown.

Various authors have investigated the kinematic regions in which the
effective field theory approach for WIMP pair production breaks
down~\cite{Buchmueller:2013dya,Busoni:2013lha,Busoni:2014sya,Busoni:2014haa}. The
problem is addressed in detail in~\ref{sec:valid}, where the
region of validity of this approach is probed for various assumptions
about the  underlying unknown new physics. 
Here, the EFT framework is
used as a benchmark to convert the measurement, and in the 
absence of any deviation from the SM backgrounds, to a limit on the
pair production of DM (with the caveat of not complete
validity in the full kinematic phase space). These are the central
values of the observed and expected limits in
Fig.~\ref{fig:wimp:ms}. A basic demonstration of the validity issue is
also included in the figure. This is done by relating the suppression
scale \MS{} to the mass of the new particle mediating the interaction,
\MMed{}, and the coupling constants of the interaction,
$g_i$ by 
\begin{equation*}
\MMed{} = f(g_i, \MS{}) \, .
\end{equation*}
For such a relation, an assumption has to be made about the 
interaction structure connecting the initial state to the final state
via the mediator particle. The simplest interaction structures are
assumed in all cases. The form of the function $f$ connecting \MMed{}
and \MS{} depends then on the operator (see~\ref{sec:valid}). For a given operator, one possible validity
criterion is that the momentum transferred in  the hard interaction, $Q_{\rm tr}$, is below
the mediator particle mass: $Q_{\rm tr} < \MMed{}$. 
According to this criterion, events are omitted where the interaction energy scale exceeds the mediator
particle mass.  This depends  on the values adopted for the couplings.  Two values 
(one and the maximum possible value for the interaction to  remain perturbative) are used. 
After reducing the signal cross section to the fraction of
remaining events, the mass suppression scale \MS{} can be rederived 
yielding potentially two additional expected truncated limit lines
in Fig.~\ref{fig:wimp:ms}. The truncated limits fulfil the respective
validity criteria wherever the lines are drawn in the figure. For D9
for example, the maximum couplings criterion is fulfilled for all WIMP
masses, the coupling equal to one criterion is fulfilled for WIMP masses up to
200~GeV. For C5 on the other hand, the validity criterion for a
coupling value of one is violated over almost the whole WIMP mass range, and 
a truncated limit line is only drawn up to a WIMP mass of 10 GeV.

Figure~\ref{fig:wimp:ms} also includes thermal relic
lines~(taken from Ref.~\cite{Goodman:2010ku}) that correspond to a
coupling, set by $M_*$, of WIMPs to quarks or gluons such that WIMPs
have the correct relic abundance as measured by the WMAP
satellite, in the absence of any 
interaction other than the one considered. 
The thermal
  relic line for D8 has a bump feature at the top-quark mass where the
  annihilation channel to top quarks opens. 
Under the assumption that DM is
entirely composed of thermal relics, the limits on $M_*$ which are
above the value required for the thermal relic density exclude the
case where DM annihilates exclusively to SM particles via the
corresponding operator. Should thermal relic WIMPs exist in these
regions (above the thermal relic line), there would have to be other
annihilation channels or annihilation via other operators in order to
be consistent with the WMAP measurements.

\begin{figure}[!h]
\centering
\subfigure[]{\includegraphics[width=0.485\textwidth]{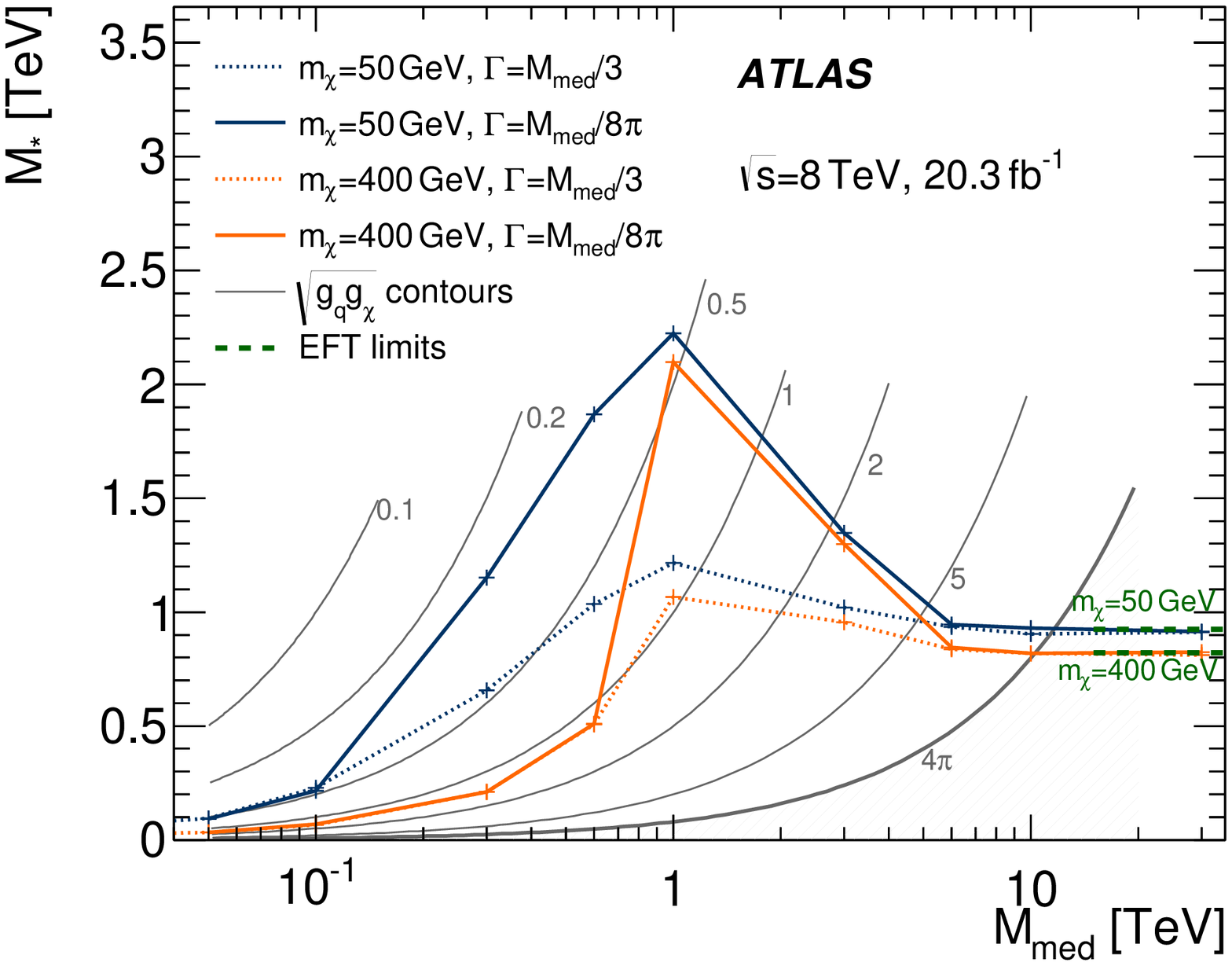}}
\subfigure[]{\includegraphics[width=0.485\textwidth]{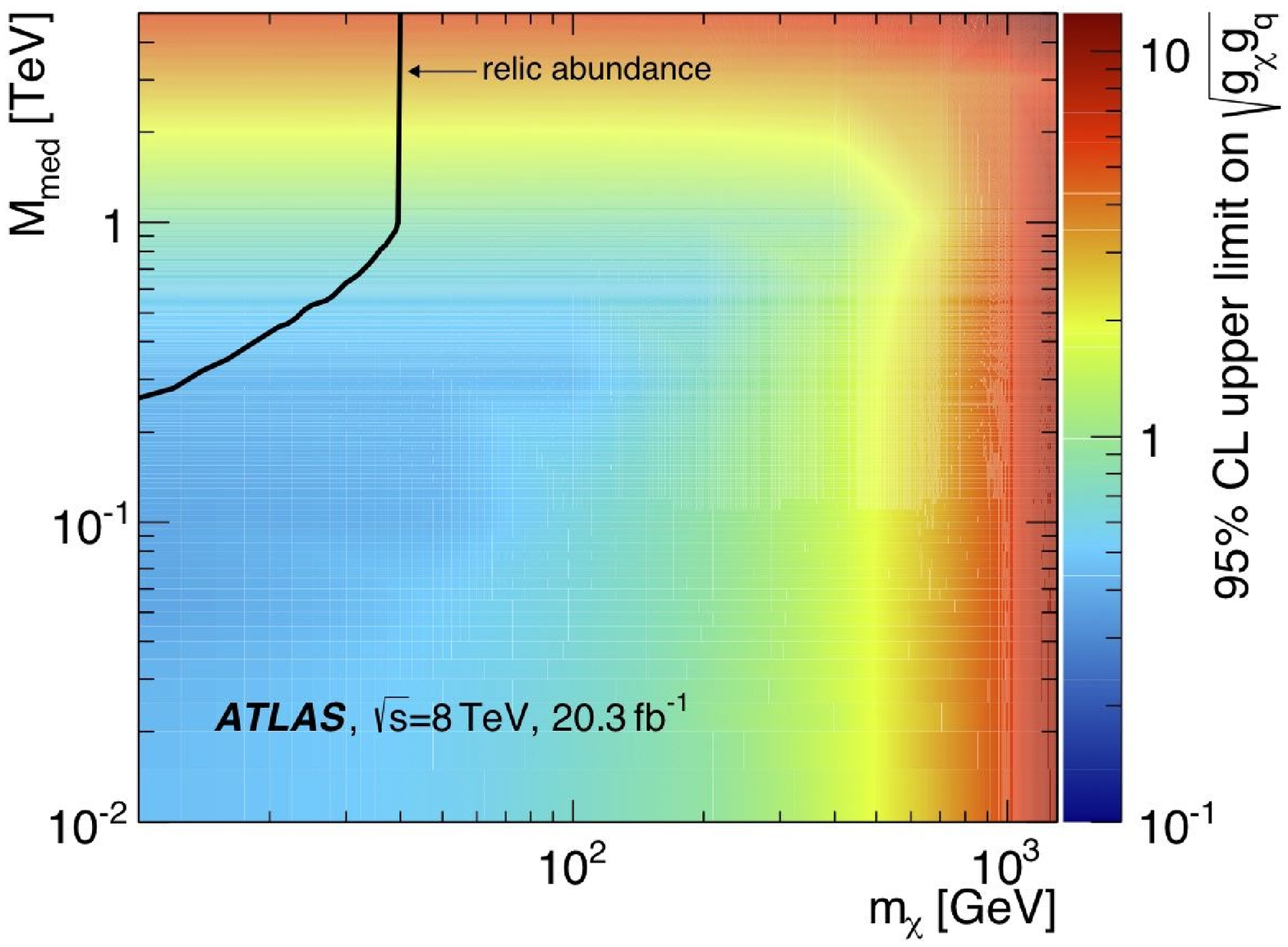}}
\caption{(a) Observed 95$\%$ CL limits on the suppression scale $M_*$ as a function of the mediator mass
  \MMed{}, assuming a $Z^\prime$-like boson in a simplified
  model and a DM mass of 50 GeV and 400 GeV. The width of the
  mediator is varied between $\MMed/3$ and $\MMed/8\pi$. The
  corresponding limits from EFT models are shown as dashed lines;  
  contour lines indicating a range of values of the product of the
  coupling constants ($\sqrt{\mathrm{g}_q\, \mathrm{g}_\chi}$) are
  also shown.
  (b) Observed 95\% CL upper limits on the product of
  couplings of the simplified model vertex in the plane of
  mediator and WIMP mass (\MMed{} versus $m_\chi$). Values leading to
  the correct relic abundance~\cite{WMAP9} are shown by the black solid 
  line. 
}
\label{fig:wimp:simpl}
\end{figure}

Another way to avoid the validity issues discussed above is to use a
simplified model to explicitly parameterize the interaction of quarks
or gluons with WIMP pairs via generic interactions with real mediator
particles. With this approach, the coupling of pairs of Dirac fermion
WIMPs to quarks via a vector mediator particle (such as a
$Z^\prime$ boson, corresponding to the operator D5) of a given mass
and width (\MMed{} and $\Gamma$, respectively) is probed. Given the
cross-section limit and using simulations at fixed values of \MMed{}
and $\Gamma$, the product of the coupling constants of the $Z^\prime$ boson 
to quarks and WIMPs, $\sqrt{\mathrm{g}_q\, \mathrm{g}_\chi}$, can be
constrained. This constraint corresponds to one value in the
\MS{}--\MMed{} plane as shown in Fig.~\ref{fig:wimp:simpl}(a), 
since the mass suppression scale can be calculated exactly in
this model, $\MS{} = {\MMed{}}/{\sqrt{\mathrm{g}_q\,
    \mathrm{g}_\chi}}$. The figure demonstrates how, for a given
mediator particle mass and two values of the width $\Gamma$, the real
value of the mass suppression scale would compare to the \MS{} value
derived assuming a contact interaction (shown as dashed lines in the
figure). This contact interaction regime is reached for \MMed{} values
larger than 5~TeV in the figure. In the intermediate range
($700$~GeV~$< \MMed{} < 5$~TeV), the mediator would be produced
resonantly and the actual \MS{} value is higher than in the contact
interaction regime. In this case the contact interaction limits would
be pessimistic: they would underestimate the actual values. Finally,
the small mediator mass regime below 700~GeV has very small \MS{}
limits because the WIMP would be heavier than the mediator, and WIMP
pair production via this mediator would thus be kinematically
suppressed. In this region, the contact interaction limits would be
optimistic and overestimate the actual \MS{} values.

In Fig.~\ref{fig:wimp:simpl}(b)  the observed 95\%
CL upper limits on the product of couplings of the simplified model
vertex are shown in the plane of mediator and WIMP mass (\MMed{}
versus $m_\chi$). 
Within this model, the regions to the left of the relic density line
lead to values of the relic density larger than measured and 
are excluded.

\begin{figure}[!h]
\centering
\subfigure[]{\includegraphics[width=0.485\textwidth]{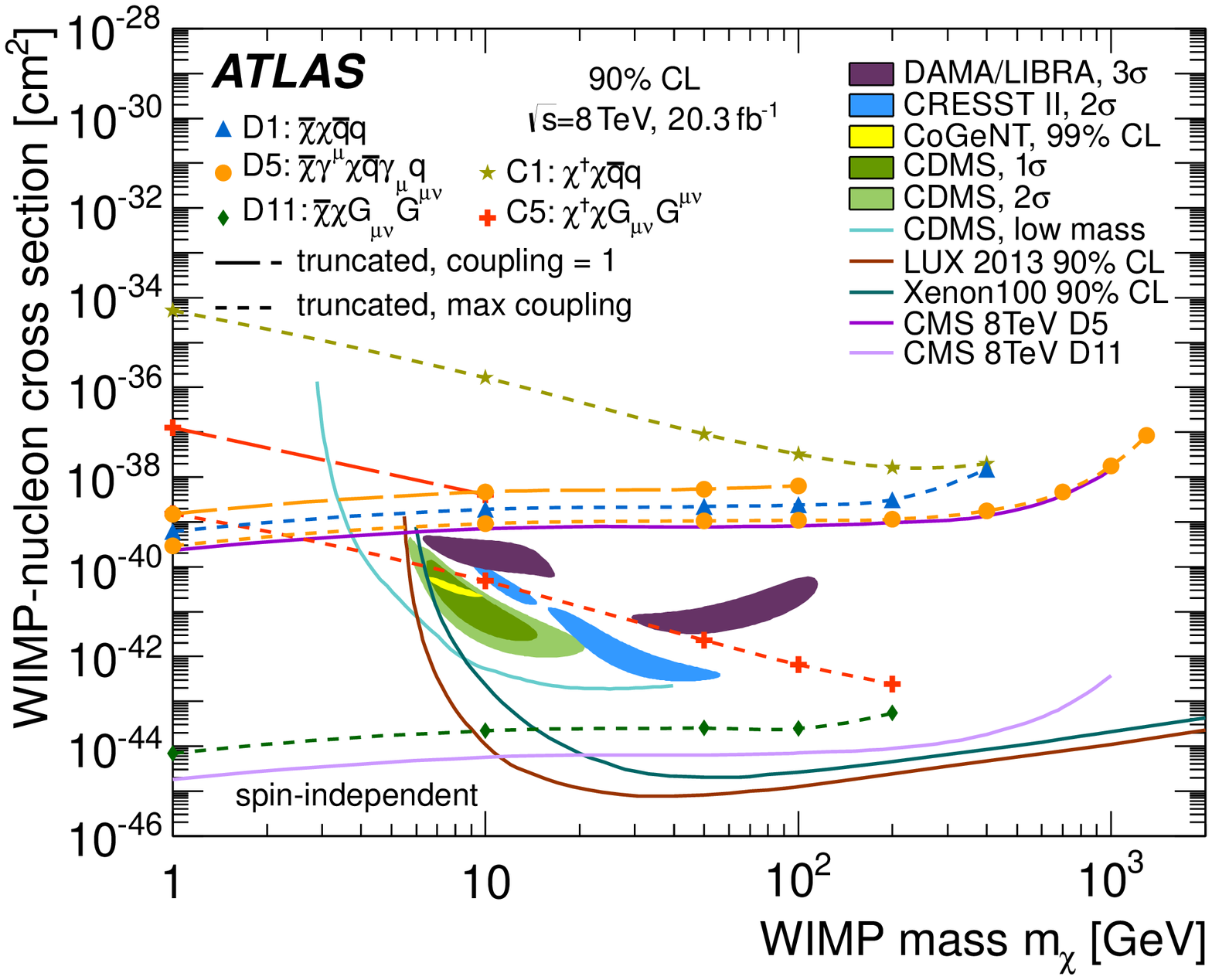}}
\subfigure[]{\includegraphics[width=0.485\textwidth]{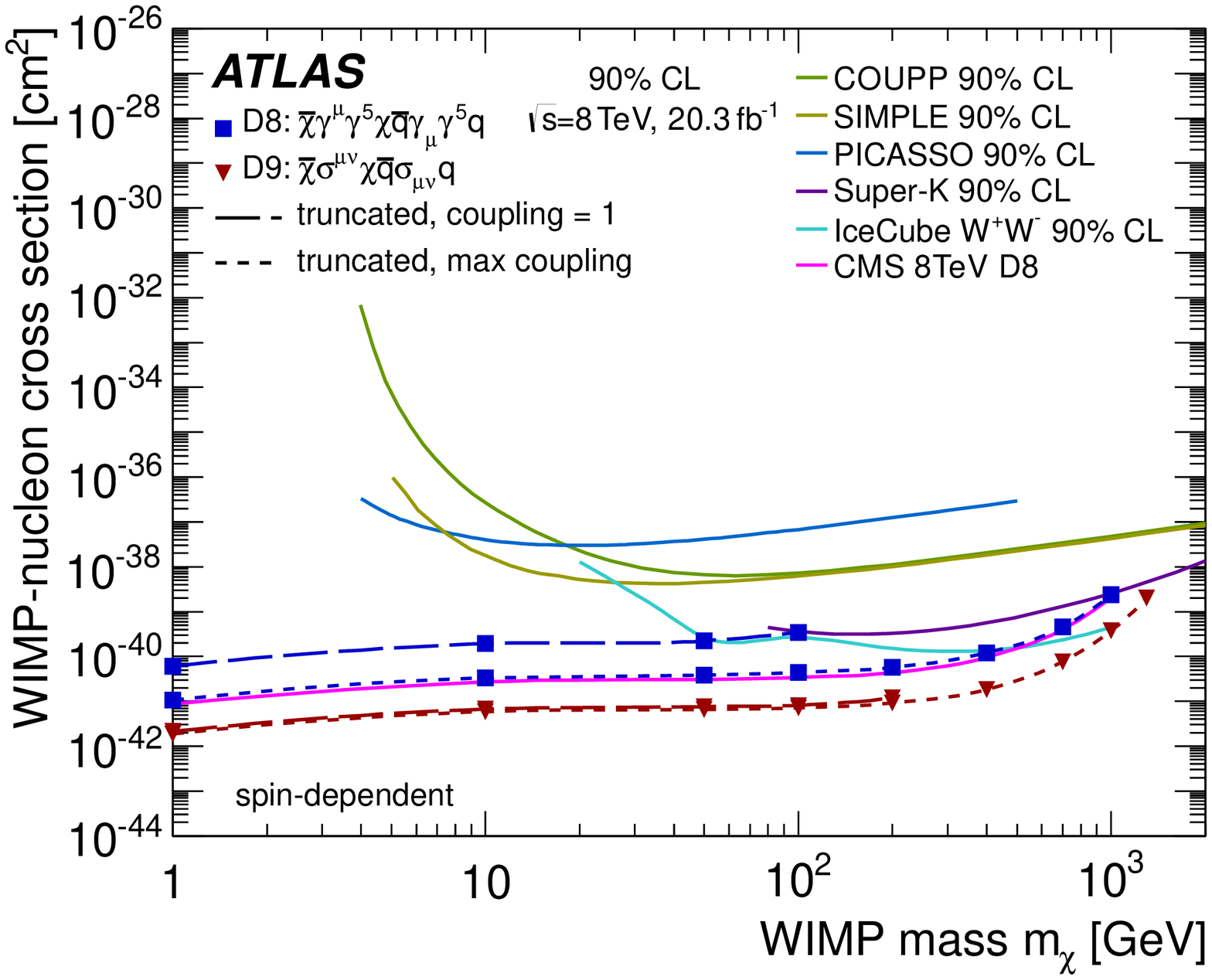}}
\subfigure[]{\includegraphics[width=0.485\textwidth]{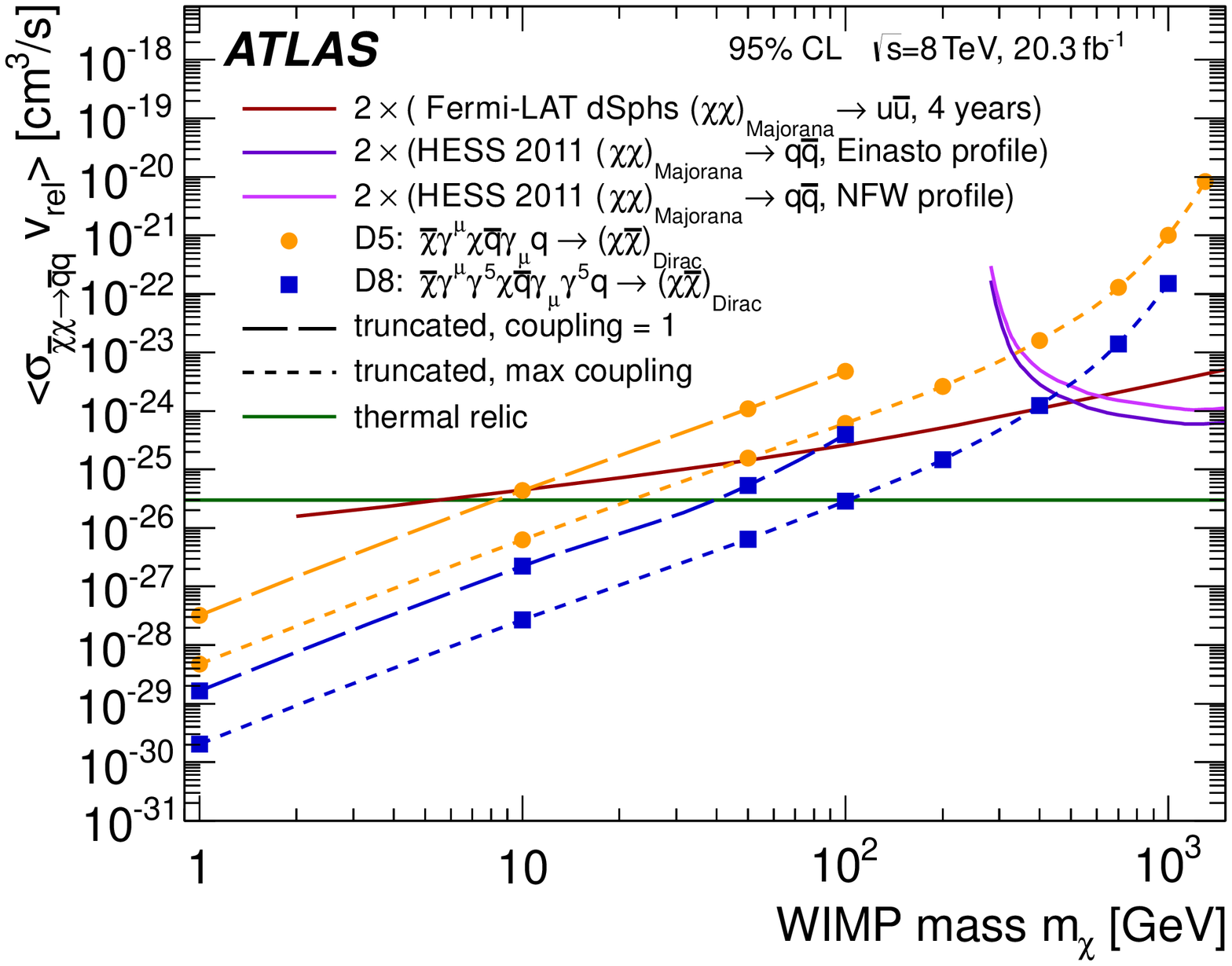}}
\caption{Inferred 90\% CL limits on (a) the spin-independent and (b) spin-dependent WIMP--nucleon scattering cross
  section as a function of DM mass $m_\chi$ for different operators (see Sect.~\ref{sec:intro}). Results from
  direct-detection experiments for the spin-independent
  \cite{Angloher:2011uu, 
    Akerib:2013tjd, 
    Agnese:2013rvf, 
    Agnese:2014aze, 
    Aalseth:2014jpa, 
    Bernabei:2008yi, 
    Aprile:2013doa} 
  and spin-dependent \cite{Archambault:2012pm, 
    Desai:2004pq, 
    Abbasi:2009uz, 
    Behnke:2010xt, 
    Felizardo:2011uw} 
  cross section, and the CMS (untruncated) results~\cite{Khachatryan:2014rra} are shown for comparison. 
  (c) The inferred 95\% CL limits on the DM annihilation rate as a function of DM mass.  The annihilation rate is
  defined as the product of cross section $\sigma$ and relative
  velocity $v$, averaged over the DM velocity distribution
  $(\left<\sigma\ v\right>)$. Results from gamma-ray
  telescopes~\cite{Ackermann:2013yva,Abramowski:2011zz} are also
  shown, along with the thermal relic density annihilation
  rate~\cite{WMAP9,Ade:2013zuv}.}
\label{fig:wimp:dm}
\end{figure}

In the effective operator approach, the  bounds on $M_*$ for a
given $m_\chi$ (see Fig.~\ref{fig:wimp:ms}) can be converted to bounds
on WIMP--nucleon scattering cross sections, which are probed by direct
DM detection experiments. These bounds describe scattering of WIMPs
from nucleons at a very low momentum transfer of the order of a
keV. Depending on the type of interaction, contributions to
spin-dependent or spin-independent WIMP--nucleon
interactions are expected. As in Ref.~\cite{ATLAS:2012ky}, the 
limits are converted here to bounds on the WIMP--nucleon scattering
cross sections and the results are displayed in
Fig.~\ref{fig:wimp:dm}. Under the assumptions made in the EFT
approach, the ATLAS DM limits are particularly relevant in the low DM
mass region, and remain important over the full $m_\chi$ range
covered. The spin-dependent limits in Fig.~\ref{fig:wimp:dm} are based
on D8 and D9, where for D8 the $M_*$ limits are calculated using the
D5 acceptances (as they are identical) together with D8 production
cross sections. Both the D8 and D9 cross-section limits are
significantly stronger than those from direct-detection experiments.

The  DM limits are shown as upper limits on the WIMP annihilation
rate, calculated using the same approach as in
Ref.~\cite{ATLAS:2012ky}, in the bottom panel of
Fig.~\ref{fig:wimp:dm}. The operators describing the vector and
axial-vector annihilations of WIMPs to the four light-quark flavours
are shown in this plot. For comparison, limits on the annihilation to
$u\bar{u}$ and $q \bar{q}$ from galactic high-energy gamma-ray
observations by the Fermi-LAT~\cite{Ackermann:2013yva} and H.E.S.S.~\cite{Abramowski:2011zz} 
telescopes are also shown. The
gamma-ray limits are for Majorana fermions and are therefore scaled up
by a factor of two for comparison with the ATLAS limits for Dirac
fermions (see Ref.~\cite{ATLAS:2012ky} and references therein for
further discussions and explanations). The annihilation rate that
corresponds to the thermal relic density measured by
WMAP~\cite{WMAP9} and PLANCK~\cite{Ade:2013zuv} satellites is also shown for
comparison in the figure.

Finally, Fig.~\ref{fig:wimp:dm} also demonstrates the impact of 
the EFT validity and the truncation procedure explained above 
on the quoted upper limits for the  WIMP--nucleon scattering and WIMP annihilation cross sections. 
The effect depends strongly on the operator  and the values  for the couplings 
considered. In general, the limits remain valid for WIMP masses up to
$O(100)$~GeV. The variation of the coupling strengths considered leads  
to changes in the quoted cross-section limits of up to one order of magnitude.

\subsection{Associated production of a light gravitino and a squark or gluino}

The results are also expressed in terms of  95$\%$ CL  limits on the cross section 
for the associated production of a gravitino and a gluino or a squark. 
As already discussed, a SUSY simplified model is used in which the 
gluino and squark decays lead to a gravitino and a gluon or a quark, respectively, producing a monojet-like  
signature in the final state. 
Squark and gluino masses up to 2.6~TeV are  considered. The 
acceptance and efficiency $A \times \epsilon$ for the SUSY signal 
depends on the mass of the squark or gluino in the final state and also on the relation between squark 
and gluino masses. 
As an example, in the case of  squarks and gluinos  degenerate in mass $(\mglu = \msqu)$, the signal $A \times \epsilon$ 
for the SR7 (SR9) selection criteria is in the range 25$\%$--45$\%$ (10$\%$--35$\%$) for squark and gluino masses of about 1--2~TeV.

The systematic uncertainties on the SUSY signal yields are determined as in the case of the ADD and WIMP models. 
The uncertainties related to the jet and $\met$
scales and resolutions introduce uncertainties in the signal yields which vary between 2$\%$ and 16$\%$
for different selections and  squark and gluino masses.  The uncertainties on the proton beam energy introduce 
uncertainties on the signal yields which vary between 2$\%$ and 6$\%$ 
with increasing squark and gluino masses. 
The uncertainties related to the modelling of 
initial- and final-state gluon radiation   
translate into a 10$\%$ to 15$\%$ 
uncertainty on the signal yields,  depending on the selection and the squark and gluino masses.
The uncertainties due to PDF
result in uncertainties on the signal yields 
which vary between 5$\%$ and 60$\%$  for squark and gluino masses increasing from 50~GeV and 2.6~TeV.
Finally, the variations of the renormalization and factorization scales 
introduce a 15$\%$ to 35$\%$ uncertainty on the signal yields with increasing squark and gluino masses.

\begin{figure}[htpb]
\begin{center}
\mbox{
\subfigure[]{  \includegraphics[width=0.485\textwidth]{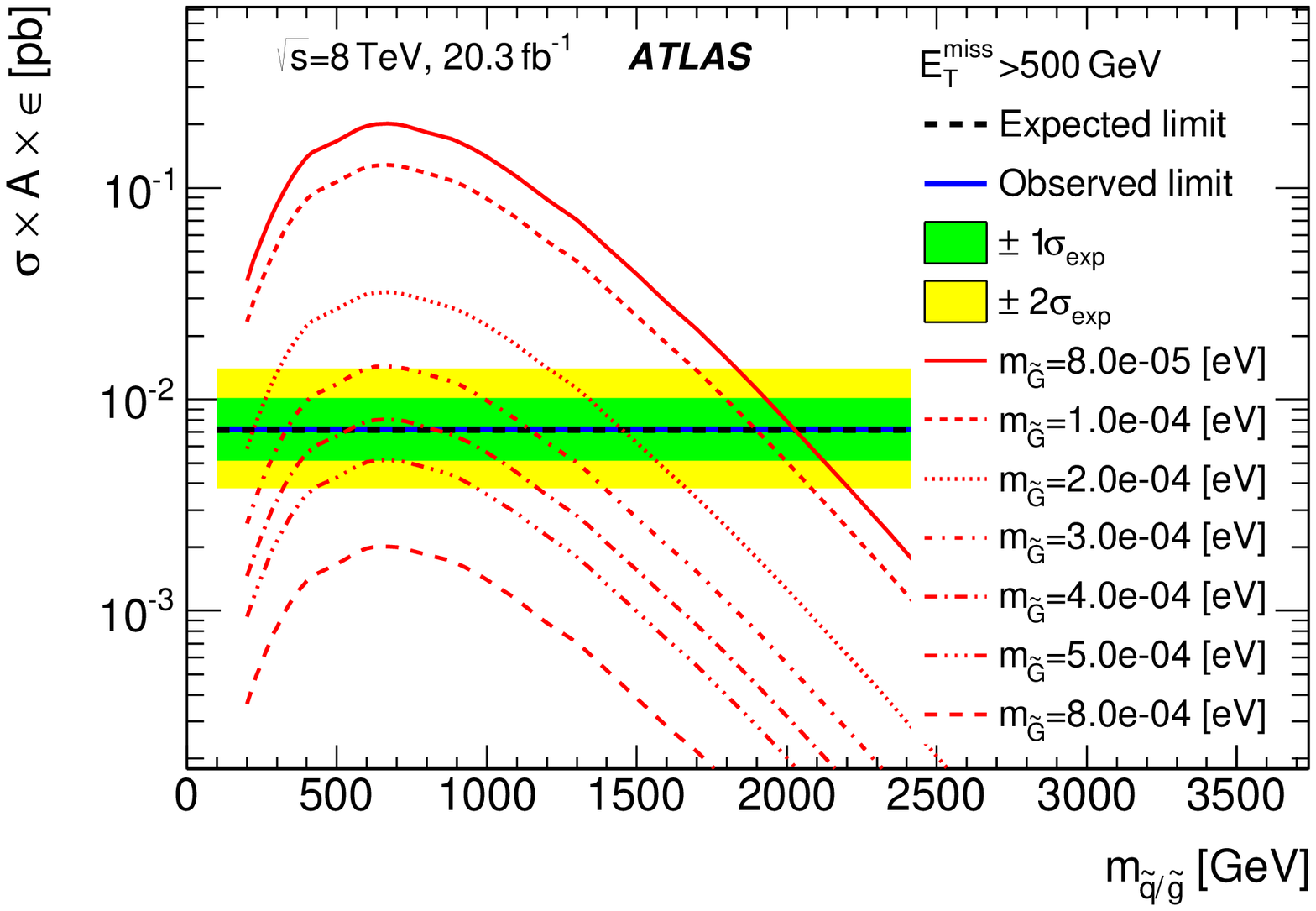}}
\subfigure[]{  \includegraphics[width=0.485\textwidth]{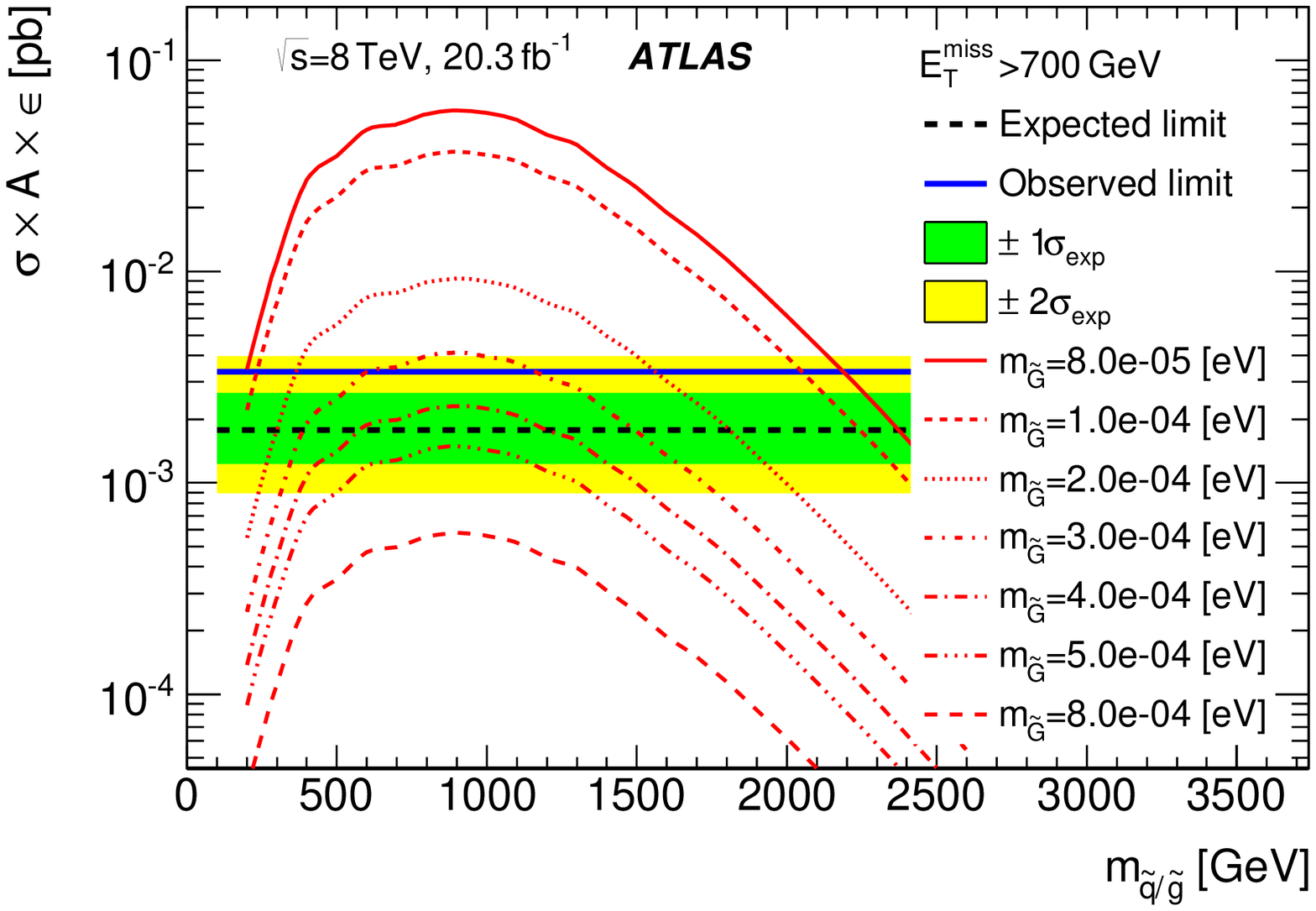}}
}
\end{center}
\caption{
Cross section times acceptance times efficiency $\sigma \times A \times \epsilon$ for gravitino+squark/gluino production as a function of the 
squark/gluino mass $\msqgl$ in the case of degenerate squarks and gluinos and different gravitino masses  
for (a) SR7 and (b) SR9,  compared with the corresponding model-independent limits.
}
\label{fig:grav_one}
\end{figure}

\begin{figure}[htpb]
\begin{center}
  \includegraphics[width=0.6\textwidth]{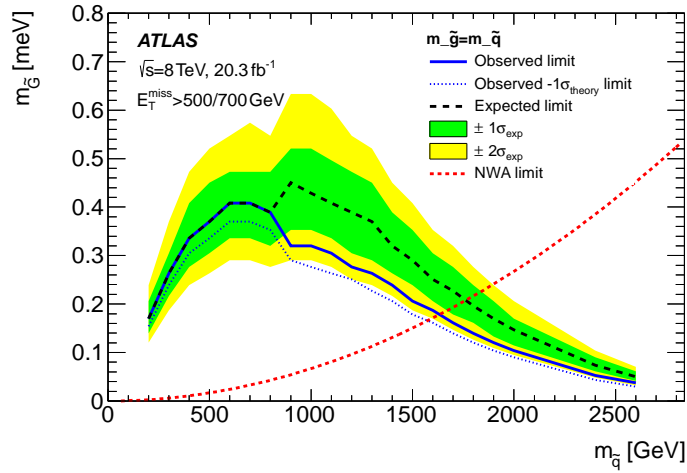}
\end{center}
\caption{Observed (solid line) and expected (dashed line) 95$\%$ CL lower limits on the gravitino mass  $\mgra$  
as a function of the squark mass $\msqu$ for degenerate squark/gluino masses.
The corresponding dotted line indicates the impact on the observed limit  
of the $-1\sigma$ LO theoretical uncertainty.
The shaded bands around the expected limit indicate the expected $\pm 1\sigma$ and $\pm 2\sigma$  ranges of limits in the absence of a
signal. The region above the red dotted line defines the validity of the narrow-width approximation (NWA) 
for which the decay
width is smaller than 25$\%$ of the squark/gluino mass.
}
\label{fig:grav_two}
\end{figure}

\begin{figure}[htpb]
\begin{center}
\mbox{
\subfigure[]{\includegraphics[width=0.45\linewidth]{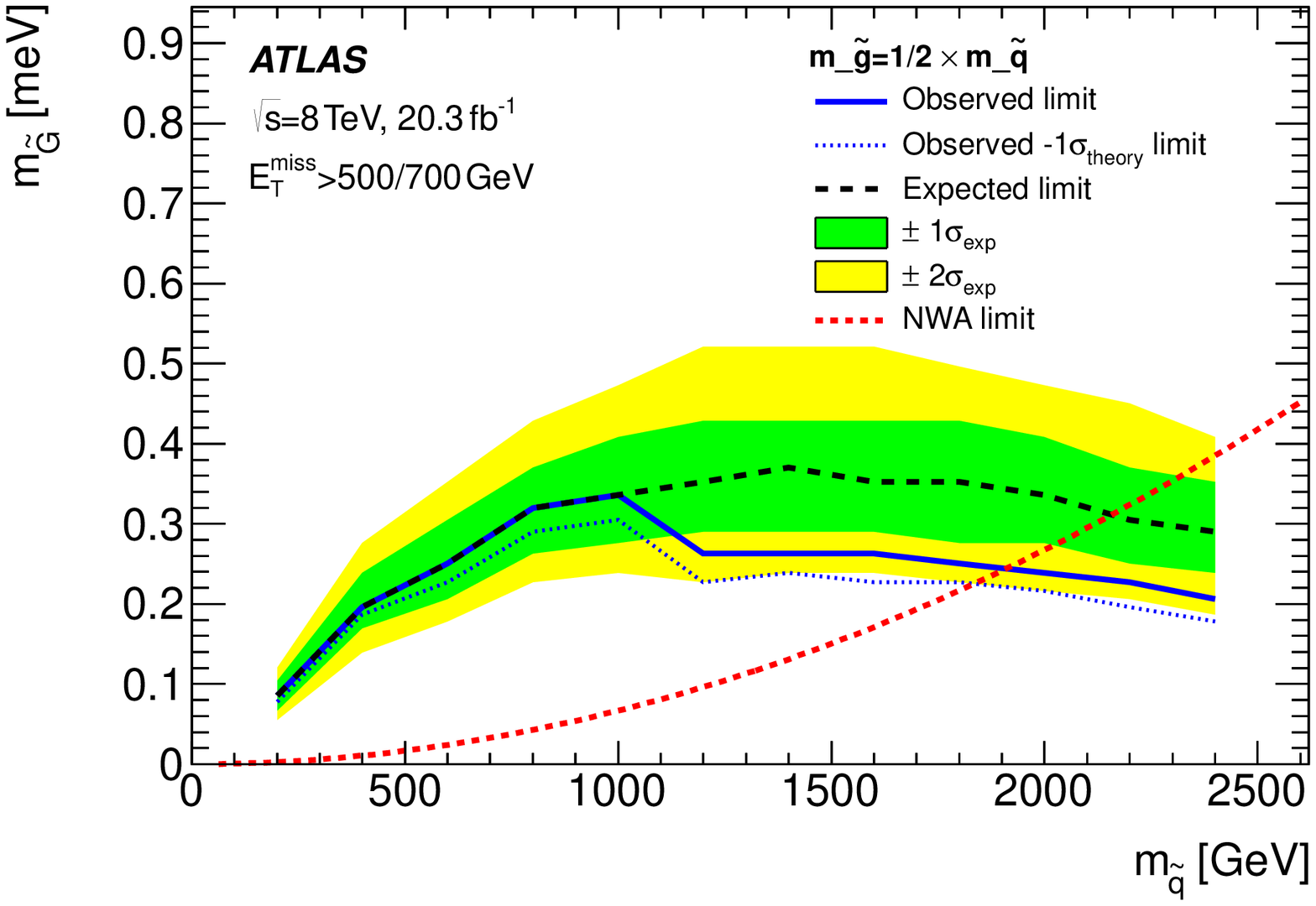}}
\subfigure[]{\includegraphics[width=0.45\linewidth]{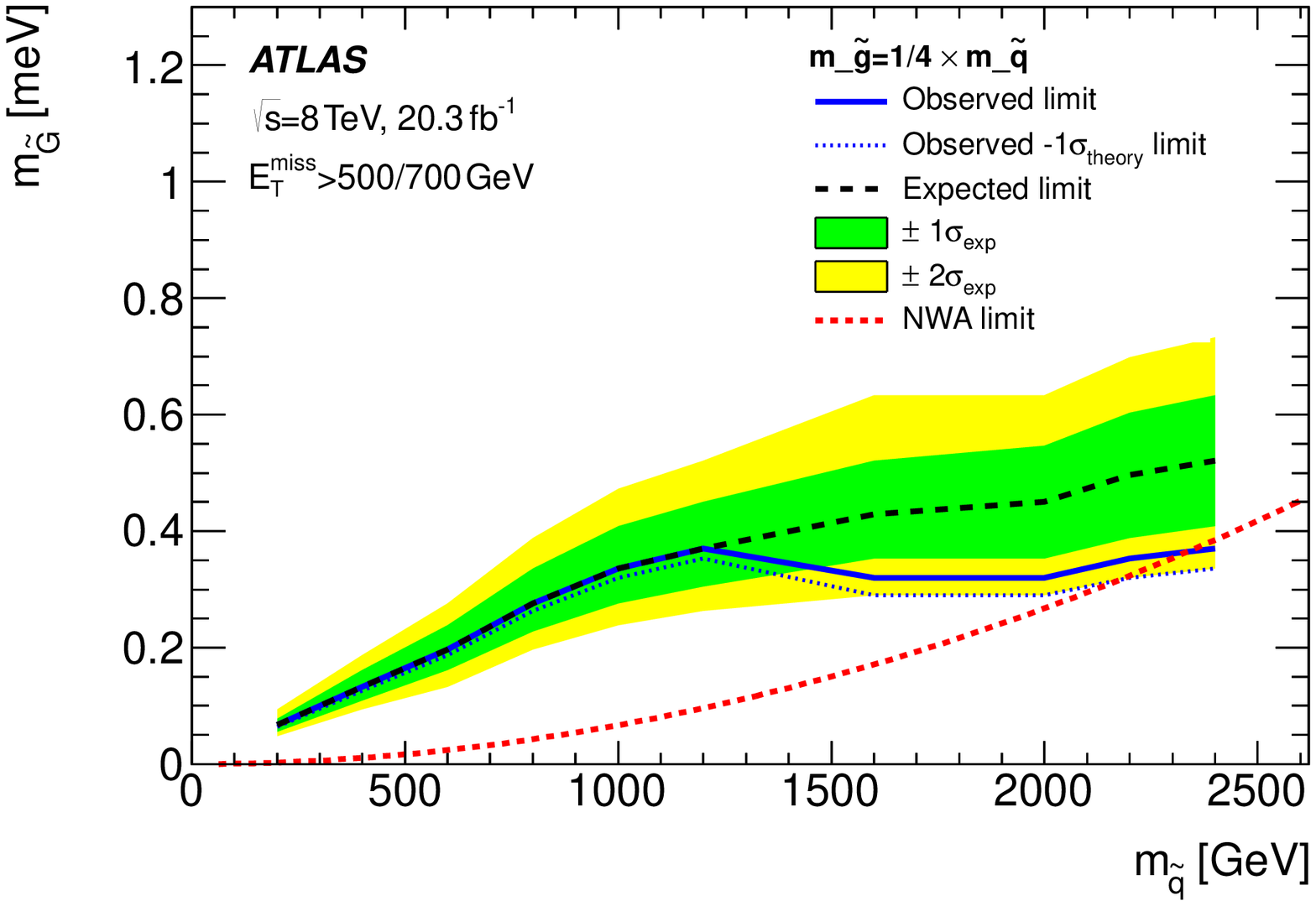}}
}
\mbox{
\subfigure[]{\includegraphics[width=0.45\linewidth]{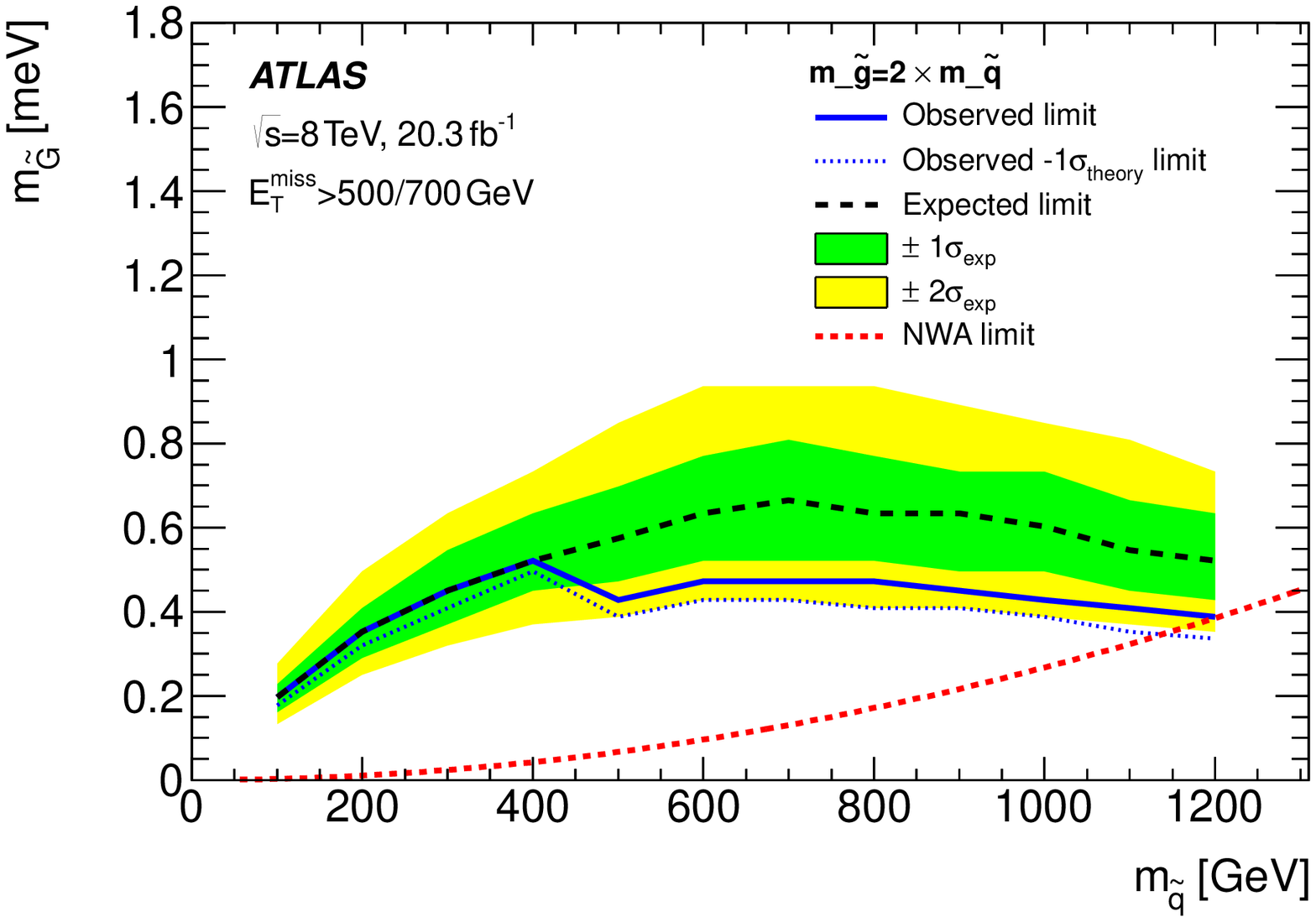}}
\subfigure[]{\includegraphics[width=0.45\linewidth]{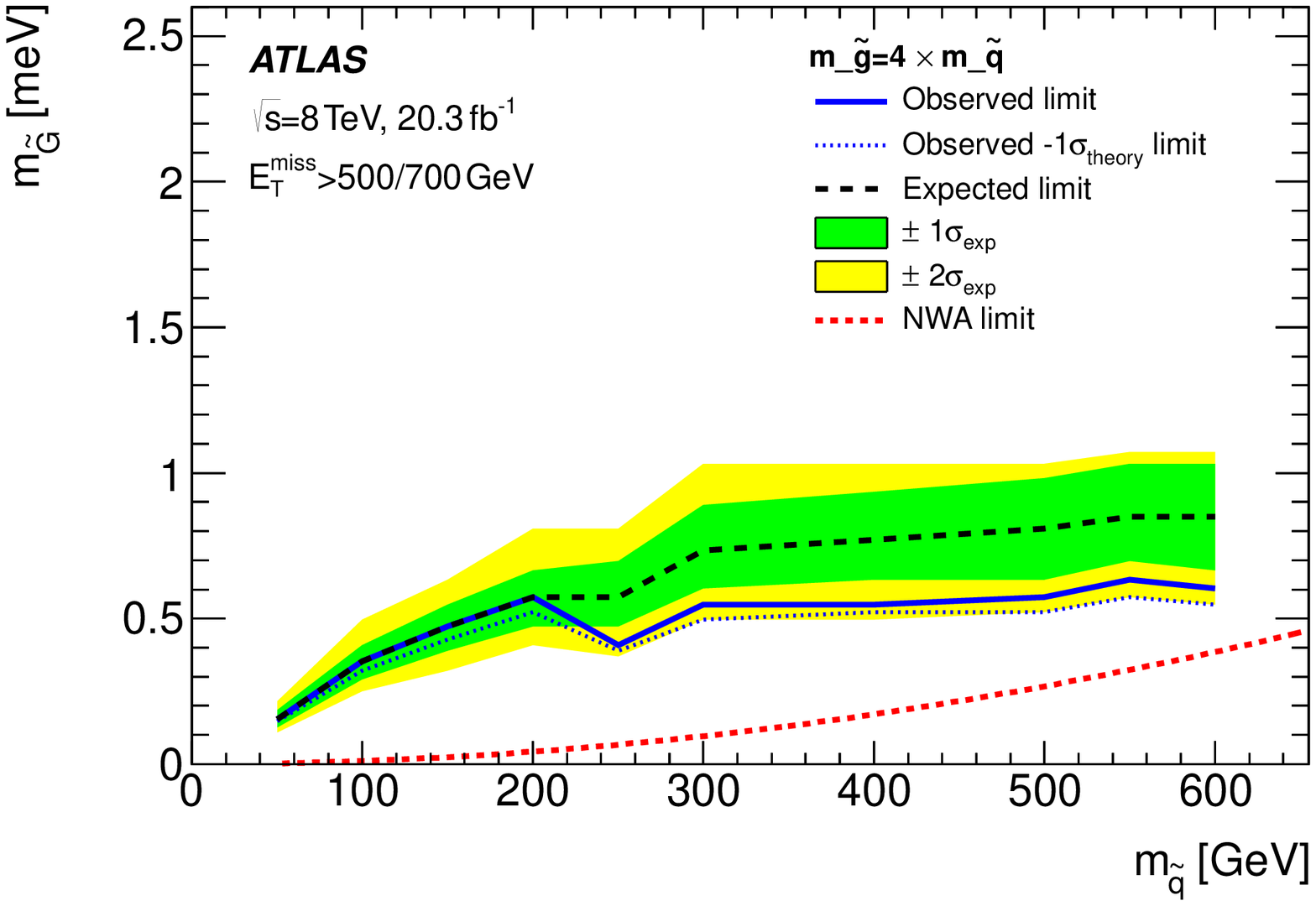}}
}
\caption{
Observed (solid line) and expected (dashed line) 95\% CL lower limits on the gravitino mass $\mgra$ as a function of the squark mass $\msqu$ for non-degenerate  squark/gluino masses
and  different squark/gluino mass configurations. 
The dotted line indicates the impact on the observed limit
of the $-1\sigma$ LO theoretical uncertainty.
The shaded bands around the expected limit indicate the expected $\pm 1\sigma$ and $\pm 2\sigma$  ranges of limits in the absence of a
signal. 
The region above the red dotted line defines the validity of the narrow-width approximation (NWA)
for which the decay
width is smaller than 25$\%$ of the squark/gluino mass.
}
\label{fig:grav_three}
\end{center}
\end{figure}

Figure~\ref{fig:grav_one} presents, for the SR7 and SR9 selections and in the case of 
degenerate squarks and gluinos, $\sigma \times A \times \epsilon$   
as a function of the squark/gluino mass for
different gravitino masses. For comparison,  the model-independent 95$\%$ CL limits  are shown.
For each SUSY point considered in the gravitino--squark/gluino mass plane, 
observed and expected 95$\%$ CL limits are computed using the same procedure as in the case of the ADD and WIMPs models.
This is done separately
for the different selections, and the one with the most stringent expected limit is 
adopted as the nominal result.  In the region with squark/gluino masses below 800~GeV,  SR7 provides
the best sensitivity while SR9 provides the most stringent expected limits for heavier squark/gluino masses.  
Figure~\ref{fig:grav_two} presents the final results.   
Gravitino masses below $3.5 \times 10^{-4}$~eV, $3 \times 10^{-4}$~eV, and $2 \times 10^{-4}$~eV 
are excluded at 95$\%$ CL for squark/gluino masses of 500~GeV, 1~TeV,  and  1.5~TeV, respectively. 
The observed limits decrease by about $9\%$--$13\%$ after considering
the $-1\sigma$ uncertainty from PDF and scale variations in the theoretical predictions.
These results are significantly better than previous  results at LEP~\cite{lepsusy} and the Tevatron~\cite{Affolder:2000ef}, 
 and constitute the most stringent bounds on the gravitino mass to date.
For very high squark/gluino masses,  the partial width
for the gluino or squark to decay into a gravitino and a parton becomes more than 25$\%$ of its mass and 
the narrow-width approximation employed is not valid any more. 
In this case,  
other decay channels for the gluino and squarks should be considered, leading to a different final state. 
The corresponding region of validity of this approximation is indicated in the figure. 
Finally, limits on the gravitino mass are also computed in the 
case of non-degenerate squarks and gluinos (see Fig.~\ref{fig:grav_three}).  
Scenarios with $\mglu = 4  \times \msqu$, $\mglu = 2  \times \msqu$, $\mglu = 1/2  \times \msqu$, and $\mglu = 1/4  \times \msqu$ 
have been considered.  
In  this case,  95$\%$  CL lower bounds on the  gravitino 
mass in the range between $1 \times 10^{-4}$~eV and $5 \times 10^{-4}$~eV are set 
depending on the squark and gluino masses.  

\subsection{Invisibly decaying Higgs-like boson}

The results are translated into 95$\%$ CL limits on the production cross section times  the branching ratio for a Higgs  
boson decaying into invisible particles as a function of the boson mass.  The SR3 selection provides  the best sensitivity to the 
signal and it is used for the final results. 
The  $A \times \epsilon$  of the selection criteria depends on the production 
mechanism and the boson mass considered.  In the case 
of the $gg \to H$ process, the $A \times \epsilon$ varies between 0.1$\%$ and 0.7$\%$ with increasing boson mass from 115~GeV to 300~GeV.   
It varies between 1$\%$ and 2$\%$ for the $VV \to H$ production process, and varies between 1$\%$ and 12$\%$ in the $VH$  case.
The $gg \to H$ process dominates the signal yield 
and constitutes more than 52$\%$ and 67$\%$ of the boson signal for a boson mass of 125~GeV and 300~GeV, respectively. 

The  uncertainties related to the jet and $\met$
scales and resolutions introduce uncertainties in the signal yields for the SR3 signal region which vary between 10$\%$ and 6$\%$
for the $gg \to H$ and $VV \to H$ processes as the boson mass increases.  Similarly, in the case of $VH$ production processes, these  
uncertainties vary between 8$\%$ and 4$\%$ with increasing mass. 
The variations of the renormalization and factorization scales 
introduce a 8$\%$ to 6$\%$, 0.2$\%$ to 0.8$\%$, and 1$\%$ to 3$\%$   uncertainty on the boson signal yields for 
$gg \to H$, $VV \to H$, and $VH$ processes, respectively, as the mass  increases.  
The uncertainties due to PDF result in uncertainties on the signal yields which vary between 
7$\%$ and 8$\%$, 2$\%$ and 4$\%$, and 2$\%$ and 4$\%$  for $gg \to H$, $VV \to H$,      and $VH$ processes, respectively.
The  uncertainty on the parton shower modelling 
results in a 7$\%$ uncertainty in the signal yields for the different channels.     

\begin{figure}[htpb]
\begin{center}
  \includegraphics[width=0.6\textwidth]{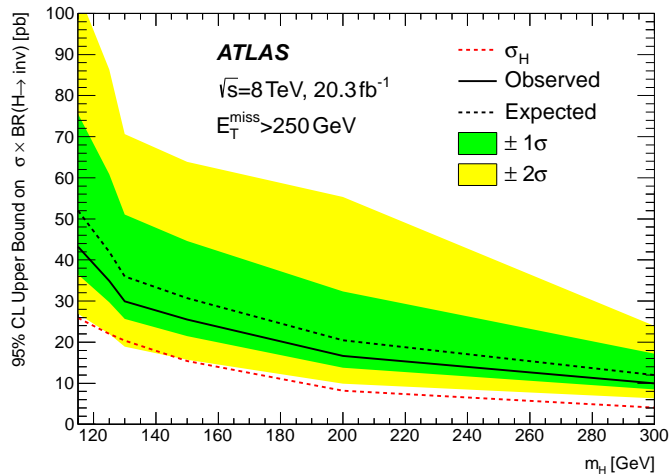}
\end{center}
\caption{
The observed (solid line) and expected (dashed line) 95$\%$ CL upper limit on 
$\sigma \times {\rm{BR}}(H \to {\rm{invisible}})$ as a function of the boson mass $\mHH$.
The shaded areas around the expected limit  
indicate the expected $\pm 1\sigma$  and $\pm 2\sigma$ ranges of limits in the absence of a signal.
The expectation for a Higgs boson with ${\rm{BR}}(H \to {\rm{invisible}}) = 1$, $\sigma_{H}$,   
is also shown. }
\label{fig:hinv}
\end{figure}

Figure~\ref{fig:hinv} shows the observed and expected 95$\%$ CL limits on the cross section times branching ratio
$\sigma \times {\rm{BR}}(H \to {\rm{invisible}})$ as a function of the  boson mass, for masses in the range between 115~GeV and
300~GeV.  Values for $\sigma \times {\rm{BR}}(H \to {\rm{invisible}})$ above  44~pb for $m_H = 115$~GeV and 10~pb
for $m_H = 300$~GeV are excluded.  This is compared with the  expectation for a Higgs  boson with ${\rm{BR}}(H \to {\rm{invisible}}) = 1$. 
For a mass of 125~GeV,  values for $\sigma \times {\rm{BR}}(H \to {\rm{invisible}})$   1.59 times larger than the SM predictions are excluded at 95$\%$ CL,
with  an expected sensitivity of 1.91 times the SM predictions. 
This indicates that, for a mass of 125~GeV, this result is less 
sensitive than that in Ref.\cite{Aad:2014iia} using $ZH (Z \to \ell^+ \ell^-)$ final states, and it
does not yet have the
sensitivity to probe the SM Higgs boson couplings to invisible particles.  
Nevertheless, for a Higgs boson mass above 200~GeV this analysis gives comparable results.

\section{Conclusions}
\label{sec:sum}
In summary, results are reported from a search for new phenomena in 
events with an energetic jet and large missing transverse momentum 
in proton--proton collisions at $\sqrt{s}=8$~TeV at the LHC, 
based on ATLAS data corresponding to an integrated luminosity of 20.3~fb${}^{-1}$. 
The measurements are in agreement with the 
SM expectations.  The results are translated into 
model-independent 90$\%$ and 95$\%$ confidence-level upper limits
on $\sigma \times A \times \epsilon$ in the range 
599--2.9~fb and 726--3.4~fb, respectively, depending on the selection criteria considered. The results are presented in terms
of limits on the fundamental Planck scale, $M_D$, versus the number of extra spatial dimensions 
in  the ADD LED model,  upper limits on the spin-independent and spin-dependent
contributions to the 
WIMP--nucleon elastic cross section as a function of the WIMP mass, and upper limits on 
the production of very light gravitinos in gauge-mediated supersymmetry.
In addition, the results are interpreted in terms of the  
production of an invisibly decaying Higgs boson 
for which the analysis shows a limited sensitivity.


\section*{Acknowledgements}


We thank CERN for the very successful operation of the LHC, as well as the
support staff from our institutions without whom ATLAS could not be
operated efficiently.

We acknowledge the support of ANPCyT, Argentina; YerPhI, Armenia; ARC,
Australia; BMWFW and FWF, Austria; ANAS, Azerbaijan; SSTC, Belarus; CNPq and FAPESP,
Brazil; NSERC, NRC and CFI, Canada; CERN; CONICYT, Chile; CAS, MOST and NSFC,
China; COLCIENCIAS, Colombia; MSMT CR, MPO CR and VSC CR, Czech Republic;
DNRF, DNSRC and Lundbeck Foundation, Denmark; EPLANET, ERC and NSRF, European Union;
IN2P3-CNRS, CEA-DSM/IRFU, France; GNSF, Georgia; BMBF, DFG, HGF, MPG and AvH
Foundation, Germany; GSRT and NSRF, Greece; ISF, MINERVA, GIF, I-CORE and Benoziyo Center,
Israel; INFN, Italy; MEXT and JSPS, Japan; CNRST, Morocco; FOM and NWO,
Netherlands; BRF and RCN, Norway; MNiSW and NCN, Poland; GRICES and FCT, Portugal; MNE/IFA, Romania; 
MES of Russia and ROSATOM, Russian Federation; JINR; MSTD, Serbia; MSSR, Slovakia; ARRS and MIZ\v{S}, Slovenia; DST/NRF, South Africa;
MINECO, Spain; SRC and Wallenberg Foundation, Sweden; SER, SNSF and Cantons of
Bern and Geneva, Switzerland; NSC, Taiwan; TAEK, Turkey; STFC, the Royal
Society and Leverhulme Trust, United Kingdom; DOE and NSF, United States of
America.

The crucial computing support from all WLCG partners is acknowledged
gratefully, in particular from CERN and the ATLAS Tier-1 facilities at
TRIUMF (Canada), NDGF (Denmark, Norway, Sweden), CC-IN2P3 (France),
KIT/GridKA (Germany), INFN-CNAF (Italy), NL-T1 (Netherlands), PIC (Spain),
ASGC (Taiwan), RAL (UK) and BNL (USA) and in the Tier-2 facilities
worldwide.

\clearpage
\bibliographystyle{atlasnote.bst}
\bibliography{mono}

\clearpage
\appendix
\section{On the validity of the effective field theory used to
  describe dark-matter pair production}
\label{sec:valid}

\newcommand\Qtr{\ensuremath{Q_{\rm{tr}}}}
\newcommand\sHat{\ensuremath{\sqrt{\hat{s}}}}
\newcommand\gSM{\ensuremath{g_\mathrm{q}}}
\newcommand\gDM{\ensuremath{g_\chi}}
\newcommand\ySM{\ensuremath{y_\mathrm{q}}}
\newcommand\aSM{\ensuremath{a}}
\newcommand\lDM{\ensuremath{\lambda_\chi}}
\newcommand\vevT{\ensuremath{\nu_\lambda}}
\newcommand\vevScaleT{\ensuremath{\zeta_\lambda}}
\newcommand\sqrtGG{\ensuremath{\sqrt{\gSM{}\gDM{}}}}
\newcommand\sqrtYG{\ensuremath{\sqrt{\ySM{}\gDM{}}}}
\newcommand\RtotI{\ensuremath{R_{\MMed{}}^\mathrm{\,i}}}
\newcommand\Rtot{\ensuremath{R_{\MMed{}}^\mathrm{\,tot}}}

\subsection{Introduction}

The effective field theories (EFTs) used here are based on the
assumption that a new mediator particle couples Standard Model
particles to pairs of DM particles and that the mediator  particle
mass is considerably larger than the energy scale of the interaction.
In such a case the mediator cannot be produced directly in LHC
collisions and can be integrated out with an EFT formalism. This
heavy-mediator assumption is indeed justifiable in \emph{direct
  detection} WIMP scattering experiments due to the very low momentum
exchange typically of order keV in the scattering interactions. 
This assumption is not always correct at the LHC, where the momentum
transfer reaches the \TeV{} scale~\cite{Buchmueller:2013dya,Busoni:2013lha,Busoni:2014sya,Busoni:2014haa}.

A minimal condition for the EFT to be valid is that the momentum transferred in 
the hard interaction at the LHC does not exceed the mediator
particle mass, thus ensuring that the mediator cannot be produced
directly: $\Qtr{} < \MMed{}$.
To probe this validity, further assumptions have to be made about the
actual form of the interaction vertex, and thereby about the (unknown)
interaction structure itself, connecting quarks or gluons to WIMPs.

The simplest of such assumptions are made below for all the operators
used here to derive expressions for \MMed{}, \MS{}, and the
interaction coupling constants, to probe the minimal validity
criterion.

\subsection{Connecting \MS{} to \MMed{}}
\label{sec:validity_mediators}

The simplest interaction structure for the operators D5, D8, and D9 is
an $s$-channel diagram, where the mediator particle couples to the
initial-state quarks and the final-state WIMPs. This interaction is
described by three parameters, the mediator mass \MMed{}, the
quark--mediator coupling constant \gSM{}, and the mediator--WIMP
coupling constant \gDM{}. The relation of these parameters to \MS{}
is 
$$\MMed{} = \sqrtGG{}\MS{} \ .$$   
 
The theory is no longer in the perturbative regime if the couplings 
are outside of the range $0<\sqrtGG{}<4\pi$. 

The simplest $s$-channel diagram for the operators D1 and C1 
involves the exchange of a scalar mediator particle where
the quark--mediator coupling constant is a Yukawa coupling
\ySM{}. In this case, the mediator particle masses can be expressed as:

\begin{align*}
  & \bf D1 & & \bf C1
  \\
  \frac{m_q}{\MS^3} &= \frac{\ySM{}\gDM{}}{\MMed^2} & \frac{m_q}{\MS^2} &= \frac{\ySM{}\lDM{}\vevT{}}{\MMed^2}
  \\
   & & & \mbox{Let }\vevT{} = \vevScaleT{}\MMed{}
  \\
  \MMed^{\rm D1} &= \sqrt{\ySM{}\gDM{}}\cdot\sqrt{\MS^3/m_q} & \MMed^{\rm C1} &= \ySM{}\lDM{}\vevScaleT{}\cdot\MS^2/m_q
\end{align*}

In the above, \lDM{} is used for scalar coupling strengths.  The
vacuum expectation value (VEV) of the trilinear scalar vertex is represented by $\vevT{}$.
The VEV is then related to the mediator mass scale by $\vevT{}=\vevScaleT{}\MMed{}$, where
the common assumption of $\vevScaleT{}\approx1$ is used.  The perturbative range is then
$0<\sqrt{\ySM{}\gDM{}}<4\pi$ for D1 and $0<\ySM{}\lDM{}\vevScaleT{}<(4\pi)^2\vevScaleT{}$
for C1.

The operators D11 and C5 describe gluons coupling to WIMPs through a
loop diagram, requiring different expressions relating \MS{} to
\MMed{}:

\begin{align*}
  & \bf D11 & & \bf C5
  \\
  \frac{\alpha_{\rm s}}{4\MS^3} &= \frac{\alpha_{\rm s}\gDM{}}{\MMed^2 \Lambda_{\rm s}} & \frac{\alpha_{\rm s}}{4\MS^2} &= \frac{\alpha_{\rm s}\lDM{}\vevT{}}{\MMed^2 \Lambda_{\rm s}}
  \\
  \MMed{} &= \sqrt[3]{\frac{4\gDM{}}{b}}\,\MS{} & \MMed{} &= \sqrt[3]{\frac{4\lDM{}\vevScaleT{}}{b}}\,\MS{}
  \\ &\mbox{Let }\aSM{}=4b^{-1} & &\mbox{Let }\aSM{}=4b^{-1}
  \\
  \MMed^{\rm D11} &= \sqrt[3]{\aSM{}\gDM{}}\,\MS{} & \MMed^{\rm C5} &= \sqrt{\aSM{}\lDM{}\vevScaleT{}}\,\MS{}
\end{align*}

$\Lambda_{\rm s} = b\MMed{}$ ($b>1$) is another mass suppression scale of
the loop connected to the initial-state gluons. The coupling terms
differ from the other operators, as $b>1 \ \implies \ 0<\aSM{}<4$.
As before, $\vevT{}=\vevScaleT{}\MMed{}$, and the assumption of $\vevScaleT{}\approx1$
is used for C5.
The perturbative range for the gluon operators is thus
$0<\sqrt[3]{\aSM{}\gDM{}}<\sqrt[3]{16\pi}$ or $0<\sqrt{\aSM{}\lDM{}\vevScaleT{}}<4\sqrt{\pi\vevScaleT{}}$
for D11 and C5 respectively.

A summary of the different relations between \MMed{} and \MS{} for
each operator of interest and the associated coupling ranges is
provided in Table~\ref{tab:MStoMMed}.  All of the operators
have a dependence on a coupling term, where the value of these
couplings is impossible to know without knowledge of the complete
theory.  Scans over the coupling-parameter space are therefore
performed below to quantify valid phase-space regions.

While these relations were derived for $s$-channel completions,
similar validity arguments can be applied to the $t$-channel as a sum
of $s$-channel operators (see Ref.~\cite{Busoni:2014haa} for further
details and caveats).

\begin{table}
  \caption{Relations between the mediator mass \MMed{} and the
    suppression scale \MS{} for the simplest interaction vertices
    matching the EFT operators considered here.}
  \centering
  \begin{tabular}{c c c} \\ \hline
    Operator(s) & Relation between \MMed{} and \MS{} & Coupling term range
    \\\hline
    D1 & $\MMed{} = \sqrt{\ySM{}\gDM{}}\ \sqrt{\MS^3/m_q}$ & $0<\sqrt{\ySM{}\gDM{}}<4\pi$
    \\
    C1 & $\MMed{} = \ySM{}\lDM{}\vevScaleT{}\ \MS^2/m_q$ & $0<\ySM{}\lDM{}\vevScaleT{}<(4\pi)^2\vevScaleT{}$
    \\
    D5, D8, D9 & $\MMed{} = \sqrtGG{}\ \MS{}$ & $0<\sqrtGG{}<4\pi$
    \\
    D11 & $\MMed{} = \sqrt[3]{a\gDM{}}\ \MS{}$ & $0<\sqrt[3]{a\gDM{}}<\sqrt[3]{16\pi}$
    \\
    C5 & $\MMed{} = \sqrt{a\lDM{}\vevScaleT{}}\ \MS{}$ & $0<\sqrt{a\lDM{}\vevScaleT{}}<4\sqrt{\pi\vevScaleT{}}$ \\ \hline
  \end{tabular}
  \label{tab:MStoMMed}
\end{table}

\subsection{Regions of Validity}
\label{sec:validity_determination}

Given a relation between the mediator mass and the suppression scale,
$\Qtr{} < \MMed{}$ can be evaluated and the fraction of events
fulfilling this validity criterion can be determined. Two different
procedures are then followed (which were shown to yield the same
results). The nominal procedure is a simple truncation, in which the
signal cross section is rescaled by the fraction of valid events. With
this truncated signal cross section, new valid limits on the
mass suppression scale are derived, \MSvalid{}.

The second alternative procedure used to cross-check the simple
truncation is an iterative procedure that scans through \MS{} until a
convergence point is reached. 
\begin{enumerate}
\item The starting point is the nominal expected limit on \MS{}
  assuming 100\% validity, named $\MSexp{}$. $\MSexp{}$ is set to
  $\MS^\mathrm{in}$ before executing step 2 for the first time.
\item For each step $i$, obtain the relative fraction of valid events
  $\RtotI{}$ satisfying $\Qtr{} < \MMed^\mathrm{in}$, where
  $\MMed^\mathrm{in}$ is the mediator mass limit obtained in the
  previous step (depending on $\MS^\mathrm{in}$).
\item Truncate \MS{} following Ref.~\cite{Busoni:2014sya}:
  $\displaystyle \MS^\mathrm{out} =
  \left[R_{\MMed{}}^{\,i}\right]^{1/2(d-4)}\MS^\mathrm{in}$, noting
  that D1 and D11 are dimension $d = 7$ operators, while D5, D8, D9, C1,
  and C5 are dimension $d = 6$.

\item Go to step 2, using the current $\MS^\mathrm{out}$ as the new
  $\MS^\mathrm{in}$, repeating until the fraction of valid events at a given
  step $\RtotI{}$ reaches 0 or 1.
\item Calculate the total validity fraction $\displaystyle
  \Rtot{} = \prod_i \RtotI{}$ and the truncated limit on the suppression scale
  $\displaystyle \MSvalid{} = \left[\Rtot{}\right]^{1/2(d-4)}\MSexp{}$.
\end{enumerate}

\begin{figure}
  \centering
\subfigure[]{  \includegraphics[width=0.485\textwidth]{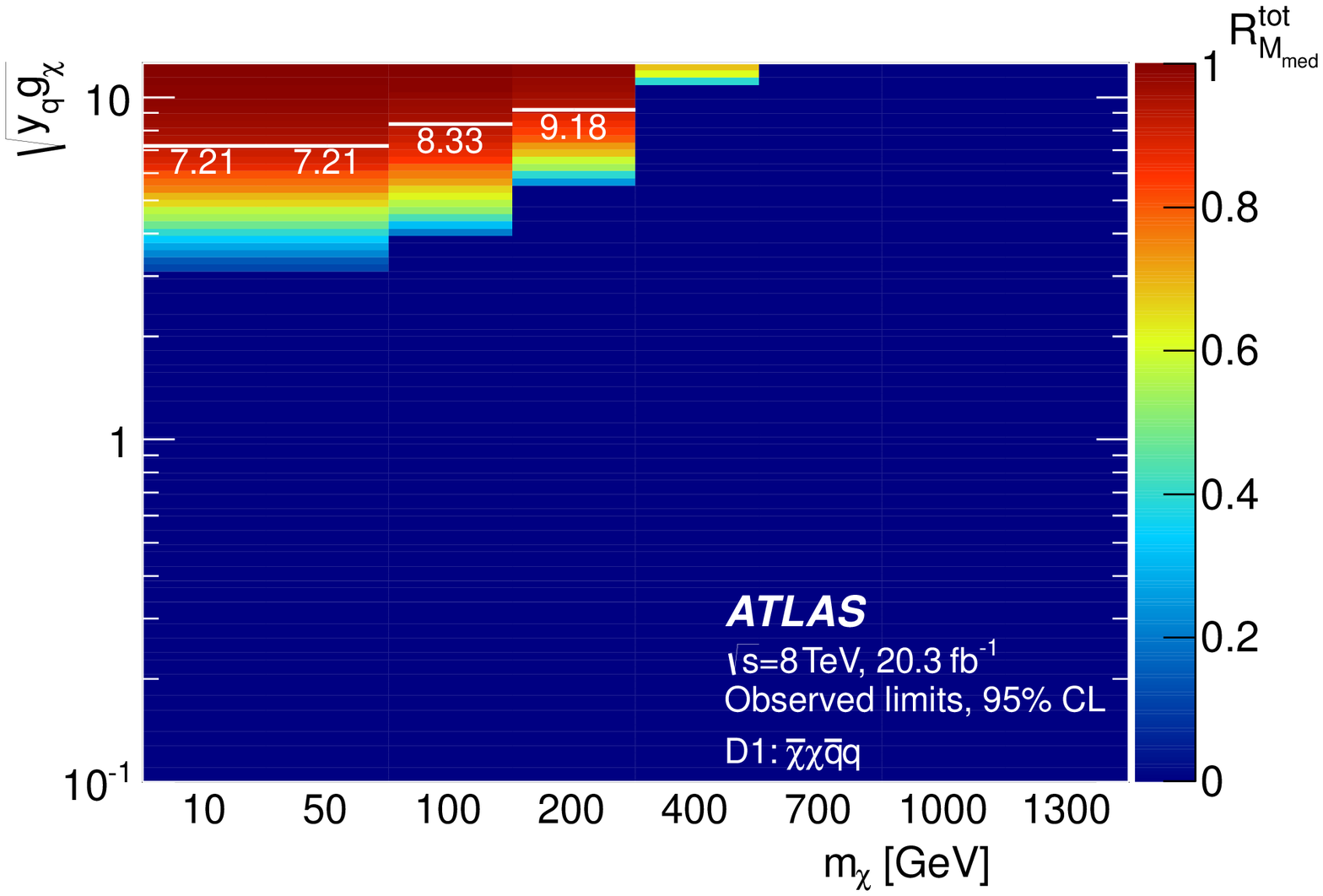}}
\subfigure[]{  \includegraphics[width=0.485\textwidth]{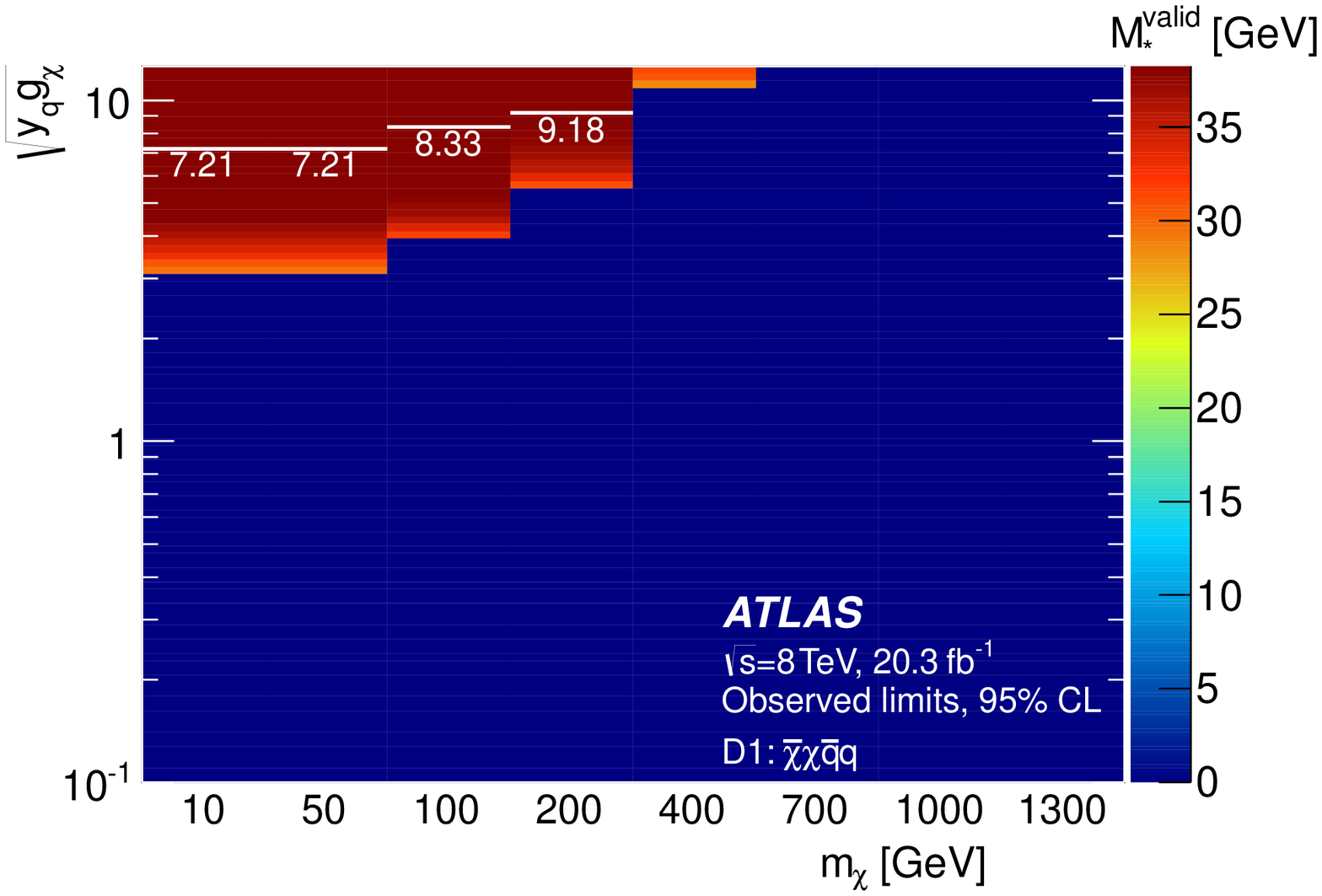}}
  \caption{(a) The fraction of valid events  and (b) truncated
    limits for D1 at 95\% CL as a function of the WIMP mass
    $m_\chi$ and couplings. The white numbers correspond to
    the minimum coupling value for which
    $M_{*}^{\mathrm{valid}}/M_{*}^{\mathrm{exp}} > 99\%$.  The upper
    perturbative coupling limit for D1 is $4\pi$.  
}
  \label{fig:validity_D1}
\end{figure}

\begin{figure}
  \centering
\subfigure[]{  \includegraphics[width=0.485\textwidth]{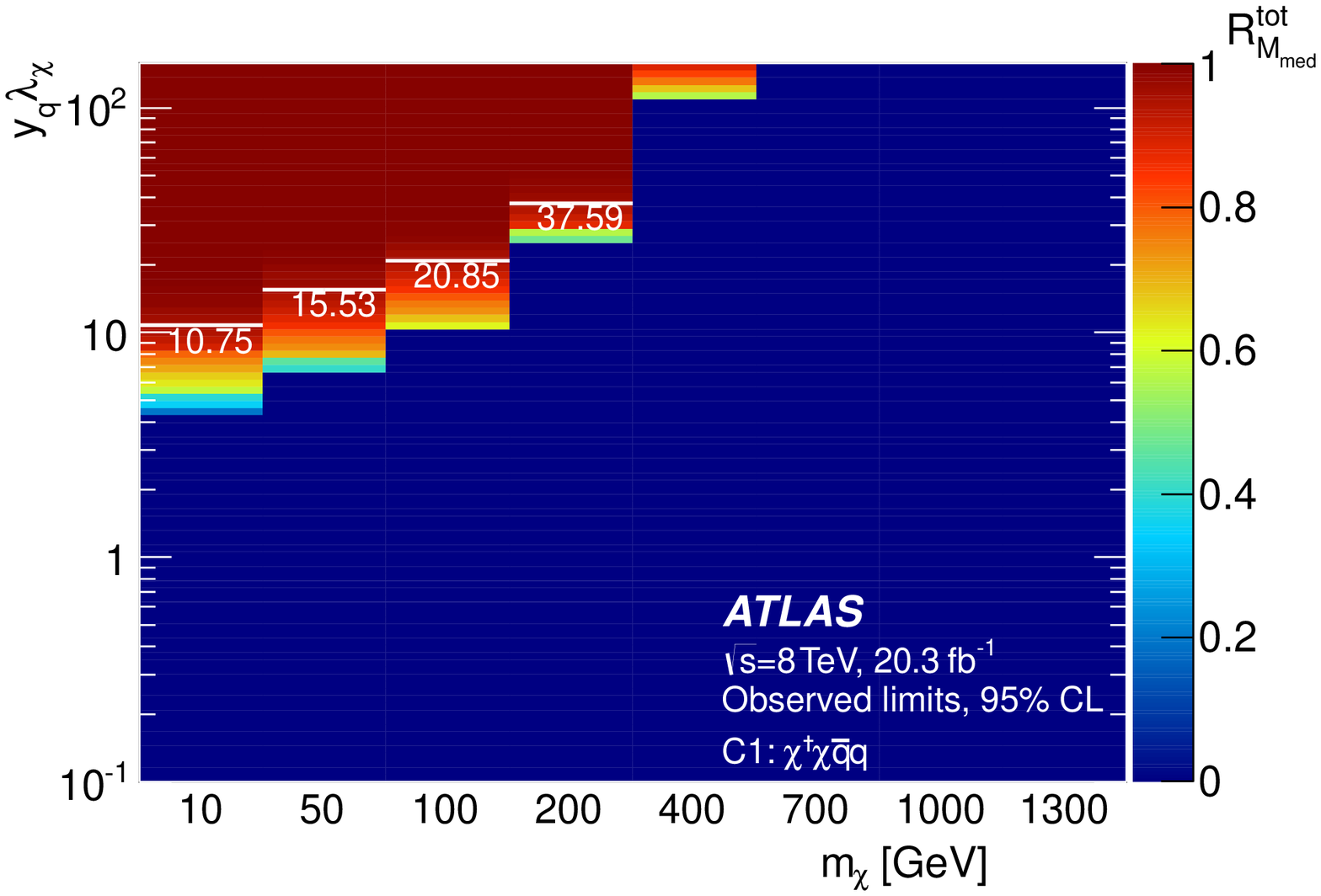}}
\subfigure[]{  \includegraphics[width=0.485\textwidth]{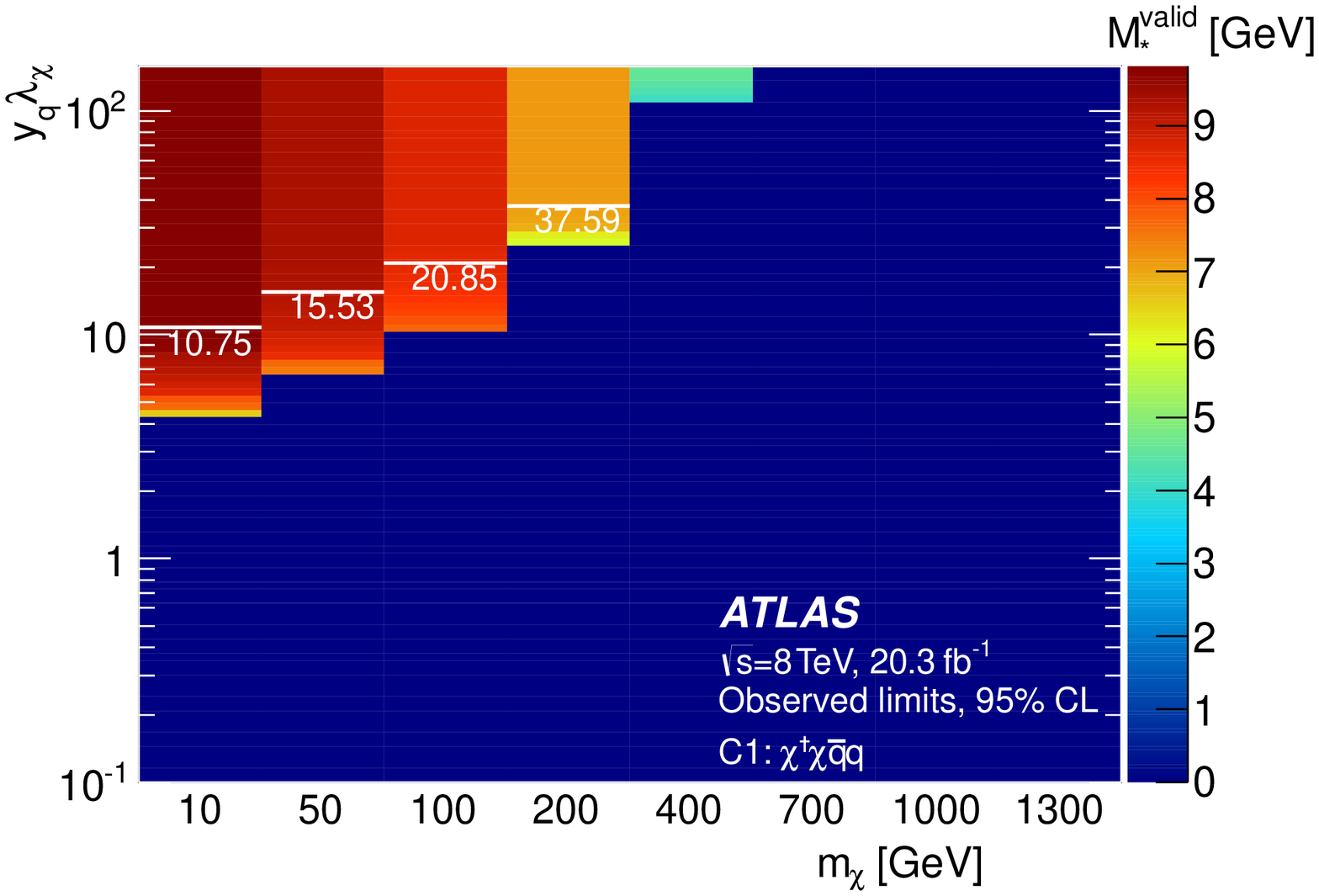}}
  \caption{(a) The fraction of valid events  and (b) truncated
    limits  for C1 at 95\% CL as a function of the WIMP mass
    $m_\chi$ and couplings. The white numbers correspond to
    the minimum coupling value for which
    $M_{*}^{\mathrm{valid}}/M_{*}^{\mathrm{exp}} > 99\%$.  The upper
    perturbative coupling limit for C1 is $(4\pi)^2\vevScaleT{}$, 
    where \vevScaleT{} is taken to be 1.  
}
  \label{fig:validity_C1}
\end{figure}

The fraction of valid events and the truncated limits on \MS{} can be
used to assess the validity of the EFT approach. In
Figs.~\ref{fig:validity_D1} and \ref{fig:validity_C1}, this is shown for D1 and C1. 
The majority of the parameter space is invalid.
The operators D1 and C1 are still 
    valid for regions of parameter space with large coupling values
    and low WIMP masses.

\begin{figure}
  \centering
\subfigure[]{  \includegraphics[width=0.485\textwidth]{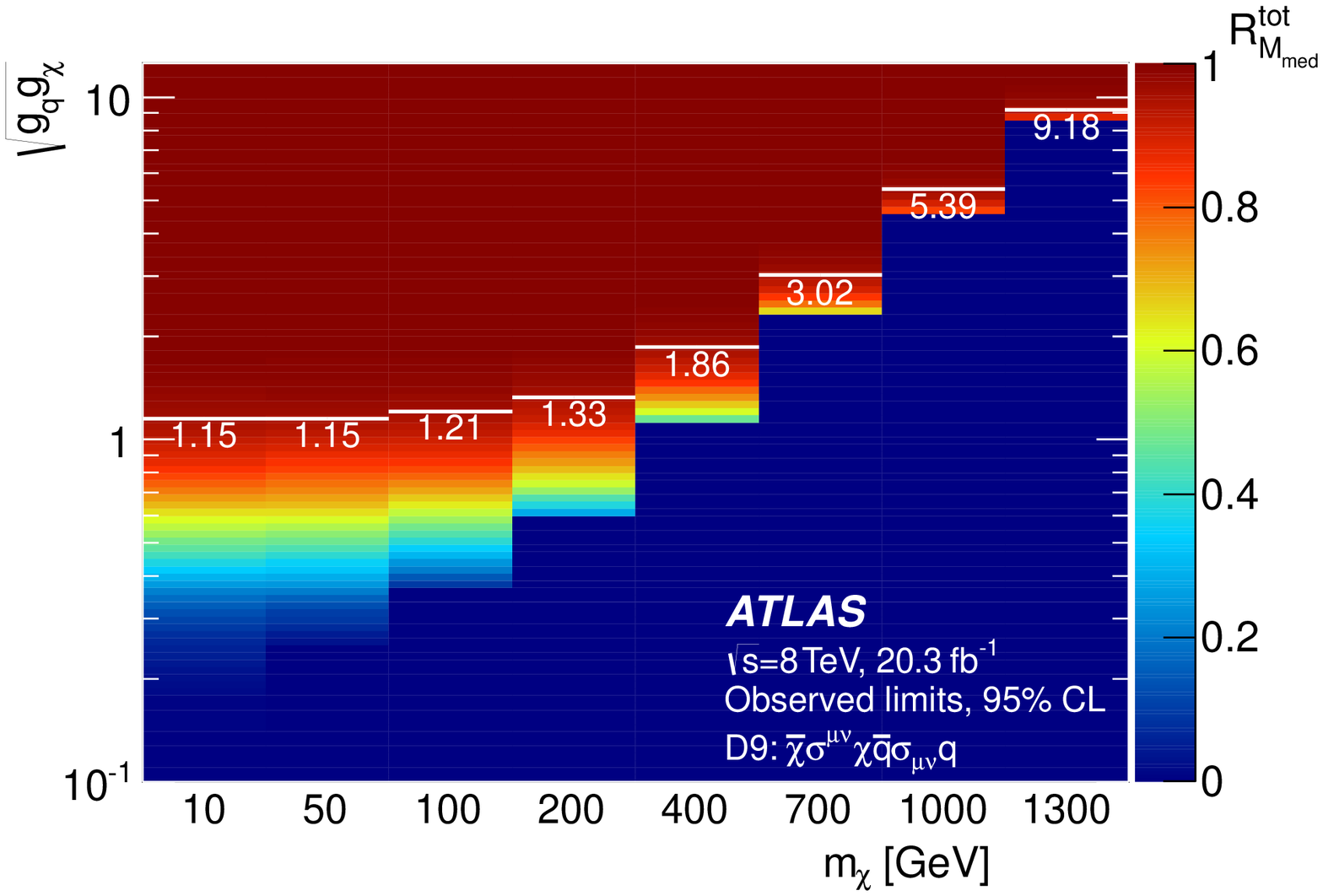}}
\subfigure[]{  \includegraphics[width=0.485\textwidth]{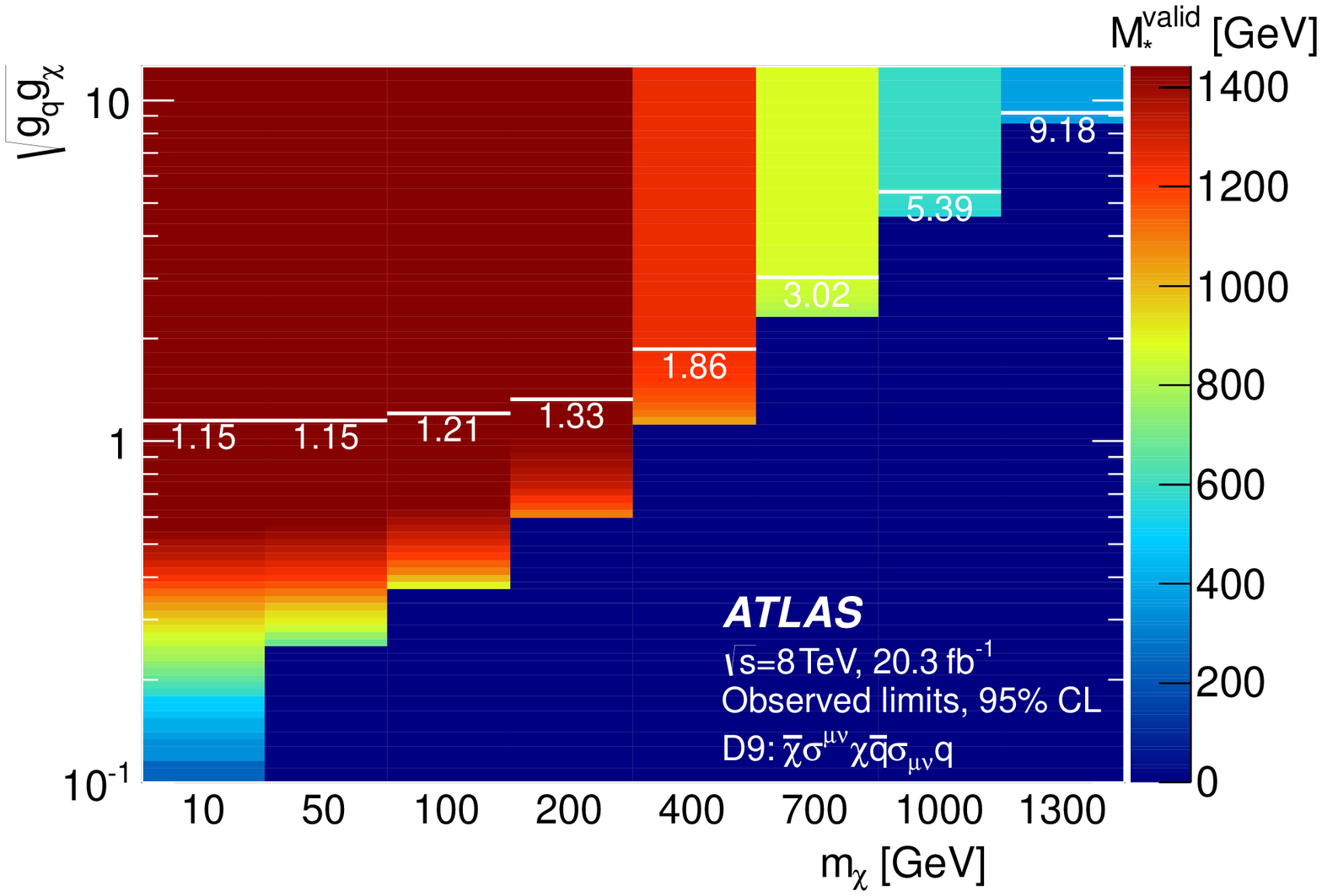}}
  \caption{(a) The fraction of valid events and (b) truncated
    limits for D9 at 95\% CL as a function of the WIMP mass
    $m_\chi$ and couplings. The white numbers correspond to
    the minimum coupling value for which
    $M_{*}^{\mathrm{valid}}/M_{*}^{\mathrm{exp}} > 99\%$. The upper
    perturbative coupling limit for D9 is $4\pi$.  
}
  \label{fig:validity_D9}
\end{figure}

The validity of the vector, axial-vector, and tensor couplings to
quarks via the D5, D8, and D9 operators, respectively, are much more justifiable, as
shown for D9 in Fig. \ref{fig:validity_D9}. 
 The operator D9 is
    valid for the majority of parameter space, across couplings and
    WIMP masses, except for the highest values of \mDM{} considered.
While only D9 is valid for
the common canonical choice of $\gSM{}=\gDM{}=1$, the other two
operators are valid for only slightly larger couplings.

\begin{figure}
  \centering
\subfigure[]{  \includegraphics[width=0.48\textwidth]{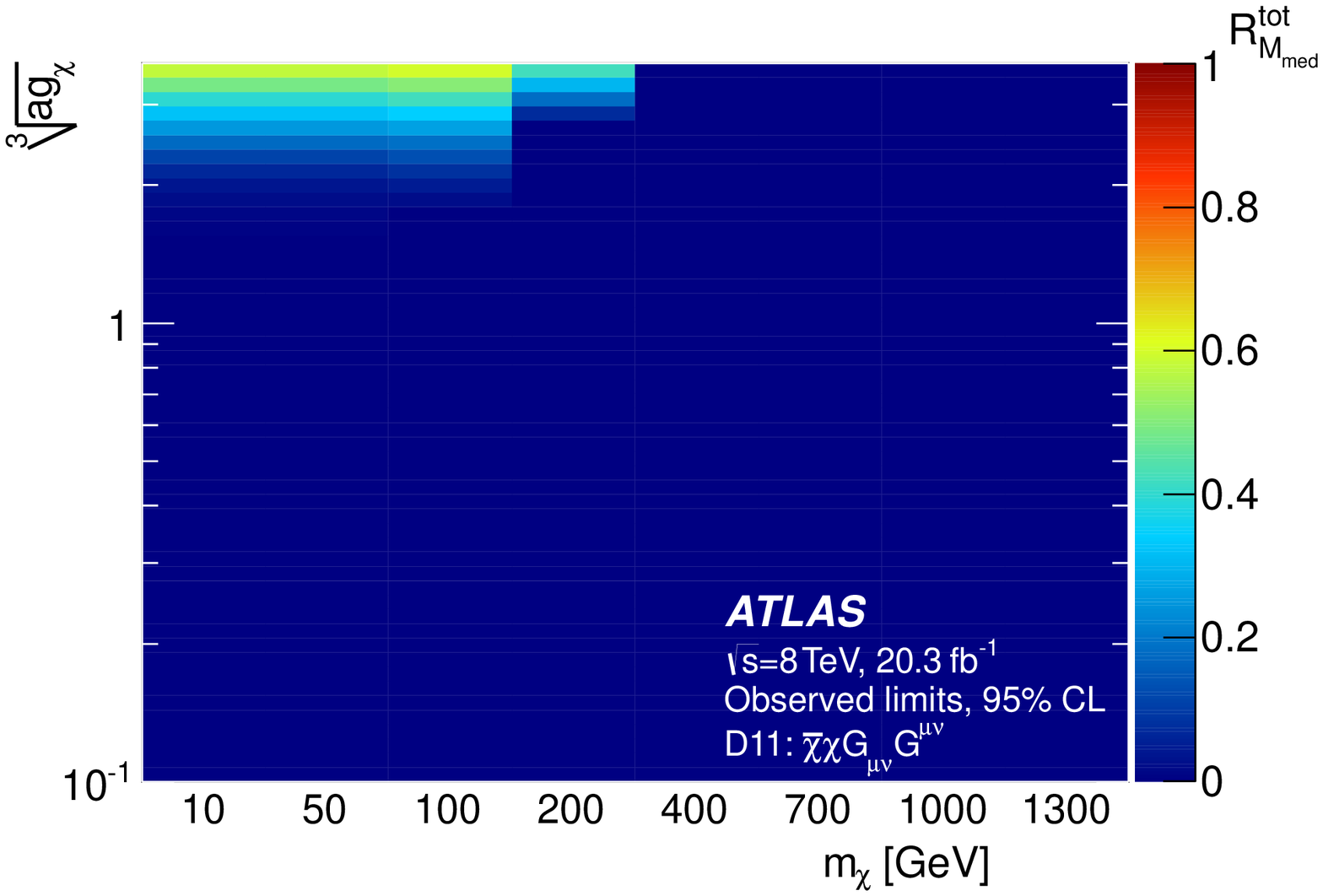}}
\subfigure[]{  \includegraphics[width=0.48\textwidth]{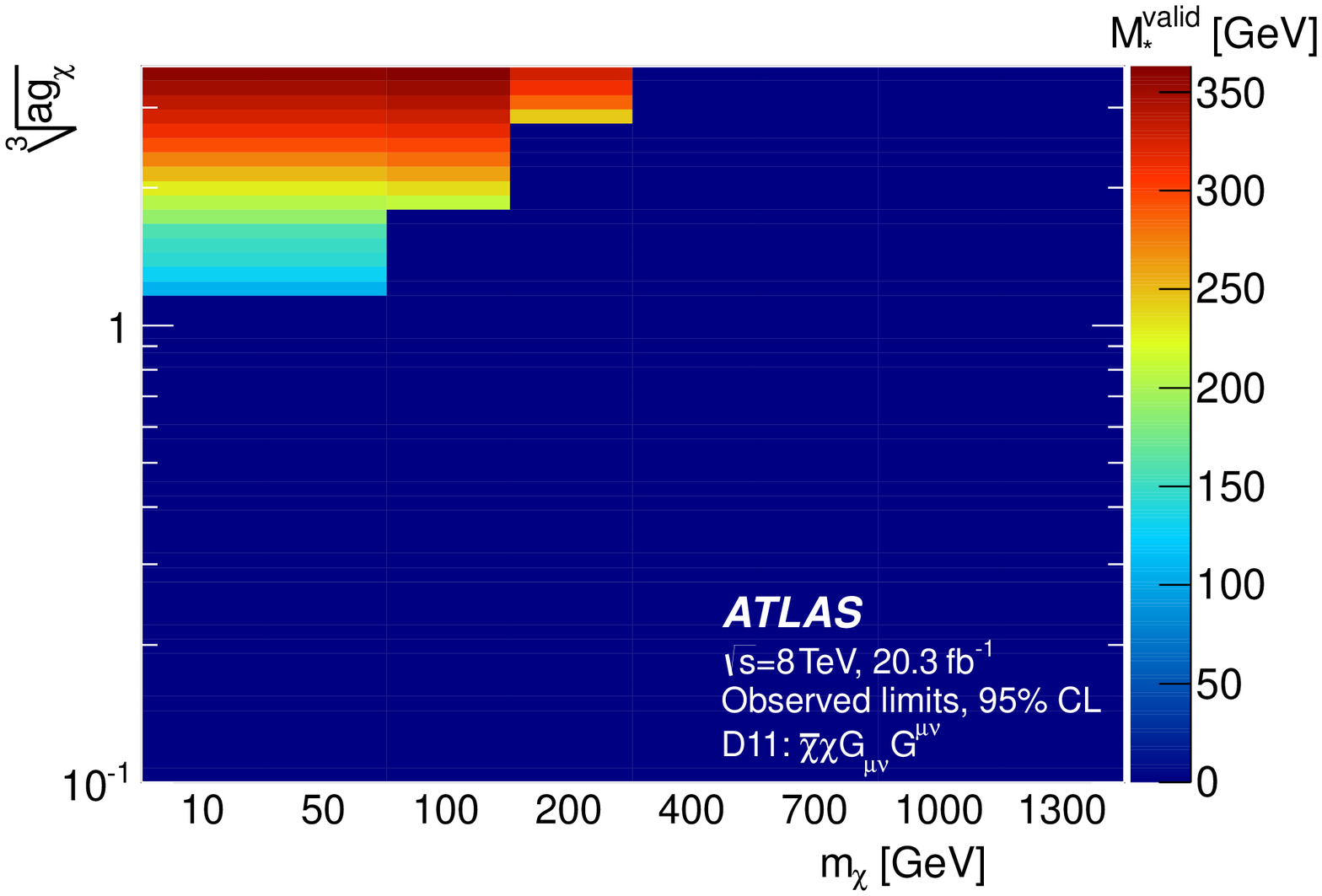}}
  \caption{(a) The fraction of valid events and (b) truncated
    limits for D11 at 95\% CL as a function of the WIMP mass
    $m_\chi$ and couplings. The white numbers correspond to
    the minimum coupling value for which
    $M_{*}^{\mathrm{valid}}/M_{*}^{\mathrm{exp}} > 99\%$. The upper
    perturbative coupling limit for D11 is $\sqrt[3]{16\pi}$.  
}
  \label{fig:validity_D11}
\end{figure}

\begin{figure}
  \centering
\subfigure[]{  \includegraphics[width=0.48\textwidth]{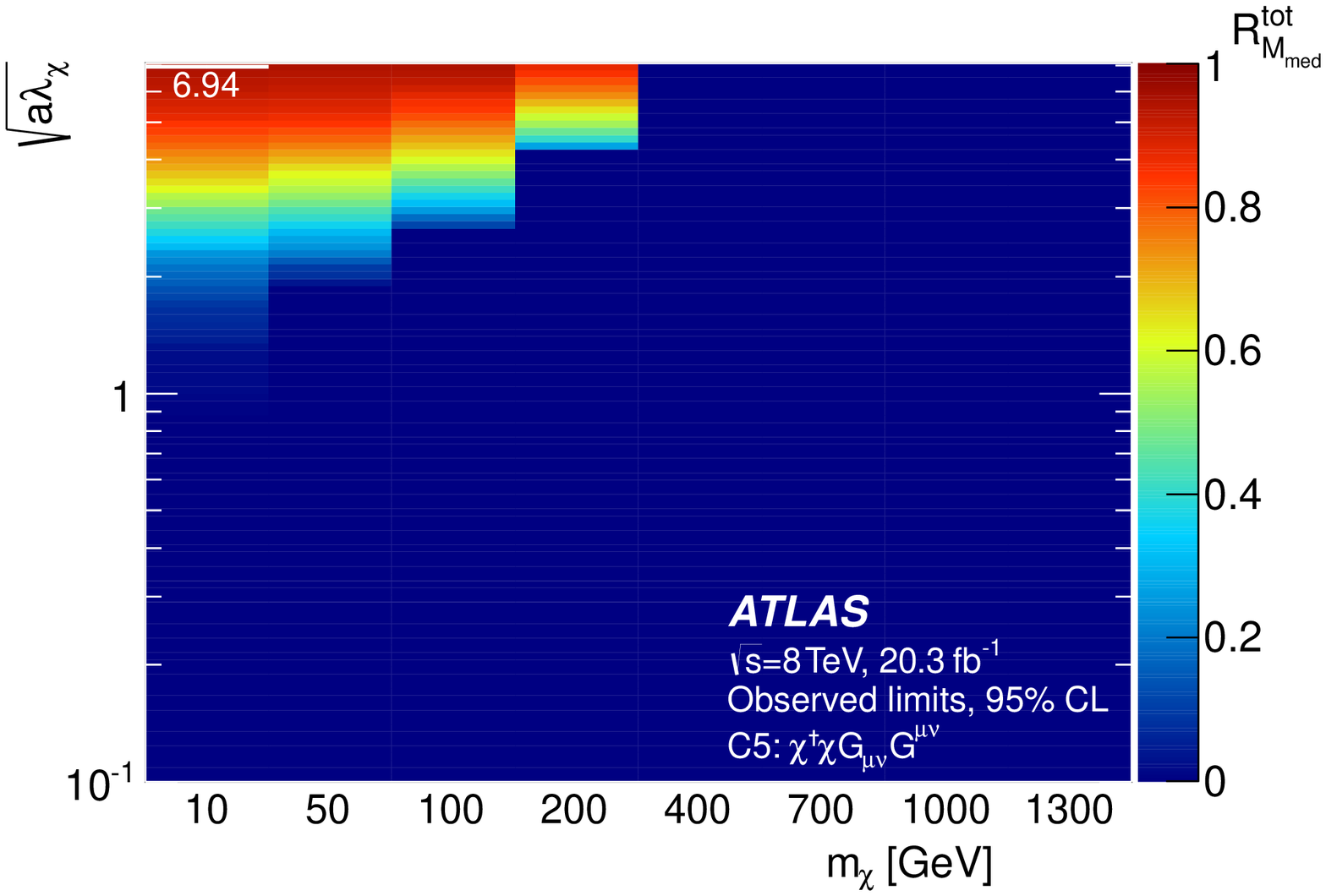}}
\subfigure[]{  \includegraphics[width=0.48\textwidth]{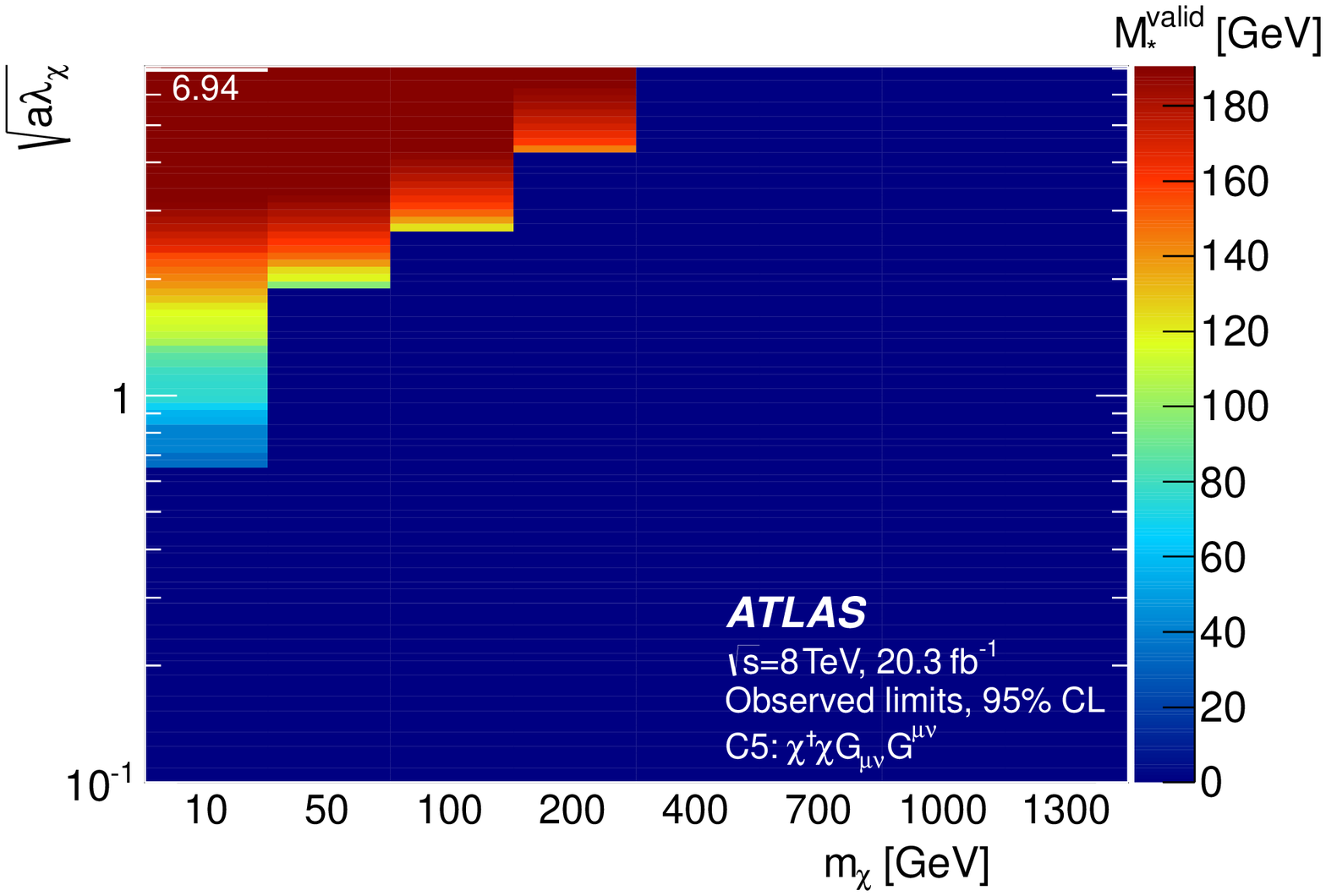}}
  \caption{(a) The fraction of valid events and (b) truncated
    limits for C5 at 95\% CL as a function of the WIMP mass
    $m_\chi$ and couplings. The white numbers correspond to
    the minimum coupling value for which
    $M_{*}^{\mathrm{valid}}/M_{*}^{\mathrm{exp}} > 99\%$. The upper
    perturbative coupling limit for C5 is $4\sqrt{\pi\vevScaleT{}}$, 
    where \vevScaleT{} is taken to be 1.  
}
  \label{fig:validity_C5}
\end{figure}

An assessment of the validity of the gluon EFT operators requires the
most assumptions, and has a very different coupling range under the
assumptions discussed in~\ref{sec:validity_mediators}.  Under
these assumptions, D11 and C5 operators are  valid for regions of parameter space with large
    coupling values and low WIMP masses, as 
shown in Figs.~\ref{fig:validity_D11} and \ref{fig:validity_C5}, 
respectively. 

In general, the validity of the EFT operators is better for low WIMP
masses.  This is important, as collider searches are most competitive
with other types of experiments at low \mDM{}.  Additionally,
Figs.~\ref{fig:validity_D1}--\ref{fig:validity_C5} 
show how the truncated limit \MSvalid{} quickly
approaches the nominal limit \MSexp{}.  
Some operators have larger validity regions than others because the \MS{}
limits are larger, and it is thus more likely that $\Qtr{} <
\MMed{}$.  Stronger limits therefore remain strong, while weak limits
are in fact even further diminished by validity considerations.

Truncated limits are more conservative than the corresponding
simplified model used for the completion, so long as \MMed{} in the
model is greater than or equal to the value used for the truncation.
This can be seen comparing the D5 operator in Fig.~\ref{fig:wimp:ms}
with the corresponding simplified model in Fig.~\ref{fig:wimp:simpl}.

%
%


\clearpage 
\begin{flushleft}
{\Large The ATLAS Collaboration}

\bigskip

G.~Aad$^{\rm 85}$,
B.~Abbott$^{\rm 113}$,
J.~Abdallah$^{\rm 152}$,
S.~Abdel~Khalek$^{\rm 117}$,
O.~Abdinov$^{\rm 11}$,
R.~Aben$^{\rm 107}$,
B.~Abi$^{\rm 114}$,
M.~Abolins$^{\rm 90}$,
O.S.~AbouZeid$^{\rm 159}$,
H.~Abramowicz$^{\rm 154}$
% ...
\bigskip
\\
% ...
$^{11}$ Institute of Physics, Azerbaijan Academy of Sciences, Baku, Azerbaijan\\
% ...
$^{85}$ CPPM, Aix-Marseille Universit{\'e} and CNRS/IN2P3, Marseille, France\\
% ...
$^{90}$ Department of Physics and Astronomy, Michigan State University, East Lansing MI, United States of America\\
% ...
$^{107}$ Nikhef National Institute for Subatomic Physics and University of Amsterdam, Amsterdam, Netherlands\\
% ...
$^{113}$ Homer L. Dodge Department of Physics and Astronomy, University of Oklahoma, Norman OK, United States of America\\
$^{114}$ Department of Physics, Oklahoma State University, Stillwater OK, United States of America\\
% ...
$^{117}$ LAL, Universit{\'e} Paris-Sud and CNRS/IN2P3, Orsay, France\\
% ...
$^{152}$ Institute of Physics, Academia Sinica, Taipei, Taiwan\\
% ...
$^{154}$ Raymond and Beverly Sackler School of Physics and Astronomy, Tel Aviv University, Tel Aviv, Israel\\
% ...
$^{159}$ Department of Physics, University of Toronto, Toronto ON, Canada\\
% ...
\end{flushleft}


\end{document}